\documentclass[11pt]{article}
\usepackage[english]{babel}
\usepackage[latin1]{inputenc}
\usepackage{latexsym,amsfonts,amsmath,amssymb,euscript,epsfig,dsfont,graphicx,color,float}
\usepackage[active]{srcltx}
\usepackage[T1]{fontenc}
\title{ Approachability, Regret and Calibration \\ implications and equivalences. }
\author{Vianney Perchet}

 \DeclareMathOperator{\argmax}{argmax}\DeclareMathOperator{\argmin}{argmin}  \DeclareMathOperator{\BR}{BR}\DeclareMathOperator{\co}{co} \DeclareMathOperator{\NC}{NC} \DeclareMathOperator{\diag}{diag}  \DeclareMathOperator{\Chi}{\mathcal{X}}

\newcommand\cA{\mathcal{A}}\newcommand\cB{\mathcal{B}}\newcommand\cC{\mathcal{C}}\newcommand\cD{\mathcal{D}}\newcommand\cE{\mathcal{E}}\newcommand\cF{\mathcal{F}}\newcommand\cG{\mathcal{G}}\newcommand\cH{\mathcal{H}}\newcommand\cI{\mathcal{I}}\newcommand\cK{\mathcal{K}}\newcommand\cL{\mathcal{L}}\newcommand\cN{\mathcal{N}}\newcommand\cS{\mathcal{S}}\newcommand\cT{\mathcal{T}}\newcommand\Tau{\mathcal{T}}
\newcommand\cU{\mathcal{U}}\newcommand\cV{\mathcal{V}}\newcommand\cX{\mathcal{X}}\newcommand\cY{\mathcal{Y}}

\newcommand\bA{\mathbf{A}}\newcommand\bG{\mathbf{G}}\newcommand\bH{\mathbf{H}}\newcommand\bU{\mathbf{U}}
\newcommand\ba{\mathbf{a}}\newcommand\be{\mathbf{e}}\newcommand\bg{\mathbf{g}}\newcommand\bp{\mathbf{p}}

\newcommand\tR{\widetilde{R}}\newcommand\tU{\widetilde{U}}
\newcommand\tg{\widetilde{g}}

\newcommand\hg{\widehat{g}}\newcommand\hz{\widehat{z}}

\newcommand\oG{\overline{G}}\newcommand\oL{\overline{L}}\newcommand\oR{\overline{R}}\newcommand\oU{\overline{U}}\newcommand\oX{\overline{X}}\newcommand\oY{\overline{Y}}\newcommand\oZ{\overline{Z}}
\newcommand\oa{\overline{a}}\newcommand\ob{\overline{b}}\newcommand\og{\overline{g}}\newcommand\oh{\overline{h}}\newcommand\op{\overline{p}}\newcommand\oq{\overline{q}}\newcommand\orr{\overline{r}}\newcommand\os{\overline{s}}\newcommand\oz{\overline{z}}

\newcommand\N{\mathds{N}}\newcommand\R{\mathds{R}}\newcommand\Z{\mathds{Z}}\renewcommand\P{\mathds{P}}\newcommand\E{\mathds{E}}

\newcommand\qed{$\hfill \blacksquare$ \vspace{1cm}}
\newcommand\LT{L_2(\Omega, \mu, \cF)}

\begin{document}
\maketitle

\newcounter{theorem} \newcounter{hypothese}\newcounter{proposition}\newcounter{example}\newcounter{definition}
\newtheorem{proposition}[proposition]{Proposition}
\newtheorem{theorem}[theorem]{Theorem}
\newtheorem{lemma}[proposition]{Lemma}
\newtheorem{corollary}[proposition]{Corollary}
\newtheorem{hypo}[hypothese]{Assumption}
\newtheorem{definition}[definition]{Definition}
\newtheorem{remark}[example]{Remark}
\newtheorem{example}[example]{Example}

\numberwithin{theorem}{section}
\numberwithin{proposition}{section}
\numberwithin{hypothese}{section}
\numberwithin{example}{section}
\numberwithin{definition}{section}

\begin{abstract}
Blackwell approachability, regret minimization and calibration are three criteria evaluating a strategy (or an algorithm) in different sequential decision problems, or repeated games between a player and Nature. Although they have at first sight nothing in common, links between have been discovered: both consistent and calibrated strategies can be constructed by following,  in some auxiliary game, an approachability strategy.

We gathered famous or recent results and provide new ones in order to develop and generalize Blackwell's elegant  theory. The final goal is to show how it can be used as a basic powerful tool to exhibit a new class of intuitive algorithms, based on simple geometric properties. In order to be complete, we also prove that approachability can be seen as a byproduct of the very existence of consistent or calibrated strategies.
\end{abstract}

\subsection*{Introduction}
Sequential decision problems can be represented as repeated games between a player and Nature. At each stage the player (also called agent, decision maker or   predictor depending on the context) chooses an element of his decision set. At the same time, Nature chooses on her side a state of the world.  Those sequences of choices generate a sequence of \textsl{outcomes} that induces an overall \textsl{payoff} to the player.

The opponent is called \textsl{Nature} as we do not precise her payoff, her objectives or her rationality; absolutely no assumptions is made on her behavior, and future states of the world cannot be inferred from the past. Typically the environment is not stochastic or Bayesian but \textsl{adversarial}; for instance, Nature can represent one malignant opponent, or a set of independent (or correlated) players. A crucial requirement of these model is that a strategy of the player must be \textsl{good} (i.e., it must fulfill some exogenous \textsl{criterion}) against every possible sequence of states of the world (or simply against any strategy of Nature).

\medskip

Depending on the structure of outcomes mappings, overall objectives of the player might vary. Hannan~\cite{Han57}  studied the case where an outcome is actually a real payoff. The player's goal is to maximize his average (or cumulative) payoff. As we made no assumption on Nature's behavior, a player can not ensure to himself a given exogenous amount, unlike in traditional zero sum game where a value can be guaranteed:  assume for instance that Nature decides to give a payoff of zero (or one, minus one, etc) to the player at each stage, no matter what he does.

The criterion Hannan  introduced is called \textsl{regret} and measures the difference between the average payoff the player got and what \textsl{he would have got if he had chosen the same action repeatedly}. It is somehow  related to convex optimization (if Nature chooses repeatedly the same loss function), or more precisely to \textsl{online convex optimization}.

Main results of Hannan~\cite{Han57} are that such  a \textsl{consistent} strategy, i.e., a strategy without regret exists, and he constructed one.
This  has been widely refined and improved using different techniques and ideas by notably (providing an exhaustive list seems almost impossible as the subject has been developed by many different communities) Foster \& Vohra~\cite{FosVoh97}, Hart \& Mas-Colell~\cite{HarMas00}, Fudenberg \& Levine~\cite{FudLev99}, Lehrer~\cite{Leh03}, Auer, Cesa-Bianchi \& Gentile~\cite{AueCesGen02}, Cesa-Bianchi \& Lugosi~\cite{CesLug06} (see also references therein), Sorin~\cite{Sor08}...

\bigskip

When outcomes are \textsl{vectorial} (and not scalar) \textsl{payoffs}, the problem is closely related to \textsl{multicriteria optimization}, each coordinate representing a different sub-objective.  
Instead of  considering some exogenous convex combination of these objectives or optimizing them in a given order (to encompass this framework into the precedent one),  Blackwell~\cite{Bla56} introduced another concept. He considered that some \textsl{target set} is given and the player's goal is that the average outcome converges to it; on the contrary, Nature tries to push it away. Formally, a given closed set is \textsl{approachable}, if the player has a strategy such that the average payoffs remains, after some maybe large stage, arbitrarily closed to this target set, no matter the sequence of moves of Nature.

Blackwell's approachability theory is quite elegant as it relies on simple geometric properties. They allowed him to characterize explicitly approachable convex sets and to provide a simple sufficient approachability condition for non-convex set (such sets are called, in reference to Blackwell, \textsl{$B$-sets}). Spinat~\cite{Spi02} proved later that this was in fact almost a necessary condition. 

Maybe the first and most important use of this whole theory is due to Kohlberg~\cite{Koh75}. He constructed, using this simple tool, an optimal strategy for the uninformed player in zero-sum games with incomplete information, introduced by Aumann and Maschler~\cite{AumMas95} (see for instance Mertens, Sorin \& Zamir~\cite{MerSorZam94} and references given for more details on this subject).  Approachability gained also a recent interest, both from the game theory and machine learning community, with works of -- again non-exhaustively --  Vieille~\cite{Vie92},  Hart \& Mas-Colell~\cite{HarMas01a}, Spinat~\cite{Spi02}, Lehrer~\cite{Leh02}, Bena\"im, Hofbauer \& Sorin~\cite{BenHofSor06}, Mannor \& Shimkin~\cite{ManShi08}, Lehrer \& Solan~\cite{LehSol06,LehSol09}, As Soulaimani, Quincampoix \& Sorin~\cite{As-QuiSor09}, Mannor \& Tsitsiklis~\cite{ManTsi09}, Perchet~\cite{Per10,Per11a}, Rakhlin, Sridharan \& Tewari~\cite{RakSriTew11}, Perchet \& Quincampoix~\cite{PerQui12}...

\bigskip

Another (and the last to be considered here) criterion is \textsl{calibration}, written within this framework by Dawid~\cite{Daw82} and extended thereafter by, in many others, Foster \& Vohra~\cite{FosVoh98},  Fudenberg \& Levine~\cite{FudLev99a},  Lehrer~\cite{Leh01}, Sandroni, Smorodinsky \& Vohra~\cite{SanSmoVoh03}, Sorin~\cite{Sor08}, Perchet~\cite{Per10},  Foster, Rakhlin, Sridharan \& Tewari~\cite{FosRakSri11}, and so on.

Here, a stage outcome is not some payoff (either scalar or vectorial) but the actual \textsl{state of the world} chosen by Nature. The overall objective of the player is to predict, sequentially, the whole sequence of states  so that the average prediction and the empirical distribution of states are asymptotically arbitrarily close. Without any other restrictions, this is in fact fairly easy: one just has to predict at some stage the outcome of the precedent one.

Additional requirements can be, for instance, that predictions can only belong to some finite (yet maybe large) set and that the empirical distribution of states \textsl{on the set of stages where a specific prediction is made} is closer to this prediction than to any other possible one. An usual and celebrated example consists in a meteorologist that predicts, each day, the probability of rain the following day. Predictions belongs to 0\%, 10\%, 20\%, etc. and it is asked that  that when a meteorologist says that the probability of rain is, say, 30\%, it rains in average  between 35\% and 45\% of the times.

Oakes~\cite{Oak85} and Dawid~\cite{Daw85} proved that no deterministic algorithm can be calibrated (yet this strong assessment could be discussed) while random algorithm can, as proved by Foster \& Vohra~\cite{FosVoh98}. The existence of such algorithms can be seen as a negative result, as it claims that a strategic non-informed meteorologist can mimic an expert one (that knows the true underlying process, if it exists); a whole literature studied this aspect and  recent results are gathered into the survey of Olszewski~\cite{Ols12}. On the other hand, it can also be seen as a positive result, as it states that the long term behavior of Nature can asymptotically be predicted, and this might lead to another class of algorithms and results, as in Foster \& Vohra~\cite{FosVoh97} or Perchet~\cite{Per09, Per11b}.

\bigskip

A common feature of regret minimization and calibration is that they can be written as a specific case of approachability of a well chosen target set in some auxiliary vectorial payoff game . The first to notice this property is Blackwell~\cite{Bla56a} (this idea is already mentioned at the end of the seminal paper of Hannan~\cite{Han57} or in Luce \& Raiffa~\cite{LucRAi57}) and then by Foster~\cite{Fos99},  Hart \& Mas-Colell~\cite{HarMas01a}, Lehrer \& Solan~\cite{LehSol07}, Sorin~\cite{Sor08}, Perchet~\cite{Per10}, Mannor \& Stoltz~\cite{ManSto10}, Abernathy, Barltlett \& Hazan~\cite{AbeBarHaz11}...

\bigskip

We assumed implicitly that the player observes the sequence of states of the world; this is in fact a crucial hypothesis here, sometimes referred to as \textsl{full monitoring}. In particular, we will not consider the case of partial monitoring (or \textsl{bandit} problems), or stochastic games (where, for instance, the whole sequence of outcomes could depend of a unique choice at some stage). Those are also interesting subjects, yet far from the current scope.

\bigskip

\textbf{Objectives and Structure of the paper}.

\noindent Describing explicit \textsl{interactions} and \textsl{equivalences} between the notions of approachability, calibration and regret is the central point of this paper, the final argument being that that explicit constructions of consistent and calibrated strategies (even for more precise or refined notions that the ones introduced here) are possible and provided thanks to approachability theory. The remaining is organized as follows:

\medskip
In Section~\ref{SE:Approachability}, we introduce  the concept of \textsl{approachability},  centerpiece of this work. 

We first recall (in Subsection~\ref{SE:ApproachabilityArbitrary}) a sufficient and necessary condition under which an arbitrary set is approachable. The specific case of convex sets, for which a complete characterization is available, is studied in Subsection~\ref{SE:AppproachabilityConvex}. First extensions and generalizations of the framework (e.g., in infinite dimension, with variable stage durations, unbounded payoffs, etc.) are given in Subsection~\ref{SE:ApproachGeneralization}. Last Subsection~\ref{SE:ApproachabilityAlternative} is concerned with other possible proofs and techniques of approachability. In particular, we show that approachability with respect to the supremum norm can be achieved using some potential minimization, generalizing the \textsl{exponential weight algorithm}; we also prove that the usual Euclidian (or Hilbertian) framework is not necessary for approachability.

Proofs are almost always provided, as long as they bring something new to the literature (yet some technical lemmas are delayed to the Appendix).

\medskip
Regret minimization is introduced in Section~\ref{SE:Regret}. Several refinements are introduced and links with game theory (as well as famous algorithms called \textsl{exponential weight algorithm} and \textsl{follow the perturbed leader}) are given in Subsection~\ref{SE:RegretLinks}. Since our purpose is to provide reduction to some auxiliary approachability problems, proofs are only sketched in this section and delayed to the last one. An example of regret minimization, with \textsl{expert advice} is given for illustration at the end; however, this subject is very well studied in the book of Cesa-Bianchi \& Lugosi \cite{CesLug06}. 

\medskip
Calibration and its generalizations are formalized in  Section~\ref{SE:Calibration}; for the same reasons, proofs are essentially delayed to the last section. We provide there a discussion on wether calibration (yet a weaker but maybe more intuitive notion) can or can not be obtained using deterministic algorithms. 

\medskip
Final Section~\ref{SE:Equivalences} contains all the reductions to approachability. We prove (or recall) how regret minimizations (either with finite or infinite action spaces) and calibration (either finite or with checking rules) can be obtained using approachability results from the first section. 

\medskip

Maybe the most general results are, on regret minimization, Theorems~\ref{TH:PhiRegretInvariant} and ~\ref{TH:RegMinInfinCase}  that provide (explicit for the first one) strategy minimizing swap regret if action space are, respectively, finite or infinite. Proposition~\ref{PR:RegBlac},  due to Blackwell \cite{Bla56a} himself, shows how minimization of the supremum norm of regret is exactly approachability.

Concerning calibration, most striking results might be Proposition~\ref{PR:EpsCalInde},  its consequence Theorem~\ref{TH:ManSto} and Theorem~\ref{TH:CalibViaAppro}. They refine and generalize recent results of Mannor \& Stoltz \cite{ManSto10} as well as Rakhlin, Sridharan and Tewari \cite{RakSriTew11}.

We conclude this Section by explaining how the circle is complete: if regret minimization and calibration can be seen as specific instances of approachability, the converse is also true. Indeed, using some generalized notions of regret and/or calibration, one can construct approachability strategies (in the case of convex sets).

\section{Blackwell's approachability}\label{SE:Approachability}

\subsection{Approachability of arbitrary sets}\label{SE:ApproachabilityArbitrary}

Consider a two-person repeated game between a player and Nature. Their actions set are respectively denoted by $\cA$ and $\cB$ (of respective cardinality $A$ and $B$) and payoffs are defined trough  some vectorial mapping $ g : \cA \times \cB \to \R^d$. The game is repeated in discrete time, and we denote actions chosen at stage $n \in \N$ by $a_n \in \cA$ and $b_n \in \cB$; they induce a payoff $g_n:=g(a_n,b_n) \in \R^d$. Formally, $a_n$ and $b_n$ are functions of the  history, \textsl{i.e.}, the past observations $h^{n-1}=(a_1,b_1,\ldots,a_{n-1},b_{n-1}) \in (\cA \times \cB)^{n-1}=: H_{n-1}$.

Explicitly, a strategy $\sigma$ of the player  is a mapping from $H:= \bigcup_{n \in \N} H_n$, the set of finite histories,  into $\Delta(\cA)$, the set of probability distributions over $\cA$. Similarly, a strategy $\tau$ of Nature is a mapping from $H$ into $\Delta(\cB)$.  Kolmogorov's extension theorem implies that a pair $(\sigma,\tau)$ induces a probability distribution $\P_{\sigma,\tau}$ over $\cH=(\cA \times \cB)^\N$, the set of infinite histories of the   game embedded with the product topology.

\medskip
Before defining the concept of approachability, we introduce some notations. Given a closed set $\cE \subset \R^d$, we denote by $d_\cE(x) = \inf_{z \in \cE} \{\|x-z\|\}$ the distance from $x$ to $\cE$, by $\cE^\delta=\{z \in \R^d\ \text{s.t.}\ d_\cE(x) < \delta \}$ the $\delta$-open neighborhood of $\cE$, and by $\Pi_\cE(x)= \{ z \in \cE\ \text{s.t.}\ \|x-z\| = d_\cE(x) \}$ the projection of $x$ onto $\cE$, which is in general non single-valued. We also denote by $\co \Big( \cE \Big)$ the convex hull of a set. The mapping $g$ defined on $\cA \times \cB$ (and more generally any such mapping) is extended to $\Delta(\cA)\times \Delta(\cB)$ by $g(x,y)=\E_{x \otimes y}\Big[g(a,b)\Big]$. The average of a sequence $s =\{s_m\}_{m \in \N}$ up to stage $n \in \N$ is  denoted by $\os_n:= \sum_{m=1}^n s_m/n$.

\begin{definition}
A closed set $\cE \subset \R^d$ is approachable by the player if he has a strategy~$\sigma$ ensuring, for every $\varepsilon > 0$, the existence of some integer $N_\varepsilon \in \N$ such that,  no matter the  strategy $\tau$ of Nature,
\begin{equation}\label{EQ:DefApproach}
\sup_{n \geq N_\varepsilon} \E_{\sigma,\tau}\Big(d_\cE(\og_n)\Big) \leq \varepsilon \quad \text{and}\quad \P_{\sigma,\tau} \left(\sup_{n \geq N_\varepsilon} d_\cE(\og_n) \geq \varepsilon \right)\leq \varepsilon\, .
\end{equation}
A set $\cE$ is excludable by Nature if she can approach the complement of  $\cE^\delta$ for some $\delta>0$.
\end{definition}
Informally, a given set $\cE \subset \R^d$ is approachable by the player if he has a strategy such that the average payoff converges almost-surely to $\cE$, uniformly with respect of the strategies of Nature. The right hand side of Equation \eqref{EQ:DefApproach} clearly implies the first one, which is actually the most commonly used (and rates of convergences, i.e.\ smallest mappings $\varepsilon \mapsto N_\varepsilon$ satisfying each condition,  might differ).

\subsubsection{Approachable arbitrary set : Blackwell's sufficient condition}\label{SE:AppArb}

Blackwell~\cite{Bla56} provided a simple geometrical condition under which a set $\cE$ is approachable. This sufficient condition is in fact almost necessary (as proved in Section~\ref{SE:NecApp}, following Spinat~\cite{Spi02}).

\begin{definition}\label{DF:BSet}
A closed  set $\cE \subset \R^d$ is  a B-set if for every $z \in \R^d$, there exists a projection $\pi \in \Pi_\cE(z)$ and $x:=x(z) \in \Delta(\cA)$ such that the hyperplane perpendicular to $z -\pi$ at $z$ separates $z$ from $\Big\{g(x,y) \, , \, y \in \Delta(\cB) \Big\}$, or formally:
\begin{equation}\label{EQ:DefBSet}
\forall\, z \in \R^d, \exists\, \pi \in \Pi_\cE(z), \exists\, x \in \Delta(\cA), \forall\, y \in \Delta(\cB): \quad \langle g(x,y)-\pi, z - \pi \rangle \leq 0\, .
\end{equation}
\end{definition}
Blackwell ~\cite{Bla56} proved that being a B-set is sufficient for approachability ; he also exhibited a specific  strategy,  from now on referred to as  Blackwell (approachability) strategy.
\begin{theorem}\label{TH:ApproachBlackwell}
If $\cE$ is a B-set, then $\cE$ is approachable by the player. Moreover, the strategy $\sigma$ defined by $\sigma(h^n)=x(\og_n)$ ensures that, for every $\eta >0$ and against any strategy $\tau$ of Nature:
\begin{equation}\label{EQ:RateDoob}
\E_{\sigma,\tau} \Big[d_{\cE} (\og_n)\Big] \leq 2\sqrt{\frac{\kappa_0}{n}} \quad \text{and} \quad \P_{\sigma,\tau}\left(\sup_{m \geq n} d_\cE(\og_m) \geq \eta \right) \leq \frac{8}{\eta^2}\frac{\kappa_0}{n}\, ,
\end{equation}
where $\kappa_0=\|g\|_\infty:= \sup_{x,y}\E_{x,y}\Big[\|g(a,b)\|^2\Big]=\sup_{a,b} \|g(a,b)\|^2$.
\end{theorem}
Blackwell~\cite{Bla56} and Mertens, Sorin \& Zamir~\cite{MerSorZam94} obtained respectively the bounds in expectation and in probability.  The very definition of $\|g\|_\infty$ allows each $g(a,b)$ to be random variables with bounded second moment.

We propose in the following Corollary~\ref{CR:ApproachBlackwellImproved} a slight variant that improves the constants (in the deterministic case or when $\cE$ is compact);  for instance, they are divided  by two if $\cE=\{0\}$, as in Section~\ref{SE:AppLargNum}.

\begin{corollary}\label{CR:ApproachBlackwellImproved}
A closed set  $\cE$ is approachable if and only if $\cE_g := \cE \cap  \co\big\{g(a,b)\, ; a \in \cA, b \in \cB \big\}$ is also approachable.  Blackwell's strategy applied to $\cE_g$  ensures that
\[\E_{\sigma,\tau} \Big[d_{\cE} (\og_n)\Big] \leq \sqrt{\frac{\kappa}{n}}  \quad \text{and} \quad \P_{\sigma,\tau}\left(\sup_{m \geq n} d_\cE(\og_m) \geq \eta \right) \leq \frac{2}{\eta^2}\frac{\kappa}{n}\, ,\]
 where  $\kappa=(\|g\|_\infty+ \|\cE_g\|)^2$ and $\|\cE_g\|:=\sup \Big\{\|z\|\, ; \, z \in \cE_g\Big\}$ is smaller than $\|g\|_\infty$.
\end{corollary}
\textbf{Proof:} An approachability strategy of $\cE$ ensures that any accumulation point of $\og_n$ must belong to both the closed set $\cE$ and to the compact set $\co\big\{g(a,b)\, ; a \in \cA, b \in \cB \big\}$, hence to $\cE_g$. Reciprocally,  any approachability strategy of $\cE_g$ approaches its super-set  $\cE$.

\medskip

Let $\sigma$ be  Blackwell's strategy applied to $\cE_g$,  define $\delta_n:=d_{\cE}(\og_n)$ and denote by $\pi_n$ any element of $\Pi_\cE(\og_n)$ given by Equation~\eqref{EQ:DefBSet}. Definition of $d_\cE$ implies that
\begin{align*}\delta_{n+1}^2 &\leq \left\|\og_{n+1}-\pi_n\right\|^2 = \left\|\frac{n}{n+1}(\og_n-\pi_n)+\frac{1}{n+1}(g_{n+1}-\pi_n)\right\|^2\\
&=\frac{n^2}{(n+1)^2}\delta_n^{2}+\frac{1}{(n+1)^2}\left\|g_{n+1}-\pi_n\right\|^2+\frac{2n}{(n+1)^2}\left\langle\og_n-\pi_n, g_{n+1}-\pi_n\right\rangle.
\end{align*}
Conditioning on the finite history $h^n$ and using Equation \eqref{EQ:DefBSet} as well as the definitions of $\|g\|_\infty$ and $\|\cE_g\|$, the last inequality becomes
\[ \E_{\sigma,\tau}\Big[\delta_{n+1}^2\, \Big|\, h^n \Big] \leq \frac{n^2}{(n+1)^2} \delta_n^2+ \frac{(\|g\|_\infty+\|\cE_g\|)^2}{(n+1)^2}
\]
and, with a simple induction, $\E_{\sigma,\tau}[\delta_n^2]\leq \kappa/n$. Thus $\og_n$ converges in probability towards $\cE$. The almost sure convergence is a consequence of the facts that \[Z_n:=\delta_n^2+\E_{\sigma,\tau}\left[\sum_{k=n}^\infty \frac{\|g_{k+1}-\pi_k\|^2}{(k+1)^2}\right]\ \text{is a supermartingale and} \ \E_{\sigma,\tau}[Z_n] \leq \frac{2 \kappa}{n}.\] Indeed, Doobs' inequality (see Neveu~\cite{Nev72}, prop. IV.5.2)  implies then that
\[\P_{\sigma,\tau}(\exists m \geq b \ \text{s.t.}\  Z_m \geq \eta^2)  \leq \frac{\E_{\sigma,\tau}[Z_n]}{\eta^2} \leq \frac{2 \kappa}{\eta^2 n},
\]
which gives the result. \qed

Blackwell's strategy depends only on the sequence $\{g_n\}_{n \in \N}$ so these results do not require the finiteness of $\cB$ or $\cA$, nor that Nature's actions are observed. In fact, we could as well assume the following model that we call the \textsl{compact case} (in opposition to the finite case).

Action sets are compact and convex sets,  denoted by $\cX \subset \R^A$ and   $\cU \subset  \left(\R^d\right)^A$. At stage $n \in \N$, Nature  chooses  an \textsl{outcome} $U_n=\left(U_n^a\right)_{a \in \cA} \in \left(\R^d\right)^A$ in $\cU$ and the player chooses $x_n \in \cX$. Those choices incur the vector payoff $g_n=x_n.U_n \in \R^d$, the standard inner product between $x_n$ and $U_n$.
Condition \eqref{EQ:DefBSet} that defines $B$-set becomes then
\[\forall\, z \in \R^d,\ \exists\, \pi \in \Pi_\cE(z),  \quad \inf_{x \in \cX} \sup_{U \in \cU} \ \langle x.U-\pi, z - \pi \rangle \leq 0\, .
\]
It is also possible to incorporate randomness in this model. The compact and convex sets  $\cX$ and $\cU$ can be sets of probability distribution (this was the case when $\cX=\Delta(\cA)$) and in that case $x.U$ is the expectation of a random payoff associated with $x$ and $U$ (that must have a second moment).

\subsubsection{Equivalent formulations and necessary condition}\label{SE:NecApp}

Blackwell defined geometrically a $B$-set from  outside. As Soulaimani, Quincampoix \& Sorin~\cite{As-QuiSor09} noticed that it can also be defined similarly from inside. Informally, one can interpret these definitions slightly differently: instead of viewing approachability as the convergence of average payoffs to $\cE$, it can be understood as preventing average payoffs from escaping $\cE$.

First, we need to recall the notion of proximal normals  to $\cE$.
\begin{definition}\label{DF:ProxNorm}
The set of normal proximal  to some closed set $\cE \subset \R^d$ at $e \in \cE$ is denoted by $NP_\cE(e) \subset \R^d$ and is defined by:
\[NP_\cE(e) :=  \Big\{ p \in \R^d, \ d_\cE(e+p)=\|p\| \Big\}=\Big\{ p \in \R^d, \ B\big(e+p,\|p\|\big) \cap \cE=\emptyset \Big\}, \]
where $B\big(e+p,\|p\|\big)$ is the open ball of center $e+p$ and radius $\|p\|$.
\end{definition}

The equivalent definition of a $B$-set, which is closely related to the notion of discriminant set in differential games,  is given by the following  lemma whose proof is immediate and omitted.
\begin{lemma}
A set $\cE$ is a $B$-set if and only if:
\begin{equation}\label{EQ:BsetEq} \forall\, e \in \cE,\ \forall p \in NP_\cE(e),\quad \min_{x \in \Delta(\cA)} \max_{ y \in \Delta(\cB)} \langle p, g(x,y)-e\rangle \leq 0.
\end{equation}
\end{lemma}

Interesting results on a slightly different (but equivalent as we shall see) notion of approachability that can be found in the literature can be easily derived from this alternative definition of $B$-set.

\begin{definition}
Given $\varepsilon>0$, a closed set $\cE \subset \R^d$ is $\varepsilon$-approachable by the player if he has a strategy $\sigma_\varepsilon$ ensuring that, after some stage $N_\varepsilon \in \N$,  no matter the  strategy $\tau$ of Nature,
\begin{equation}\label{EQ:Def0Approach}
\sup_{n \geq N_\varepsilon} \E_{\sigma_\varepsilon,\tau}\Big(d_\cE(\og_n)\Big) \leq \varepsilon \quad \text{and}\quad \P_{\sigma_\varepsilon,\tau} \left(\sup_{n \geq N_\varepsilon} d_\cE(\og_n) \geq \varepsilon \right)\leq \varepsilon\, .
\end{equation}
And a set $\cE$ is $0$-approachable if it is $\varepsilon$-approachable for every $\varepsilon>0$.
\end{definition}

The difference between approachability and $\varepsilon$-approachability is wether the strategy can depend on $\varepsilon$ or not.
It is clear that an approachable set is $0$-approachable but the converse is not immediate. It is easier to show -- following Spinat~\cite{Spi02} and  thanks to  Lemma ~\ref{LM:SpinatWrong} -- that a 0-approachable set must contain a $B$-set and so both notions  coincide.

\begin{lemma}\label{LM:SpinatWrong}
Let $\{\cE_n\}_{n \in \N}$ be a decreasing sequence of compact non-empty 0-approachable sets, then $\cE_\infty:=\cap_{n \in \N} \cE_n$ is also a compact non-empty 0-approachable set.
\end{lemma}
\textbf{Proof:} One just has to notice that, for every $\varepsilon >0$, the $\varepsilon/2$ neighborhood of $\cE_\infty$ is included in some $\cE_n$ which is $\varepsilon/2$-approachable. And an $\varepsilon/2$-approachability strategy of $\cE_n$ will $\varepsilon$-approach $\cE$. \qed

Lemma~\ref{LM:SpinatWrong} is not trivially true for approachability\footnote{S.\ Mannor  pointed out this interesting property.}. Indeed, one must find an approachability strategy that is independent of $\varepsilon$ and a simple concatenation of  those $\sigma_\varepsilon$ might not work (except in the specific case of convex sets).
\begin{proposition}\label{PR:Cond0Appr}
If a closed set $\cE$ is 0-approachable,  it contains a $B$-set.
\end{proposition}
We only provide a sketch of the proof, complete details can be found in Spinat~\cite{Spi02}.

\textbf{Proof:}
Consider the family of every compact subset of $\cE$ that are 0-approachable. It is a non-empty family, ordered by  inclusion and, because of Lemma~\ref{LM:SpinatWrong}, every fully ordered subset has a minorant (the intersection of all elements of this subset) which belongs to this family. Thus Zorn's lemma yield that a minimal element $\cE_\infty$ exists and we claim that $\cE_\infty$ is a $B$-set.

Indeed, assume the converse: condition \eqref{EQ:BsetEq} does not hold for some  $e \in \cE_\infty$ and some proximal normal $p \in NP_{\cE_\infty}(e)$. So there  exists $y_0 \in \Delta(\cB)$  such that
\[ 0 < \min_{x \in \Delta(\cA)} \max_{y \in \Delta(\cB)} \big\langle p, g(x,y)-e \big\rangle = \max_{y \in \Delta(\cB)}   \min_{x \in \Delta(\cA)}\big \langle p, g(x,y) -e\big\rangle=:\min_{x \in \Delta(\cA)}\big \langle p, g(x,y_0) -e \big\rangle.
\]
In particular,  Definition~\ref{DF:ProxNorm} of proximal normals implies that, at least for some small $\lambda \in (0,1)$, $(1-\lambda)e +\lambda g(x,y_0)$ belongs, for every $x \in \Delta(\cA)$, to $B\left(e+p,\|p\|\right)$. Therefore, \begin{equation}\label{EQ:CondNecAppr}\exists\, \delta>0,\ \forall x \in \Delta(\cA),\quad d_{\cE_\infty}\Big( (1-\lambda) e + \lambda g(x,y_0) \Big) \geq \delta\, .\end{equation} By continuity, Equation \eqref{EQ:CondNecAppr} holds (up to $\delta/2$ instead of $\delta$) on a small open neighborhood $V$ of $e$.  We shall prove that this implies that  $\cE_\infty \backslash V$ is still 0-approachable ;  it is a contradiction with the minimality of $\cE_\infty$ which must  therefore be a $B$-set.

Assume that at some stage $n \in \N$, $\og_n$ belongs to $V$ and that Nature plays repeatedly accordingly to $y_0$ after. Then if $n$ is large enough, there exists some large  $m \in \N$ such that  $\frac{m}{n+m}$ and $\og_{n+m}$ are, respectively and with arbitrarily high probability, arbitrarily close to $\lambda$ and  to some $(1-\lambda) \og_n + \lambda g(x,y_0)$, which is at $\delta/2$ from $\cE_\infty$.

Consider a $\delta/4$-approachability strategy of $\cE_\infty$ denoted by $\sigma_{\delta/4}$. For some large $N \in \N$ independent of $\tau$, the $\P_{\sigma_{\delta/4},\tau}$-probability that $\og_n$ belongs to $V$ for some $n \geq N$ must therefore be smaller than $\delta/4$. In particular, this implies that $\og_n$ stays within $\delta$ of $\cE_\infty\backslash V$ with probability greater than $1- \delta$. Thus, for every $\delta>0$, there exists a $\delta$-approachability strategy of $\cE\infty\backslash V$.
\qed

A direct consequence of Theorem~\ref{TH:ApproachBlackwell} and Proposition~\ref{PR:Cond0Appr} is the characterization of approachable sets.

\begin{theorem}\label{TH:CNSapproach}
A closed set $\cE$ is approachable if and only if it contains a $B$-set.
\end{theorem}
\subsection{Specific case of convex sets}\label{SE:AppproachabilityConvex}
In the specific case of convex sets, there exists a \textsl{dual and complete characterization} of approachability and excludability due to Blackwell~\cite{Bla56}. It is somehow a consequence of the fact that, for any $z$ in some closed and convex set  $\cC \subset \R^d$ one has:
\begin{equation}\label{EQ:DefNormalCone} NP_\cC(z) = \left\{ p \in \R^d \ \mbox{s.t.}\ \langle p, c-z \rangle \leq 0, \ \forall c \in \cC\right\};
\end{equation}
in particular this implies that $NP_\cC(z)$ is a cone, referred to as the normal cone.

\subsubsection{Complete characterization of approachable convex set.}

\begin{theorem}\label{TH:CondConvSet}
A closed and convex set $\cC \subset \R^d$ is  approachable by the player if and only if:
\begin{equation}\label{EQ:CondConvSet}
\forall\, y \in \Delta(\cB),\ \exists\, x \in \Delta(\cA),\quad g(x,y) \in \cC.
\end{equation}
And a convex set is either approachable by the player or excludable by Nature.
\end{theorem}
\textbf{Proof:}
Let $\cC \subset \R^d$ be a convex set  and $p \in \R^d$ be a normal proximal of $\cC$ at some $z \in \cC$. Because of  Property \eqref{EQ:DefNormalCone},   Condition \eqref{EQ:CondConvSet} can be immediately rewritten into
\[ \max_{y \in \Delta(\cB)} \min_{x \in \Delta(\cA)} \langle p, g(x,y) - z \rangle \leq 0.
\]
The mapping $(x,y)\mapsto \langle p, g(x,y) - z \rangle$ is linear in both of its argument, so von Neumann minmax theorem implies that operator $\min$ and $\max$ can be switched, i.e.,
\begin{equation}\label{EQ:BlacStratConv} \min_{x \in \Delta(\cA)}  \max_{y \in \Delta(\cB)}\langle p, g(x,y) - z \rangle = \max_{y \in \Delta(\cB)} \min_{x \in \Delta(\cA)} \langle p, g(x,y) - z \rangle \leq 0,
\end{equation}
thus $\cC$ is a $B$-set and is approachable by the player.

\medskip

On the contrary, if Condition \eqref{EQ:CondConvSet} is not satisfied, there exists some $y_0 \in \Delta(\cB)$ such that $g(x,y_0) \not \in \cC$ for every $x \in \Delta(\cA)$. By continuity, there exists $\delta>0$ such that $d_\cC(g(x,y_0)) \geq \delta$. If Nature plays repeatedly accordingly to $y_0$, then the law of large numbers implies that $\og_n$ converges uniformly to the set of $\{g(x,y_0), \ x \in \Delta(\cA) \}$ which is included in the complement of  $\cC^\delta$. So $\cC$ is excludable by Nature and, of course, is not approachable by the player. \qed

Proof of Theorem~\ref{TH:CondConvSet} relies on the Hilbertian structure of $\R^d$. However, using different arguments, it can be generalized to any normed space, see Theorem~\ref{TH:AppGene}.

\begin{remark}
In the specific case of a convex set, Blackwell strategy at stage $n+1 \in \N$ can be decomposed as follows:
\begin{itemize}
\item[i)]{Given $\og_{n} \in \R^d$, compute its projection $\Pi_{\cC}(\og_n)$ on the closed and convex set $\cC$;}
\item[ii)]{Solve the \textsl{projected zero-sum game} defined by Equation \eqref{EQ:BlacStratConv}, \textsl{i.e.}, find $x_{n+1} \in \Delta(\cA)$ that minimizes this problem and choose $a_{n+1}$ accordingly to it.}
\end{itemize}
These  steps ensure that $x_{n+1}=x(\og_n)$ as introduced in Definition~\ref{DF:BSet}. So Blackwell strategy reduces to a projection onto a convex set and the resolution of some linear program (solving a zero-sum game can be reduced to the latter, see  Sorin~\cite{Sor02}, appendix A).

On the other hand, checking wether a convex set is approachable or not, i.e., if it satisfies Condition \eqref{EQ:CondConvSet} (or equivalently the more complicate  Condition \eqref{EQ:DefBSet}) is NP-hard, even with $\cC=\{0\}$. Mannor \& Tsilikis~\cite{ManTsi09}  has indeed reduced this to the 3-SAT problem.
\end{remark}

In the compact case where action set are $\cX \subset \R^A$ and $\cU\subset \left([0,1]^d\right)^A$, a closed convex $\cC \subset \R^d$ is approachable if and only if
\[ \forall\, U \in \cU,\ \exists\, x \in \Delta(\cA),\ x.U \in \cC.
 \]
\subsubsection{Sharper high probability bounds}\label{SE:SharpHighProbaBound}

In this section, we use the convexity of $\cC$ to exhibit high probability bounds improving Corollary~\ref{CR:ApproachBlackwellImproved}.

\begin{corollary}
If $\cC \subset \R^d$ is a closed and convex approachable set, Blackwell strategy ensures that for every $\eta >0$ and against any strategy $\tau$ of Nature :
\begin{equation}\label{EQ:RateGauss}
  \P_{\sigma,\tau}\left(\sup_{m \geq n} d_\cC(\og_m) -\frac{2\|g\|_\infty}{\sqrt{m}} \geq \eta \right) \leq  4\exp\Big(-\frac{\eta^2n}{32\|g\|_\infty^2}\Big) .\end{equation}
\end{corollary}
\textbf{Proof:}
Distance to a convex set is Lipschitz and  convex, so 
\begin{align*} \left\{\sup_{m \geq n} d_\cC(\og_m) -\frac{2\|g\|_\infty}{\sqrt{m}}\geq \eta\right\}  & \subset  \left\{\sup_{m \geq n} d_\cC\Big(\E\big[\og_m\big]\Big)+\Big\|\og_m-\E\big[\og_m\big]\Big\|-\frac{2\|g\|_\infty}{\sqrt{m}} \geq \eta\right\}\\
& \subset  \left\{\sup_{m \geq n} \E\Big[d_\cC\big(\og_m\big)\Big]+\Big\|\og_m-\E\big[\og_m\big]\Big\| -\frac{2\|g\|_\infty}{\sqrt{m}} \geq \eta \right\}\\
& \subset  \left\{\sup_{m\geq n} \Big\|\og_m-\E\big[\og_m\big]\Big\| \geq \eta\right\},
\end{align*}
where the third inclusion is a consequence of the rate of convergence of Blackwell strategy. We conclude using  Lemma~\ref{LM:peelingFin}.
\qed

This result must be put in perspective with Corollary~\ref{CR:ApproachBlackwellImproved} that states that, for any arbitrary approachable  set $\cE$ and every $\eta>0$,  $  \P_{\sigma,\tau}\left(\sup_{m \geq n} d_\cE(\og_m) \geq \eta \right) \leq (\eta^2n/8\|g\|_\infty^2)^{-1}$.

\subsubsection{Biased approachability}

We assume in this section that the closed and convex set $\cC \subset \R^d$ is not approachable by the player. In that case, the natural extension of Blackwell strategy would be defined by  $\sigma(h^n)=x_{n+1} \in \Delta(\cA)$, where $x_n$  is optimal in the projected zero-sum game with payoffs
\[ \langle g(x,y) -\Pi_\cC(\og_n), \og_n -\Pi_\cC(\og_n) \rangle. \]
\begin{corollary}\label{CR:BiasApproach}
Even if a closed and convex set $\cC \subset \R^d$ is not approachable by the player, Blackwell's strategy $\sigma$ ensures that
\[ \E_{\sigma,\tau} \Big[d_\cC\left(\og_n\right)-\delta\Big] \leq \sqrt{\frac{\kappa}{n}}+\frac{\delta}{\sqrt{n}}, \ \mbox{where}\ \delta = \sup_{y \in \Delta(\cB)} \inf_{x \in \Delta(\cA)} d_\cC(g(x,y).
\]
\end{corollary}
\textbf{Proof:}
We only need to prove that $\sigma$ is in fact exactly Blackwell's approachability strategy of the closure of $\cC^\delta$ (the $\delta$-neighborhood of $\cC$) which is by definition and Condition~\eqref{EQ:CondConvSet} approachable. This is simply due to the fact that: \[\forall z \not \in \cC^{\delta}, \quad \Pi_{\cC^\delta}(z)=\Pi_{\cC}(z)+\delta\frac{z-\Pi_\cC(z)}{\|z-\Pi_\cC(z)\|}.\]
Indeed,   $\cC^\delta=\cC+\delta B(0,1)$, so $\Pi_{\cC^\delta}(z)$ minimizes $\|z-(c+\delta e)\|^2=\|z-c\|^2-2\delta \langle z-c,e \rangle +\delta^2 $ over $(c,e) \in \cC \times B(0,1)$. And necessarily, one must have $e=(z-c)/\|z-c\|$ and $c = \Pi_\cC(z)$.

The results follows from the fact that $d_{\cC^\delta}(z) \leq d_{\cC}(z)+\delta$ and $\|\cC^\delta\|\leq \|\cC\|+\delta$.\qed

\medskip

The key ingredient of Corollary~\ref{CR:BiasApproach} is not the rates of convergence (which are  a direct consequence of the fact that $\cC^\delta$ is approachable), but the fact that it does not require the computation of $\delta$ and $\cC^\delta$ (we recall that determining if a convex set is NP-hard, thus determining the smallest approachable extension is even more complex). Notice that if $\cC$ is approachable, rates of Condition~\ref{EQ:CondConvSet} and Corollary~\ref{CR:BiasApproach} and of Theorem~\ref{TH:ApproachBlackwell} match.

\bigskip

This result has to be put in perspective with the following proposition that also deals with biased approachability, yet on different level.
\begin{proposition}\label{PR:BiaAppEps} Assume that player and Nature strategies generates a sequence of payoffs such that, at every stage $n$,
 \[ \langle \og_n -\pi_\cE(\og_n), \E_{\sigma,\tau}\big[g_{n+1}\big|h^n\big]-\pi_{\cE}(\og_n)\rangle \leq \varepsilon_{n},\]
 for some sequence $\varepsilon_n$. Then
 \[ \E_{\sigma,\tau}\Big[d_\cE(\og_n)^2\Big] \leq \frac{\kappa}{n}  + \frac{2\sum_{m=1}^n m\varepsilon_m}{(n+1)^2}  \ \text{and}
 \]
\[ \P_{\sigma,\tau}\left(\sup_{m \geq n} d_\cE(\og_m) \geq \eta \right) \leq \frac{2\kappa+2\sum_{m=n}^\infty\frac{\varepsilon_m}{m} }{\eta^2n}\, .\]

In particular, if $\varepsilon_n$ converges to 0, then $\og_n$ converges in expectation to $\cE$; the convergence is almost sure as soon as $\sum_{n \in \N} \frac{\varepsilon_n}{n} < \infty$.\end{proposition}
\textbf{Proof:} The proof is identical to the one of Corollary~\ref{CR:ApproachBlackwellImproved}. \qed

Actually, the result is stated for arbitrary sets and holds for non-deterministic sequences of $\varepsilon_n$. On the other hand, for convex sets, concentration inequalities introduced in the previous section show that
\[\P_{\sigma,\tau} \Big\{\sup_{m \geq n} d_\cC^2(\og_m) - \frac{2\sum_{s=1}^ms\varepsilon_s}{(m+1)^2} \geq \eta \Big\} \leq 3\exp\Big(-\frac{M_{\eta,n}}{2}\Big)\]
thus $\og_n$ converges almost surely to $\cC$ as soon as $\varepsilon_n$ goes (in expectation) to 0.

\subsection{Generalizations and extensions}\label{SE:ApproachGeneralization}

\subsubsection{Deterministic approachability and procedures in law}\label{SE:AppComLaw}
As mentioned in Section~\ref{SE:AppArb}, Blackwell's approachability strategy does not use the fact that actions chosen by Nature are observed, as it is only required  to observe the sequence of payoffs. In fact,  it is not even required that the random variable $g_n=g(a_n,b_n)$ is perfectly observed.

Indeed, denote by $\gamma_n$ the observation made after stage $n$, and assume it is equal to either $g(x_n,b_n)$ or $g(x_n,y_n)$, where $x_n$ and $y_n$  are  mixed action of stage $n$ (i.e., laws of $a_n$ or $b_n$). Blackwell's strategy applied to the sequence of $\gamma_n$ ensures that the sequence of deterministic averages $\overline{\gamma}_n$ converges to $\cE$, uniformly with respect to Nature's strategy.

\medskip

To conclude that this describes an approachability strategy, it remains to notice that $d_\cE(\og_n) \leq d_\cE(\overline{\gamma}_n)+ \big\|\og_n-\overline{\gamma}_n\big\|$ and that the norm of $\og_n-\overline{\gamma}_n$ converges almost surely to zero, because it is an average of bounded martingale differences (using  classical concentration arguments to get rates of convergence independent of strategies).

\subsubsection{Approachability in infinite dimension spaces}

We assume in this section that $g$ no longer takes value in some Euclidian space. Formally, there exists a probability space $(\Omega,\mu,\cF)$  such that, for every $a \in \cA$ and $b \in \cB$, $g(a,b) \in \LT$ -- $g$ is extended to $\Delta(\cA)\times\Delta(\cB)$ as before. The finite case  can be easily embedded into this framework by defining, $\Omega=\{1,\ldots,d\}$ and $\mu=\frac{1}{d}\sum_{k=1}^d \delta_k$.

In this context,  notions of approachability slightly differ, as the uniform convergence with respect to Nature's strategy is not required:
\begin{definition}
A closed set $\cE \subset \R^d$ is approachable by the player if he has a strategy $\sigma$ ensuring that,  no matter the  strategy $\tau$ of Nature, $\og_n$ converges $\mu$-almost surely to $\cE$, for $\P_{\sigma,\tau}$-almost every histories.

A set $\cE$ is excludable by Nature if she can approach the complement of  $\cE^\delta$ for some $\delta>0$.
\end{definition}

Lehrer~\cite{Leh02} has proved that the natural inner product of $\LT$ allows to extend the definition of $B$-sets and Blackwell's characterization of approachable convex sets still holds (Equation \eqref{EQ:CondConvSet}, in the previous section).
\begin{theorem}\label{TH:AppInfDim}
A closed convex set $\cC$ is approachable if and only if
\begin{equation}\forall\, y \in \Delta(\cB),\ \exists\, x \in \Delta(\cA),\quad g(x,y) \in \cC.
\end{equation}
\end{theorem}
The proof relies on the following geometric principle, adapted from Lehrer~\cite{Leh02}.
\begin{lemma}\label{LM:AppInfDim}
Let $\cC$ be a closed convex subset of $\LT$. If, for every $n \in \N$, $g_n$ is bounded  $\mu$-as by $M \in \LT$ and  $\langle \og_n -\Pi_{\cC}(\og_n), g_{n+1}-\Pi_{\cC}(\og_n)\rangle \leq 0$ , then $\og_n$ converges $\mu$-as to $\cC$.
 \end{lemma}
\textbf{Proof:}
Let us denote $f_n=\og_n-\Pi_{\cC}(\og_n)$. The finite dimensional arguments of the proof of Corollary~\ref{CR:ApproachBlackwellImproved} imply that $\Big\|f_n\Big\|\leq 2\|M\|/\sqrt{n}$ thus $\og_n$ converges in probability to $\cC$.

The almost sure convergence is a consequence of the fact that \[\Big\|f_{n+1}-f_n\Big\| \leq \Big\| \Big(\og_{n+1}-\og_n\Big) - \Big(\Pi_{\cC}(\og_{n+1})-\Pi_{\cC}(\og_{n})\Big) \Big\| \leq 2\Big\|\og_{n+1}-\og_n\Big\| \leq \frac{4\|M\|}{n+1}
\]
so $f_n$ has small increments and we conclude using the technical Lemma~\ref{LM:PsConv}.
\qed

Convexity of $\cC$ is only used to get a Lipschitzian projection.

\medskip

\textbf{Proof of Theorem~\ref{TH:AppInfDim}:} Every arguments behind the proof of Theorem~\ref{TH:CondConvSet} hold in $\LT$. Therefore, a closed convex set satisfying Blackwell condition remains a $B$-set with respect to the natural inner product of $\LT$.

Assume that $\cC$ is a $B$-set and consider Blackwell's strategy, denoted as usual by $\sigma$ (and $\tau$ is Nature's strategy). Let $\mu \otimes \P_{\sigma,\tau}$ be the product measure on $\Omega \times \cH$ on which we define the random variable $\tg_n$ by $\tg_n[\omega,h]=\tg_n(a_n,b_n)[\omega]$ where $(a_n,b_n)$ is the pair of actions played at stage $n$ accordingly to $h$. Since $\cA$ and $\cB$ are finite, $g_n$ and $\tg_n$ are uniformly bounded and the sequence $\tg_n$ satisfies the geometric principle.

As a consequence, $\tg_n$ converges $\mu \otimes \P_{\sigma,\tau}$-as to $\cC$ which is therefore approachable. \qed

\subsubsection{Approachability with infinite action space -- non-linear approachability}

It is also possible to generalize the previous results when actions spaces are not necessarily finite but  two subsets of a given topological space, denoted by $\cX$ and $\cY$. Payoff mapping $g$ is now a function  from $ \cX \times \cY$ into $\LT$. In particular, it is not required in this section that $g$ is linear in each of its variable.

\begin{theorem}
Assume the following regularity assumptions on $g$:
\begin{itemize}
\item[a)] there exists   $M  \in \LT$ such that $g(x,y) \leq M$, $\mu$-as, for every $x,y \in \cX \times \cY$;
\item[b)] for every $y \in \cY$, $\cG(y)$, the closure of $\left\{g(x,y), x \in \cX \right\}$,   is a compact and convex set.
\item[c)] for every $u \in \LT$ such that $\sup_{c \in \cC} \langle c, u \rangle < +\infty$, the zero-sum game with payoffs defined by $\langle u, g(x,y) \rangle$ has a value.
\end{itemize}
Then it holds that
\begin{itemize}
\item[i)] Blackwell's characterization of convex approachable set holds :
 \[ \cC\ \text{is approachable (in pure strategy)  if and only if}\ \ \forall\, y \in \cY,\ \cG(y) \cap \cC \neq \emptyset ;\]
 \item[ii)] $\cC$ is approachable if and only if for every $z \in \LT$: \[ \sup_{y \in \cY} \inf_{x \in \cX} \langle z - \Pi_{\cC}(z), g(x,y)-\Pi_{\cC}(z)\rangle \leq 0.\]
\item[iii)] If there exists $y_0$ such that $\cG(y_0) \cap \cC =\emptyset$, then $\cC$ is excludable by Nature;
\end{itemize}
\end{theorem}
\textbf{Proof:}
The deterministic approachability strategy associated with Blackwell's characterization is defined as follows. Denote as before by $\og_n \in \LT$ the average payoff up to stage $n$. Since  $\Pi_{\cC}$ is the projection onto a convex set, one has \[\sup_{c \in \cC}\, \langle c, \og_n - \Pi_{\cC}(\og_n) ,  \rangle \leq \langle \Pi_{\cC}(\og_n),  \og_n - \Pi_{\cC}(\og_n)\rangle < +\infty.\]
Assumption \textsl{c)} ensures that  the game with payoff $\langle g(x,y) - \Pi_{\cC}(\og_n), \og_n - \Pi_{\cC}(\og_n) \rangle$ has a value which is, using Blackwell characterization, less or equal than 0. The approachability strategy consists in playing $x_n \in \cX$, any $2^{-n}$-optimal strategy of the latter game, i.e.,
\[\sup_{y \in \cY}\, \langle g(x_n,y) - \Pi_{\cC}(\og_n) ,  \og_n - \Pi_{\cC}(\og_n)  \rangle \leq\frac{1}{2^n}.\]
The fact that this describes an approachability strategy follows from   arguments used in the proof of Corollary~\ref{CR:ApproachBlackwellImproved} and  technical Lemma~\ref{LM:PsConv}.

Assume that Blackwell's condition does not hold, i.e., there exists $y_0$ such that $\cG(y_0) \cap \cC =\emptyset$;  Nature, by playing repeatedly $y_0$, can ensure that $\og_n$ belongs to  $\cG(y_0)$. The intersection between  the closed convex set $\cC$  and the  compact convex set $\cG(y_0)$ is empty, so they can be strictly separated. Since Nature can approach $\cG(y_0)$, $\cC$ is excludable, thus not approachable. \qed

Assumption \textsl{b)} is required to get point \textsl{iii)}.  Second conditions of $i)$ and $ii)$ are  sufficient for approachability (but not necessary).

\medskip

When actions sets $\cA$ and $\cB$ are finite, the projected game with payoff $\langle g(a,b), u\rangle$ typically does not have a value for some $u \in \LT$; so we considered instead mixed actions and strategies. This can be generalized when actions space are two measurable sets $(\cA,A)$ and $(\cB,B)$,  using the same tools as for procedures in law, see Section~\ref{SE:AppComLaw}.

Denote by $\cX$ and $\cY$  the sets of probability distributions onto $(\cA,A)$ and $(\cB,B)$, embedded with the weak-$\star$ topology ; the mapping $g$ is extended to $\cX \times \cY$ multi-linearly as usual. Then, under mild assumptions (for example if $\cA$ and $\cB$ are compact and $g$ is continuous, see e.g. Sorin~\cite{Sor02}), the projected game with payoff  $\langle g(x,y), u\rangle$  has a  value (at least for every $u$ such that  $\sup_{c \in \cC} \langle c, u \rangle < +\infty$). So $\cC$ is approachable with respect to  action sets   $\cX$ and $\cY$. In particular, there exists an approachability strategy such that the averages of observed payoffs $\gamma_n=g(x_n,b_n)$, where $x_n \in \cX$ is the action dictated to be played at stage $n$, converge to $\cC$ -- and the rate of convergence is $O\left(1/\sqrt{n}\right)$.

Similarly to Section~\ref{SE:AppComLaw}, this is an approachability strategy of $\cC$ since $\og_n- \overline{\gamma}_n$ is again an average of bounded Martingale differences, and concentration inequalities of sums of  bounded martingales differences in any Hilbert spaces, see e.g. Chen \& White~\cite{CheWhi96}, imply that, in expectation and with great probability, $\Big\| \og_n- \overline{\gamma}_n \Big\| \leq O\Big(1/\sqrt{n}\Big)$. Almost sure convergence is again a  consequence of Lemma~\ref{LM:PsConv}.

\subsubsection{Approachability with activation}
This section is concerned with the case where only a fragment of all coordinates of the payoff vector (belonging to $\LT$) are active at each stage. Formally, there exists a mapping $\Chi : H \to \LT$ such that, after any finite history $h^n=(a_1,b_1,\ldots,a_n,b_n)$, $\Chi[h^n] \in \LT$ has value in $\{0,1\}$ and only the coordinates  $\omega \in \Omega$ with $\Chi[h^n](\omega)=1$ are active. In particular, wether a coordinate is active at a stage might depend on  choices of actions of this specific stage.  We also assume that $\Chi[h^n]$ increases $\mu$-almost surely to infinity, no matter  the pair of strategies.

In this framework, we denote \textsl{tilted} averages of payoffs by
\[
\og_{\Chi,n}:=\frac{\sum_{m=1}^n \Chi[h^m] g(a_m,b_m)}{\sum_{m=1}^n \Chi[h^m]} \quad (\text{with the convention that}\ \frac{0}{0}=0).\] A set $\cE \subset \LT$ is approachable if the player has a strategy $\sigma$ such that, for any strategy $\tau$ of Nature, the sequence $\og_{\Chi,n}-\Pi_{\cE}(\og_{\Chi,n})$ converges to zero $\mu$-almost surely, for $\P_{\sigma,\tau}$-almost all infinite histories.

\bigskip

We will only focus on \textsl{product sets}, that can be described by
\[ \cC = \Big\{ f \in \LT\,, \ f_0 \leq f \ \text{on}\ \Omega_0 \ \text{and}\ f \leq f_1\ \text{on}\ \Omega_1\, \Big\}
\]
where $\Omega_0$ and $\Omega_1$ are two measurable subsets of $\Omega$ and $f_0, f_1 \in \LT$. The following theorem shows that, in this specific framework, a notion of \textsl{tilted} $B$-set is sufficient for approachability

\begin{theorem}\label{TH:AppActiv}
Let $\cC \subset \LT$ be a product set. Then any strategy $\sigma$ such that, for any strategy $\tau$ of Nature, and for $\P_{\sigma,\tau}$-almost every infinite history,
\[\left\langle \frac{\Chi[h^{n+1}]}{\sum_{m=1}^n \Chi[h^m]} \bigg(\og_{\Chi,n}-\Pi_\cC(\og_{\Chi,n})\bigg), g(x_{n+1},y_{n+1})-\Pi_\cC(\og_{\Chi,n}) \right\rangle\leq 0,
\]
where $x_{n+1}=\sigma(h^n)$ and $y_{n+1}=\tau(h^n)$, is an approachability strategy of $\cC$.
\end{theorem}
 The proof is similar to the one of Theorem~\ref{TH:AppInfDim}, except that Lemma~\ref{LM:AppInfDimAct} is used instead of  Lemma~\ref{LM:AppInfDim}, so it is omitted.

\medskip
The next proposition shows that approachability with activation of a product set $\cC= \prod_{k=1}^d \cC^k \subset \R^d$ in Euclidian spaces can actually be reduced to usual approachability. The only condition is that activation at stage $n$ depends only of current actions (i.e., $\cX[h^n]=\cX(a_n,b_n)$ where $\cX(a,b)$ might be a random variable); we also assume, without loss of generality, that the origin belongs to $\cC$ and even that $\cC=\prod_{k=1}^d [0,b^k] \subset \R^d$.

\begin{proposition}
A product set $\cC \subset \R^d$ is approachable with activation depending only on current actions  if and only if the following convex set \[\widetilde{\cC}:= \Big\{ (z,\omega) \in \R^d\times \R_+^d; \left(\frac{z^k}{\omega^k}\right)_{k \in \{1,\dots,d\}} \in \cC \Big\} \ \text{with the convention that}\ \frac{0}{0}=0
 \] is approachable in the game with payoffs  defined by
\[ \tg_{\Chi}(a,b) = \Big(g_{\Chi}(a,b) ,\Chi(a,b) \Big) \in \R^d\times \R_+^d, \ \ \text{where}\ \ g_{\Chi}(a,b)=\big(\Chi^k(a,b)g^k(a,b)\big)_{k \in \{1,\dots,d\}}.\]
Moreover, there exists a strategy such that, in expectation,
\[ d_\cC\left(\og_{\Chi,n}\right)\leq \frac{4\|g\|_\infty^2}{\sqrt{n}}\frac{1}{\underline{\Chi}_n}, \text{with} \ \underline{\Chi}_n=\inf\left\{\overline{\Chi}_n^k \ \text{ s.t.}\ \overline{\Chi}_n^k:=\frac{\sum_{m=1}^n\Chi^k(a_m,b_m)}{n}>0 \right\}.\]
\end{proposition}
\textbf{Proof:} Consider any fixed $(z,\omega) \in \R^d\times \R^d$;  we can always assume that every coordinates of $\omega$ are non equal to 0. Indeed, since $\cC$ is a product set, $d_{\cC}\left((z^k/\omega^k)_k\right) = d_{\cC}\left((\hz^k/\widehat{\omega}^k)_k\right)$ where $(\hz^k,\widehat{\omega}^k)=(z^k,\omega^k)$ if $\omega^k\neq 0$ and $(\hz^k,\omega^k)=(c^k,1)$ with $c^k$  arbitrarily chosen in $\cC^k$ if $\omega^k=0$.

Define  $(z_e,\omega_e) \in \Pi_{\widetilde{\cC}}\big(z,\omega \Big)$ and $\underline{\omega}$ the smallest coordinate of $\omega$. Since $\cC=\prod_{k=1}^d [0,b^k]$,  then $\widetilde{\cC}=\prod_{k=1}^d\Big\{(z^k,\omega^k); 0\leq z^k \leq b^k \omega^k\Big\}$, thus necessarily \[\omega^k\leq\omega_e^k \leq \omega^k\frac{\|g\|_\infty+1}{b^k+1}\leq (\|g\|_\infty+1)\omega^k.\] As a consequence,
\[ d_{\cC}\left(\frac{z}{\omega}\right)\leq \left\| \frac{z}{\omega}-\frac{z_e}{\omega_e}\right\|\leq  \frac{1}{\underline{\omega}}\left\|z-z_e\right\|+\|z\|\sup_k\left|\frac{1}{\omega^k}-\frac{1}{\omega^k_e}\right| \leq 2\|g\|_\infty\frac{d_{\widetilde{\cC}}(z,\omega)}{\underline{\omega}}.
\]
Reciprocally,
\[d_{\widetilde{\cC}}(z,\omega)\leq \Big\|(z,\omega)-\left(\left(\omega^k\, \Pi_{\cC}^k\left(\frac{z}{\omega}\right)\right)_k,\omega\right) \Big\|\leq \left\|\frac{z}{\omega}-\Pi_{\cC}\left(\frac{z}{\omega}\right)\right\| \leq  d_\cC\left(\frac{z}{\omega}\right)\, .
\]
Finally, if $\cC$ is a product set containing 0, then $\widetilde{\cC}$ is a convex cone. The result  is a consequence of Blackwell's characterization of approachable sets. \qed

Assuming that the origin belongs to the product set $\cC$ is of course non-restrictive, one can always choose to transform the origin into any point. Moreover, in some cases, product set property can be relaxed. For instance, if there exists two coordinates $\ell$ and $\ell'$ that are always active together, i.e., if $\Chi(a,b)^\ell=\Chi(a,b)^{\ell'}$  for every pair $(a,b)$, then the results holds if $\cC:=\prod_{k \not \in \{\ell,\ell'\}} \cC^k\times \cC^{\ell,\ell'}$ where the convex set $\cC^{\ell,\ell'} \subset \R^2$ does not need to be a product set.

\subsubsection{Variable  stage duration}

Cesaro averages of payoffs are considered in the usual definition of approachability. In this section, we make the implicit assumption that all stages does not have the same weights (when computing averages) or, equivalently, that they do not have the same length duration:  payoffs obtained on  long stages must have more importance than on short stages. We distinguish two classes of variable and random stage duration: wether they depend or not on the actions chosen.

\bigskip

Assume for the moment that $\omega_n$, the maybe random  length (or weight) of the $n$-th stage, is independent of  actions chosen by player and Nature.  In this context, $\sigma$ is an approachability strategy of a closed set $\cE$ if $\og_{\omega,n}:= \sum_{m=1}^n \omega_m\og_m/\sum_{m=1}^n \omega_m$ converges to $\cE$, $\P_{\sigma,\tau}$-almost surely, uniformly with respect to the strategy $\tau$ of Nature. It will be convenient to define $\Omega_n=\sum_{m=1}^n \omega_m$.

\begin{proposition}
Let $\cE \subset \R^d$ be a closed $B$-set. Then Blackwell's strategy applied to the sequence of weighted averages $\og_{\omega,n}$ ensures that for every $n \in \N$ and $\eta >0$
\[ \E_{\sigma,\tau}\Big[ d_\cE \left(\og_{\omega,n}\right)\Big] \leq \sqrt{\frac{\sum_{m=1}^{n} \omega_m^2}{\Omega^2_{n}}\kappa}  \quad \text{and}
\]
\[\P_{\sigma,\tau}\Big\{\exists m \geq n\,, \ d_\cE \left(\og_{\omega,m}\right) \geq \eta \Big\} \leq \left[\frac{\sum_{m=1}^n \omega_m^2}{\Omega_{n}^2}+\sum_{k=n+1}^\infty \left(\frac{\omega_k}{\Omega_k}\right)^2 \right]\frac{\kappa}{\eta^2}\ .
\]
\end{proposition}
The proof is absolutely identical with Cesaro averages (when $\omega_n=1$ for every $n \in \N$) thus omited. In particular, for any polynomial weights, i.e. if $\omega_n=n^\alpha$ with $\alpha > -1$, a $B$-set is  approachable at the rate of convergence of $O\left(1/\sqrt{n}\right)$, which is independent of $\alpha$ -- only the constant depends on $\alpha$, see e.g. Mannor, Perchet \& Stoltz~\cite{ManPerSto12}.

In fact, as we shall see in the following Section~\ref{SE:AppContTime}, a $B$-set is approachable as soon as the usual Robbins-Monroe assumptions are satisfied almost surely:
\[ \sum_{n \in \N} \frac{\omega_n}{\Omega_n} = +\infty \quad \text{and} \quad \sum_{n\in \N} \left(\frac{\omega_n}{\Omega_n}\right)^2 < \infty\ .
\]

\bigskip

We now turn to the case where a stage length might depend on the actions of the player and Nature. For simplicity, we assume that there exists a mapping $\omega: \cA \times \cB \to [\underline{\omega},\overline{\omega}] \subset (0,1]$ such that $\omega_n:=\omega(a_n,b_n)$. Approachability in this framework can be reduced to regular approachability, similarly to what has be done with activation.

\begin{proposition}
A closed set $\cE \subset \R^d$ is approachable with respect to weighted averages if and only if the following cone $\widetilde{\cE}$ is approachable with Cesaro averages
\[ \widetilde{\cE} = \Big\{ (z,\omega) \in \R^d \times [\underline{\omega},\overline{\omega}]\, ; \ \frac{z}{\omega} \in \cE \Big\}\, .
\]
Moreover, if $\cE$ is convex then $\widetilde{\cE}$ is also convex, thus $\cE$ is approachable with respect to weighted averages if and only if
\[ \forall\, y \in \Delta(\cB),\ \exists\, x \in \Delta(\cA),\ \frac{\E_{x,y}\Big[\omega(a,b)g(a,b)\Big]}{\E_{x,y}\Big[\omega(a,b)\Big]} \in \cE \ .
\]
\end{proposition}
\textbf{Proof:}
Let $(z,\omega) \in \R^d\times [\underline{\omega},\overline{\omega}]$ and $(z_e,\omega_e) \in \Pi_{\widetilde{\cE}}\big(z,\omega \Big)$, then
\[ d_{\cE}\left(\frac{z}{\omega}\right)\leq \left\| \frac{z}{\omega}-\frac{z_e}{\omega_e}\right\|\leq \frac{1}{\omega}\left\|z-z_e\right\|+\|z\|\left|\frac{1}{\omega}-\frac{1}{\omega_e}\right| \leq \left(\frac{1}{\underline{\omega}}+\frac{\|g\|_\infty}{\underline{\omega}^2}\right)d_{\widetilde{\cE}}(z,\omega).
\]
As before, one has reciprocally,
\[d_{\widetilde{\cE}}(z,\omega)\leq \Big\|(z,\omega)-\left(\omega\, \Pi_{\cE}\left(\frac{z}{\omega}\right),\omega\right) \Big\|\leq \overline{\omega}\left\|\frac{z}{\omega}-\Pi_{\cE}\left(\frac{z}{\omega}\right)\right\| \leq \overline{\omega}\, d_\cE\left(\frac{z}{\omega}\right)\, .
\]
If $\cE$ is convex, then $\widetilde{\cE}$ is a convex cone and the characterization of approachable convex set (in this framework due to Mannor and Shimkin~\cite{ManShi08}) is a simple consequence of Condition \eqref{EQ:CondConvSet}.
\qed

\subsubsection{Unbounded payoffs and strong law of large numbers}\label{SE:AppLargNum}

At the end of Section~\ref{SE:AppArb}, we noticed that we can assume that Nature choose outcomes $U$ in some given compact set $\cU \subset \left(\R^d\right)^A$ such that the player's payoff is, in expectation,  $x.U$. The fact that $\cU$ is a fixed compact set can be weakened (similarly to Stoltz~\cite{Sto05}), and we can assume that $U_n$ belongs to $\cU_n\subset \left(\R^d\right)^A$ as long as
\[ \sum_{n \in \N} \frac{\big\|U_n\big\|^2}{(n+1)^2} < \infty \quad \text{or even} \quad \sum_{n \in \N} \frac{\E_{\sigma,\tau} \left[ \|a_n.U_n\|^2\right] }{(n+1)^2}< \infty\ ,\]
with a convergence uniform with respect to Nature's strategy.

Indeed, under this assumption, the proof of Theorem~\ref{TH:ApproachBlackwell} does not change, i.e., if condition \eqref{EQ:DefBSet} is satisfied at every stage, then the same arguments yield that \[ \E_{\sigma,\tau}\Big[\delta^2_n\Big] \leq 2\frac{\sum_{m=1}^n \E_{\sigma,\tau}\Big[ \|a_m.U_m\|^2\Big]+\|\cE_g\|^2}{n^2}
\]
and $Z_n$ is a supermartingale such that
\[
\E_{\sigma,\tau}\Big[Z_n\Big]  \leq 4\frac{\|\cE_g\|^2}{n}+2\left(\frac{\sum_{m=1}^{n}   \E_{\sigma,\tau}\Big[ \|a_m.U_m\|^2\Big]}{n^2}+ \sum_{m=n+1}^{\infty} \frac{\E_{\sigma,\tau} \Big[ \|a_m.U_m\|^2\Big]}{m^2}\right)\, .
\]
By assumption, every  terms goes to zero uniformly with respect to Nature's strategy, hence $\cE$ is approachable.

\bigskip

This sheds new lights on approachability theory: it can be seen as a generalization of  Kolmogorov strong law of large numbers (see Feller~\cite{Fel68}, chapter X.7 and Mertens, Sorin \& Zamir~\cite{MerSorZam94}, exercice 4, page 104). Indeed, Let $\{X_n\}_{n \in \N}$ be a sequence of independent random variable in $\R^d$ and define $v_n:=\E\Big[\big\|X_n - \E[X_n]\big\|^2\Big]$. As soon as  $\sum_{n \in \N} v_n/n^2$ is bounded, $\oX_n-\E\Big[\oX_n\Big]$ converges almost-surely to 0; moreover
\[
\P\left\{\exists\, m \geq n ;  \Big\| \oX_m-\E[\oX_m]\Big\| \geq \eta \right\} \leq \frac{1}{\eta^2}\left(\frac{\sum_{m=1}^n v_m}{n^2}+ \sum_{m=n+1}^\infty \frac{v_m}{m^2}\right)\
\]
or even with an exponential decay (since $\{0\}$ is convex, see Section~\ref{SE:SharpHighProbaBound}).

Finally, the approachability bound (in expectation) matches the optimal bound in the law of large number and thus is in some sense optimal. Indeed, if $X_n$ is an i.i.d. sequence such that $X_n=\pm 1$ with probability $1/2$, then by denoting $\cE=\{0\}$, one has
\[\E\left[\left|\oX_n\right|^2\right]=\E\left[d_{\cE}(\oX_n)\right]=\frac{1}{n}=\frac{\kappa}{n}.\]

\subsubsection{Bounded memory}
Blackwell's approachability strategy does not require to know at each stage the whole sequence of past payoffs, but only the current average. Nonetheless, to update this average either stage number of the complete history must be kept in memory which  takes of course an increasing required size of memory. This is  why the question of wether it is possible to approach a closed set $\cE$ using  simpler strategies, for example implementable by a finite automata or with a finite memory, arises.

A strategy $\sigma$ has a bounded memory of size $M \in \N$ if, for every finite history $h^n \in H_n$, $\sigma(h^n)$  depends only on $\big(a_{n-M+1},b_{n-M+1},\ldots, a_n,b_n\big)$, i.e.\ the last $M$ profiles of actions played. Lehrer \& Solan~\cite{LehSol06,LehSol09} proved that an approachable convex set $\cC$  remains approachable by a player if it is  restricted  to use strategies with a bounded memory of size $M \in \N$; indeed, the average payoff converges to some   $O(1/\sqrt{M})$-neighborhood of $\cC$.

The basic idea is relatively natural; play Blackwell's strategy on a block of size $M$, then erase the memory and start over. It is only necessary to encode the beginning (and the end) of a block, but this can be done using $\sqrt{M}$ stages, for examples by playing always the same action and by ensuring that no such sequence appears in the same bloc. The average payoff on each block will be $1/\sqrt{M}$ close to $\cC$ which is convex, hence the overall average payoff is also $1/\sqrt{M}$ of $\cC$.

\medskip

On the other hand,  Zapechelnyuk~\cite{Zap08} considered  the strategy with bounded memory directly adapted from Blackwell's, that is defined by  $\sigma(h^n)=x\left(\og_n^M\right)$, where $x(\cdot)$ is given by the definition of a $B$-set and   $\og_n^M$ is the average payoff on the last $M$ stages. For instance, we are interested by this strategy in the game where payoffs of player (that chooses a row) are given by the following matrix:\begin{center}
\begin{tabular}{c|c|c|}
\multicolumn{1}{c}{}&\multicolumn{1}{c}{$L$} & \multicolumn{1}{c}{$R$}\\
\cline{2-3}
$T$ & (0,-1) & (0,1)\\
\cline{2-3}
$B$ & (1,0) & (-1,0)\\
\cline{2-3}
\end{tabular}
\end{center}
and $\cC = \R^{2}_-$. For  $M$ big enough, there exists a strategy of Nature such that the sequence  $\left(\og_n^M\right)_{n \in \N}$ enters a cycle (of length either  $2M$ or $2M+2$). Roughly speaking, this latter is defined by four successive blocks of lengths $M/2$ (or $M/2+1$) where within a block, the same pair of actions (except on at most one stage) is played. And one can show that the order of these actions is $(T,L)$, $(T,R)$, $(B,R)$ et $(B,L)$.

At the end of the blocks $(B,R)$ and $(T,L)$, $\og_n^M$ is close, respectively to  $(-1/2,1/2)$ or $(1/2,-1/2)$. So it is at a  distance of around 1/2 from $C$, and the sequence  $(\og_n^M)_{n \in \mathds{N}}$ of averages of  payoffs on the  $M$ last stages does not converge to $\cC$.

However, nothing indicates wether the sequence $\og_n$ does or does not  converge to $\cC$ (which is the case in this example).

\subsection{Alternative techniques and proofs of approachability}\label{SE:ApproachabilityAlternative}

\subsubsection{Approachability in continuous time}\label{SE:AppContTime}
Bena\"im, Hofbauer \& Sorin~\cite{BenHofSor06} noticed that Blackwell's approachability strategy of a $B$-set $\cE$ satisfies the following recurrence relation: condionnaly to $h^n$,
\[\frac{1}{1/n+1}\bigg(\E_{\sigma,\tau}\left[\left.\og_{n+1}\right|h^n\right]-\og_n\bigg) \in T(\og_n)-\og_n,\]
where $T(z)=\left\{\omega \in \R^d; \|\omega\| \leq \|g\|_{\infty}\ \text{and}\ \exists\, p \in \Pi_\cE(z), \langle z-p, \omega -p \rangle \leq 0 \right\}$. Therefore, the sequence of averages payoff $\{\og_n\}_{n \in \N}$ is a \textsl{Discrete Stochasitic Approximation} (a DSA for short) of $\bg$, solution of the associated ordinary differential inclusion
\[ \dot{\bg}\in T(\bg)-\bg, \quad \bg(0)=\bg_0 \in \R^d.\]
The derivative of the mapping $\delta(t)=d^2_\cC(\bg(t))$ satisfies $\delta'(t) \leq -2\delta(t)/t$ thus it is a Lyapounov function and $\delta(t)\leq \delta(0)t^{-2}$. As a consequence, $\bg$ converges to $\cE$ and, as an DSA, the sequence $\{\og_n\}_{n \in \N}$ converges a.s.\ to $\cE$. However, rates of convergence of DSA are usually not explicit and might not be uniform.

To circumvent this issue, one might consider procedures in law, as defined in Section~\ref{SE:AppComLaw}, that are deterministic and thus can be represented as an Euler Scheme of the associated ordinary differential inclusion. They might provide explicit rates  as the difference between the average payoff and its expectation converges to zero, and is controlled by concentration inequalities (see Sorin~\cite{Sor09} or Kwon~\cite{Kwo12}).

\bigskip

As Soulaimani, Quincampoix \& Sorin~\cite{As-QuiSor09} have considered an auxiliary differential game $\mathcal{D}$ where control spaces of the player and Nature are respectively $\cX=\Delta(\cA)$ and  $\cY=\Delta(\cB)$) and the game dynamic is given by:
\[\frac{d}{dt}\overline{\bg}(t)=\frac{-\overline{\bg}(t)+g(x(t),y(t))}{t}, \quad\overline{\bg}(0)=0.\]
The intuition is that  $\overline{\bg}(t)=\frac{1}{t}\int_0^tg(x(s),y(s))ds$ is the average  payoff at time $t$. The change of variables $t=e^s$ and $\overline{\bg}(s)=\bg(e^s)$ modifies the dynamic into
\[ \frac{d}{dt}\bg(s)=-\bg(s)+g(x(s),y(s)):=f(\bg(s),x(s),y(s)), \quad \bg(0)=\overline{\bg}(1).\]
This  transformation proves the characterization of a $B$-set given in Equation \eqref{EQ:BsetEq}. Indeed,  a set $\cE$ is approachable if the player can force the dynamic to stay within $\cE$. Therefore,  a closed set $E$ is a $B$-set if and only if it is a discriminating domain for the player with respect to the dynamic $f$, i.e.\ if
\[\forall p \in \cC, \forall q \in \NC_\cC(p), \sup_{y \in \cY} \inf_{x \in\cX}  \langle f(p,x,y),q \rangle \leq 0.\]

\subsubsection{Information-based strategies}
Blackwell's strategy is a payoff-based strategy as the running relevant state variable is the sequence of average payoffs. We develop in this section a conceptually completely different kind of strategy based on the sequence of observed profile of action played, as in Perchet \& Quincampoix~\cite{PerQui12} or Mannor, Perchet \& Stoltz~\cite{ManPerSto11}.

The basic idea follows from the following simple fact. Define $\theta_n=\delta_{a_n,b_n} \in \Delta(\cA\times\cB)$ as the Dirac mass on $(a_n,b_n) \in \cA\times\cB$ and let $\overline{\theta}_n=\sum_{m=1}^n \theta_m/n$ be their average. By definition, $\og_n=\E_{\overline{\theta}_n}[g(a,b)]$ belongs to $\cE$ if and only if $\overline{\theta}_n$ belongs to the following set
\[ \widetilde{\cE}:= \Big\{ \theta \in \Delta(\cA \times \cB)\ \text{s.t.} \ \E_\theta[g(a,b)] \in \cE \Big\} \subset \Delta(\cA\times\cB).
\]
If $\cE$ is closed and convex, then $\widetilde{\cE}$ (seen as a subset of $\R^{A\times B}$) is also closed and convex; it remains to compare distance between $\cE$ and $\widetilde{\cE}$.

\begin{lemma}
There exists $\gamma>0$ such that, for any probability measure $\theta \in \Delta(\cA\times\cB)$ and any set $\cE$
  \[ \gamma d_{\widetilde{\cE}}\Big(\theta\Big) \leq d_\cE\Big(\E_\theta[g(a,b)]\Big) \leq \|g\|_{\infty} \sqrt{AB}d_{\widetilde{\cE}}\Big(\theta\Big).\]
 \end{lemma}
\textbf{Proof:} For any $\theta \in \Delta(\cA\times \cB)$, define $g(\theta)= \E_\theta[g(a,b)]$. Let $\underline{\theta} \in \Pi_{\widetilde{\cE}}(\theta)$, so  $g(\underline{\theta}) \in \cE$ and
\begin{align*} d_\cE\Big(g(\theta)\Big) &\leq \Big\| g(\theta)-g(\underline{\theta})\Big\|=\left\|\sum_{a,b} \Big(\theta(a,b)-\underline{\theta}(a,b)\Big)g(a,b)\right\| \\
& \leq \|g\|_{\infty}\big\|\theta-\underline{\theta}\big\|_1\leq \|g\|_\infty\sqrt{AB}\big\|\theta-\underline{\theta}\big\|_2
\end{align*}
This gives  the second inequality.

For the first inequality, notice that $g : \Delta(\cA \times \cB)\subset \R^{A\times B} \to \co\{g(a,b)\}$ is a linear mapping, so its inverse $g^{-1}$ is  piecewise linear thus Lipschitz, see e.g., Billera \& Sturmfels~\cite{BilStu92}, bottom of page 530, or Walkup \& Wets \cite{WalWet69}. As a consequence, there exists $\lambda>0$ such that  for every $z, z' \in \co\{g(a,b)\}$ and any points $\theta$ such that $g(\theta)=z$, there exists $\theta'$ such that $g(\theta')=z'$ and $\|\theta-\theta'\| \leq \lambda \|z-z'\|$. In particular, for every $\theta \in \Delta(\cA\times\cB)$, if $\underline{z} \in \Pi_\cE\left(g(\theta)\right)$ there exists $\underline{\theta}$ such that $g(\underline{\theta})=\underline{z}$, thus $\underline{\theta} \in \widetilde{\cE}$ and
\[ d_{\widetilde{\cE}}(\theta) \leq \Big\| \theta - \underline{\theta}\Big\| \leq \lambda \Big\| g(\theta)-g(\underline{\theta})\Big\|=\lambda \Big\|g(\theta)-\Pi_{\cE}(g(\theta))\Big\| = \lambda d_\cE(g(\theta))\ ,
\]
and one just has to take $\gamma=1/\lambda$.
\qed

The consequence of this lemma is that  an approachability strategy for $\widetilde{\cE}$ is an approachability strategy for $\cE$ (and reciprocally); apart from the  requirement to compute $\widetilde{\cE}$, only  constants in  rates of convergence deteriorate.

The main advantage of this new kind of algorithms is that they do not rely on the observed sequences of payoffs. For example, consider the cases where payoffs are not vectors in some Euclidian space but  in some arbitrarily normed space (or even  payoffs can be  subsets of this space). If the image  space is not Hilbertian, then Blackwell's proofs do no longer hold; on the other hand, the transformation sequences of payoff into sequences of profile of action remains true.  Therefore, we get this very general version of characterization of approachable convex sets.

\begin{theorem}\label{TH:AppGene}
Let $\Big(\bH,\cN(\cdot) \Big)$ be any normed space (not necessarily Hilbertian) and $g : \Delta(\cA) \times \Delta(\cB) \to \bH$ (or $g: \cA \times \cB \to \bH$ is $\cA$ and $\cB$ are some compact convex sets) any continuous bi-linear mapping. Then Blackwell's characterization of approachable convex sets holds:
\[
\text{A convex set}\ \cC \subset \bH\ \text{is approachable if and only if}\ \forall y \in \Delta(\cB), \exists x \in \Delta(\cA), g(x,y) \in \cC.
\]
\end{theorem}
The result is already proved if $\cA$ and $\cB$ are finite. If they are some compact convex sets and $g$ is continuous, then one can discretize them to get $\varepsilon$-approachability strategy. Since $\cC$ is convex, they can be concatenate into an approachability strategy (using the doubling trick).

From the point of view of computational geometry, this result is rather intuitive. Indeed, no matter the image space, $\co \big\{ g(a,b) ; a \in \cA, b\in \cB\big\}$ is a polytope with at most $AB$ vertices which belongs to a relative space of finite dimension at most $AB-1$. Up to a renormalization, this gives Theorem~\ref{TH:AppGene}. However, in case where $\bH=\LT$, this  does not directly imply previous results as the approachability is only  in probability and not $\mu$-almost surely.

\subsubsection{Potential-based and uniform-norm approachability}

Approachability was first defined with respect to the $\ell^2$ distance. Roughly speaking, this induce a repeated game (see also the next subsection) between the player and Nature where the first player minimizes the distance to the  set $\cE$ and Nature maximizes it. This can be generalized to a more general class of  mappings $\Phi : \R^d \to \R$, called potentials, that are twice continuously differentiable (although this condition can be fairly weakened). 

An illustration of the interest of potential based approachability is given in the following Corollary \ref{CR:AppNormInf}. It yields fastest rates of convergence when distances to sets are defined with respect to  the uniform norm $\|\cdot\|_\infty$ instead of the Euclidian norm $\|.\|_2$.

Let us denote by $\delta$ the minimum level of $\Phi$ that player can guarantee in expectation if he plays second, i.e.
\[ \delta = \inf \Big\{ \lambda\in \R\ \text{s.t.}\  \forall y \in \Delta(\cB), \exists x \in \Delta(\cA), \Phi\big( g(x,y) \big) \leq \lambda \Big\}\  \ \text{and}\ \cE_\delta := \Phi^{-1}\big((-\infty,\delta]\big) .\]

\begin{theorem} Assume  that, for every $z$ outside $\cE_\delta$, the gradient  $\nabla \Phi(z)$ points sufficiently towards $z$, i.e., there exists $\beta>0$ such
\begin{equation}\label{EQ:AppPotEq} \forall z \not\in \cE_\delta, \ \exists x:=x(z) \in \Delta(\cA) \ \text{s.t.}\  \langle \nabla \Phi(z), g(x,y) -z \rangle \leq -\beta (\Phi(z)-\delta), \ \forall y \in \Delta(\cB).\end{equation}
Then, no matter the strategy of Nature, choosing $x_{n+1}=x(\og_n)$ yields, in expectation,
\[ \Phi(\og_n) -\delta  \leq \frac{\kappa_\Phi}{(\beta-1)}\frac{1}{n} \ \text{if}\ \beta > 1 \quad \text{and} \quad \Phi(\og_n) -\delta \leq   \kappa_\Phi\frac{\log(n)+1}{n^\beta} \ \text{if} \ 0 <\beta \leq 1,\]
where $\kappa_\Phi$ is a constant depending  uniquely on $\Phi$.

If $\beta=0$ but the inequality is strict in \eqref{EQ:AppPotEq}, then uniform convergence still holds yet at a non-explicit rate.
\end{theorem}
\textbf{Proof:} First, notice that we can  focus on the case where $\delta=0$. The proof follows from Hart \& Mas-Colell~\cite{HarMas01a} and Sorin~\cite{Sor08} (see also Cesa-Bianchi \& Lugosi~\cite{CesLug03,CesLug06}) and is based on a Taylor expansion of $\Phi$.   Indeed, since $\og_{n+1}=\og_n+(g_{n+1}-\og_n)/(n+1)$ and $\Phi$ is $\cC^2$, there exists some $\xi_n \in [\og_{n+1},\og_n]$ such that
\[ \Phi(\og_{n+1}) = \Phi(\og_n)+ \frac{1}{n+1}\left\langle \nabla \Phi (\og_n), g_{n+1}-\og_n\right\rangle + \frac{1}{2(n+1)^2}(g_{n+1}-\og_n)'D^2\Phi(\xi_n)(g_{n+1}-\og_n)
\]
where $\nabla \Phi$ and $D^2\Phi$ are respectively the gradient and the Hessian of $\Phi$; since the latter is $\cC^2$ and every $g_n$ belongs to the same compact set, there exists $\kappa_\Phi$ such that $(g_{n+1}-\og_n)'D^2\Phi(\xi_n)(g_{n+1}-\og_n) \leq 2\kappa_\Phi$, for every $n \in \N$. As a consequence, one has
\[ \E \Big[ \Phi(\og_{n+1}) \Big]\leq \left(1-\frac{\beta}{n+1}\right)\E\Big[\Phi(\og_n)\Big]+\frac{\kappa_\Phi}{(n+1)^2}
\]
and the result follows from simple induction when $\beta \geq 1$. When  $0<\beta < 1$, the bound is a consequence of the fact that \[ \left(1-\frac{\beta}{n+1}\right)\frac{\log(n)+1}{n^\beta} +\frac{1}{(n+1)^2} \leq \frac{\log(n+1)+1}{(n+1)^\beta}.
\]
The proof is a bit more intricate for $\beta=0$ (along with a strict inequality  in \eqref{EQ:AppPotEq}), but we can use the fact that $\og_n$ is a D.S.A.\ of the following differential inclusion
\[ \dot\bg \in A_\Phi(\bg) - \bg \ \text{with}\ A_\Phi(z)=\Big\{ \omega \in \R^d ; \|\omega\| \leq \|g\|_\infty\ \text{and}\ \langle \nabla \Phi(z), \omega - z \rangle >0 \Big\}.
\]
The mapping  $t\mapsto\Phi(\bg(t))$ is a Lyapounov function since if $\bg(t) \not\in \cC_\delta$
\[ \frac{d}{dt} \Phi(\bg(t)) = \langle \nabla \Phi(\bg(t)) , \dot\bg(t)\rangle \in  \langle \nabla \Phi(\bg(t)) , A_\Phi(\bg(t))-\bg(t)\rangle <0,
\]
therefore $\bg$ converges to $\cC_\delta$ and so does $\og_n$. \qed

If $\cC$ and $\Phi$ are convex, then Equation \eqref{EQ:AppPotEq} always holds with $\beta=1$, and we recover Theorem 7.6 of Cesa-Bianchi \& Lugosi~\cite{CesLug06} (due to Hart \& Mas-Colell \cite{HarMas01a}):
\begin{corollary}
If $\cC = \Phi^{-1}\big((-\infty,0])$ for some  convex, twice continuously differentiable mapping $\Phi$  whose Hessian is bounded in norm by $\kappa_\Phi$ on $\co\big\{g(a,b)\big\}$, there exists a strategy  such that, in expectation and no matter the strategy of Nature, $\Phi(\og_n)\leq \frac{2\kappa_\Phi(\log(n)+1)}{n}$.
\end{corollary}

The assumption that $\Phi$ is twice continuously differentiable can be easily weakened, in particular as soon as the constant $\kappa_\Phi$ exists. The next proposition is concerned with the sequence of sums of payoffs $G_n = \sum_{m=1}^n g_m$ instead of averages. It will be used, in some cases, to improve rates of convergence.

\begin{proposition}\label{PR:AppPotSum}
Assume that
\[ \forall z \not\in \cE_\delta, \ \exists x:=x(z) \in \Delta(\cA) \ \text{s.t.}\  \langle \nabla \Phi(z), g(x,y) \rangle \leq 0, \ \forall y \in \Delta(\cB)\]
and there exists $\kappa_\Phi>0$ such that $g(x,y)'D^2\Phi(z)g(x,y) \leq \kappa_\Phi$ for every $x \in \Delta(\cA)$, $y \in \Delta(\cB)$ and $z \not\in \cE_\delta$. Then, no matter Nature's strategy, choosing $x_{n+1}=x(G_n)$ yields $\E\big[\Phi(G_n)\big] \leq n \kappa_\Phi+\Phi(0)$.
\end{proposition}
\textbf{Proof:} This is a consequence of the fact that, for some $\xi_n \in [G_n,G_{n+1}]$,
\[\Phi(G_{n+1}) = \Phi(G_n) + \langle \nabla \Phi(G_n),g_{n+1}\rangle + g_{n+1}'D^2\Phi(\xi_n)g_{n+1}\]
followed by  an immediate induction. \qed

This result can be immediately extended if $\Phi$ is not $\cC^2$ but such that
\[ \Phi(G_{n+1})-\left(\Phi(G_n)+\langle \nabla \Phi(G_n), g_{n+1} \rangle \right)\leq \kappa_\Phi.
\]
As mentioned before, the following corollary shows a faster convergence if $\cC$ is an approachable cone. Proposition~\ref{PR:AppPotSum} is even used more deeply to get optimal rates of convergence (both in the number of stages and the dimension) below to obtain approachability with respect to the uniform norm.

\begin{corollary}\label{CR:BlackCone}
If $\cC$ is an approachable closed and convex cone, then Blackwell's strategy ensures that, no matter Nature's strategy and for every $N\in \N$
\[ \E_{\sigma,\tau} \left[d_\cC(\og_n)\right] \leq \frac{\|g\|_\infty}{\sqrt{n}}.
\]
\end{corollary}
\textbf{Proof:} First, if $\cC$ is a cone then necessarily $\langle z - \Pi_\cC(z), \Pi_\cC(z) \rangle \leq 0$, therefore the first condition of Proposition~\ref{PR:AppPotSum} is the characterization of the fact that $\cC$ is a $B$-set. Second, if $\Phi(\cdot)=d^2_\cC(\cdot)$  then $\Phi$ satisfies the second condition of Proposition~\ref{PR:AppPotSum} -- or at least its straightforward extension -- with $\kappa_\Phi=\|g\|^2_\infty$.

Since $\cC$ is a cone, $d^2_\cC(\og_n)=\Phi(G_{n+1})/n^2$ and the result follows. \qed

For simplicity, we will assume that $\|g(a,b)\|_\infty \leq 1$ and we only consider target sets such that, for some $b_k,c_k \leq 1$,
\[ \cC := \Big\{ z \in \R^d \ \text{s.t.} \  b_k\leq z_k \leq c_k, \forall k \in \{1,\ldots, d\} \Big\}\ .
\]
The $\ell^\infty$-distance to this set is denoted by $d_\cC^\infty$, i.e. $d_\cC^\infty(z)=\inf_{c \in \cC} \|z-c\|_\infty$.
\begin{corollary}\label{CR:AppNormInf}
There exists a strategy $\sigma$ of the player such that, against any strategy $\tau$ of Nature and every $n \in N$ and $\delta >0$, with probability at least $1-\delta$,
\[ \E_{\sigma,\tau} \Big[ d_\cC^\infty(\og_n)\Big] \leq 14\sqrt{\frac{\log(2d)}{n}}\ \text{and}\  \P_{\sigma,\tau}\Big\{d_\cC^\infty(\og_n) \geq \delta\Big\} \leq 16\sqrt{\frac{\log\left(\frac{2d}{\delta}\right)}{n}}.
\]
\end{corollary}
\textbf{Proof:} We first prove a similar result  in the specific case where $\cC=\R_-^d$ is the negative orthant and if an horizon $N$ is known in advance. Then we will use a \textsl{doubling trick} to conclude for the orthant; we will finally show how to reduce approachability of any product set $\cC$.

Let $\Phi$ be the following potential, depending on a parameter $\eta >0$ to be fixed later:
\[ \Phi(z):= \frac{1}{\eta} \log \left[\sum_{k=1}^d e^{\eta z_k}\right]\ \text{so that}\ d_\cC^\infty(z) \leq \Phi(z)\ \text{and}\ \Phi(0)\leq \log(d)\ ;\ \text{moreover}
\] 
\[ \nabla \Phi(z)^i = \frac{e^{\eta z_i}}{\sum_{k=1}^d e^{\eta z_k}}\ \text{and}\ D^2\Phi(z)= \eta\diag\left(\frac{e^{\eta z_i}}{\sum_{k=1}^d e^{\eta z_k}}\right)-\eta\nabla\Phi(z)\nabla\Phi(z)',\]
where $\diag(\lambda_i)$ is the matrix whose diagonal is $\lambda_1,\ldots,\lambda_d$ and zero everywhere. As a consequence,  since $\cC$ is approachable, the first condition of Proposition \ref{PR:AppPotSum} is satisfied and $\omega'D^2\Phi(z)\omega \leq \eta\omega'\diag(\lambda_i)\omega-\eta\|\nabla\Phi(z)\omega\|_2^2\leq \eta\|\omega\|_\infty^2$ implies the second one.

So, convexity of $d_\cC^\infty(\cdot)$, Proposition~\ref{PR:AppPotSum} and the  choice $\eta=\sqrt{\log(d)/N}$ imply that
\[\E_{\sigma,\tau}\Big[d_\cC^\infty
(\og_N)\Big]\leq \E_{\sigma,\tau}\Big[\frac{d_\cC^\infty
(\oG_N)}{N}\Big] \leq \frac{1}{N}\E[\Phi(G_N)] \leq \eta + \frac{\log(d)}{\eta N}=2\sqrt{\frac{\log(d)}{N}}.
\]

We now make appeal to the \textsl{doubling trick}, that is, we consider the strategy consisting in playing by blocks of lengths $2^k$, following the potential associated with $\eta_k:=\sqrt{\log(d)/2^k}$ on the $k$-th block and reseting everything at the beginning of a new block. A simple induction, based on the convexity of $d_\cC^\infty(\cdot)$, shows that, at the end of any block,
\[\E_{\sigma,\tau}\Big[d_\cC^\infty
(\og_{2^{k}-1})\Big]  \leq 2(1+\sqrt{2})\sqrt{\frac{\log(d)}{2^k-1}}.
\]
Hence it remains to control distances within blocks. Yet, using the previous bound obtained for ends of blocks, one has for $n=2^{k}-1+m$ with $m \leq 2^k$,
\begin{align*}
\E_{\sigma,\tau}\left[d_\cC^\infty
(\og_{n})\right] &\leq \frac{2^k-1}{n}\E_{\sigma,\tau}\left[d_\cC^\infty
(\og_{2^k-1})\right]+\frac{m}{n}\left(\eta_k+\frac{\log(d)}{\eta_k m}\right)\\
& \leq \frac{\sqrt{\log(d)}}{n}\left[2(1+\sqrt{2})\sqrt{2^k-1}+\frac{m}{\sqrt{2^k}}+\sqrt{2^k}\right]\\
&\leq (4+2\sqrt{2})\sqrt{\frac{\log(d)}{n}}.
\end{align*}
 Concentration arguments give the bound in high probability. Indeed, the union bound implies
 \[ \P \Big\{\exists k \leq d \ \text{s.t.}\ \big|\og_n^k-\E[\og_n^k]\big| \geq \varepsilon \Big\} \leq 2d \exp\left(-\frac{n\varepsilon}{2}\right),
 \]
thus the probability that  $\|\og_n-\E[\og_n]\|_\infty$ is smaller than $\sqrt{\frac{2}{n}\log\left(\frac{2d}{\delta}\right)}$ is bigger than $1-\delta$. The result for the orthant is a direct consequence of the triangle inequality.

\medskip

We no longer assume that $\cC$ is an orthant, but is defined by 
\[ \cC := \Big\{ z \in \R^d \ \text{s.t.} \  b_k\leq z_k \leq c_k, \forall k \in \{1,\ldots, d\} \Big\}\ .\]
Let $h(x,y)=\Big(g_k(x,y)-c_k,b_k-g_k(x,y)\Big)_{k \in \{1,\ldots,d} \in \R^{2d}$, then approachability of $\og_n$ to $\cC$ is equivalent to the approachability of $\oh_n$ to the negative orthant, since $d^\infty_\cC(\og_n)=d^\infty_{\R_-^{2d}}(\og_n)$ and $g(x,y) \in \cC$ if and only if $h(x,y) \in \R_-^{2d}$. The result follows from the bound exhibited for the orthant.

\qed

\subsubsection{From weak approachability to approachability}\label{SE:WeakAppr}

Recall that a closed set $\cE$ is approachable if the player has a strategy such that after some (maybe large) stage $N$, the payoffs remains in a small neighborhood of $\cE$. Similarly, it is excludable if Nature can enforce the dual: after some stage $N$, the payoffs remains outside some neighborhood of $\cE$. Blackwell proved that there exists a dichotomy for convex sets:  they are either approachable or excludable. This is not true for any set, as illustrated in the following example, due to Blackwell.

Consider the  set and payoff matrix defined by, with $\cA=\{T,B\}$ and $\cB=\{L,R\}$,
\[ \cE = \Big\{ (1/2,y), y \in \left[0,1/4 \right]\Big\} \cup \Big\{ (1,y), y \in \left[1/4,1\right]\Big\} \ \quad \text{and} \quad \begin{array}{ccc} & L & R \\
																													  T & (1,0) & (1,1) \\
																													  B & (0,0) & (0,0) \end{array}.\]
Assume that the strategy of the player dictates to play $T$ during $N$ stages (with $N$ a large even number) then to play either always $T$ or always $B$ during the following $N$ stages, depending on wether Nature has played more than half of the time $R$ during the first $N$ stages.

In the former case, the player got after $N$ stages, an average payoff of $(1,y)$ with $y \geq 1/2$ thus by  keeping to play $T$ for $N$ stages, he ensures that its average payoff after $2N$ stages is $(1,y')$ with $y' \geq y/2 \geq 1/4$. In the latter case, the payoff after $N$ stages is $(1, y)$ with $y \leq 1/2$, thus the payoff after $2N$ stages is $(1/2,y')$ with $y'=y/2 \leq 1/4$.

\medskip

As a consequence, this strategy guarantees that, after $2N$ stages, the payoff is exactly in $\cE$. So if this procedure is applied during $2N_1$ stages, then  started over for $2N_2=2e^{N_1}$ stages, then started again over for $2N_3=2e^{N_2}$ stages and so on, the payoff is infinitely often arbitrarily closed to $\cE$ which is therefore not excludable.

Unfortunately, $\cE$ is not approachable; indeed, this would imply that at least one of the two connected (and convex) component of $\cE$ is approachable. But neither of them satisfies Blackwell's characterization.

\medskip

In this example, the player cannot enforce the payoff to remain close to $\cE$, but if he knows in advance that there are only $N$ stages in the game, then he can ensure that, at the terminal stage, the payoff is in $\cE$ (or at least, for odd integer, $1/N$-close to $\cE$). A natural weaker concept of approachability emerges: a set $\cE \subset \R^d$ is weakly-approachable if, given some fixed large length of the game, the player has a strategy such that the terminal average payoff is close to $\cE$.

\begin{definition}\label{DF:WeakApproch}
A closed set $\cE\subset \R^d$ is weakly approachable if for every $\varepsilon >0$, there exists $N_\varepsilon \in \N$ such that, in any game of length $n \geq N$, the player has a strategy $\sigma_n$ such that, no matter the strategy $\tau$ of Nature, $\E_{\sigma_n,\tau}\Big[d_\cE(\og_n)\Big]\leq \varepsilon$.

Similarly, $\cE$ is weakly excludable if Nature can weakly approach the complement of $E^\delta$ for some $\delta >0$.
\end{definition}
We emphasize the fact that in weak-approachability, strategies can depend on the length of the game $n$, which is not allowed for regular approachability. The question rose by Blackwell~\cite{Bla56} and solved by Vieille~\cite{Vie92} is wether there exists a dichotomy between weakly-approachable and weakly-excludable sets.

\begin{theorem}
Any closed set is either weakly-approachable or weakly-approachable.\
\end{theorem}
\textbf{Proof:} We only sketch here the proof of Vieille\cite{Vie92}.

Consider the differential zero-sum game  where the player and Nature choose action $x(t) \in \Delta(\cA)$ and $y(t) \in \Delta(\cB)$ in continuous time (actually, even the formal definition of strategies might require precise notations and concepts). In this game, a state variable which represents the accumulated payoff, evolves following the dynamic $\dot\bG(t) =g\big(x(t),y(t)\big)$ and $\bG(0)=0$ during the time $t=0$ and $t=1$.

In this game, the overall objective of the player is to minimize the terminal payoff $d_\cE\left(\bG(1)\right)$, while Nature maximizes it. The important fact is that one can prove, using techniques and results from differential games, that this game has a value $v$. If $v=0$, then the player has a strategy such that the cumulated payoff at time $t=1$ is exactly in $\cE$ whereas if $v>0$, Nature has a strategy such that this cumulated payoff is bounded away from $\cE$.

It remains to understand that a game in discrete time with $N$ stages is a  discretization (or an approximation) of this differential game and as $N$ goes to infinity, this approximation is more and more precise. Therefore, if the player can enforce that $\bG(1)$ belongs to $\cE$, then he can ensure that $\og_N$ is arbitrarily close to $\cE$ when $N$ is large enough. The converse holds for Nature, hence the result. \qed

Actually, the focus of this section if not only this important (and elegant) result but also the following properties, inspired from Cesa-Bianchi \& Lugosi~\cite{CesLug06} or Rakhlin, Sridharan \& Tewari~\cite{RakSriTew11}. Given an approachable convex set $\cC$, let $\sigma_N$ be an optimal strategy in the $N$-stage zero-sum game with terminal payoff $\E_{\sigma,\tau}\Big[d_\cC(\og_N)\Big]$ and denote by $v_N$ the value of this game (its existence  is not difficult). 

We know that $\E_{\sigma,\tau}\Big[d_\cC(\og_n)\Big]$ can be  upper bounded, using some adequate approachability strategy, by $O\left(\|g\|_\infty/\sqrt{n}\right)$; but it is also obviously lower-bounded by $v_n$. So the computation of $v_n$ could indicate wether the rate is tight or not. On the other hand, exact computation of $v_n$  might be challenging, yet if satisfies 	\[v_n =\min_{\sigma_n}\max_\tau \E_{\sigma,\tau}\Big[d_\cC(\og_n)\Big]=\max_\tau \min_{\sigma_n}\E_{\sigma,\tau}\Big[d_\cC(\og_n)\Big]\leq \max_\tau \min_{\sigma(\tau)} \E_{\sigma,\tau}\Big[d_\cC(\og_n)\Big]
\]
where $\sigma(\tau)$ is the strategy that chooses,  given $\tau$  and after the finite history $h^n$,  $x_{n+1}=x(\tau(h^n))$. In particular,  $\E_{\tau,\sigma(\tau)} [ \hg_n]:=\sum_{m=1}^n \E_{\tau,\sigma(\tau)}[g_m|h^{m-1}]/n$ belongs to $\cC$ and therefore (removing the dependency in $\sigma$)
\[ v_n  \leq \max_{\tau} \E\Big[ \left\| \og_n - \E_{\tau} [\og_n] \right\| \Big] \leq \sup_{\tg} \E\left[ \left\| \frac{ \sum_{m=1}^n \tg_m - \E[\tg_m|h^{m-1}]}{n}\right\|\right] \leq O\left(\frac{\|g\|_\infty}{\sqrt{n}}\right)
\]
where the  supremum is taken over all sequences $\tg_m - \E[\tg_m|h^{m-1}]$ of martingale differences with $\tg_m \in \{g(a,b), a \in \cA, b \in \cB\}$. Last inequality is a consequence of Hoeffding-Azuma's inequality in Euclidian spaces.

\bigskip

A question that naturally arises is wether we can concatenate -- using the doubling trick -- optimal strategies in games of length $2^k$ to construct an approachability strategy of $\cC$ (i.e., independent of any horizon $n$). The answer is both no and yes: no with the current definition of $v_n$. Indeed the only guarantee is that  terminal payoff is $v_{2^k}$-close to $\cC$ but, for instance,  payoff at  middle stages could be arbitrarily away. 

On the other hand, since $\cC$ is a convex set, we can modify the definition of $v_n$ as follows so that the answer is yes. Define
\[ v'_n =\max_\tau \min_{\sigma_n}\E_{\sigma,\tau}\Big[\sup_{m\leq n} \frac{m}{n}d_\cC(\og_m)\Big]\leq \max_\tau \min_{\sigma(\tau)} \E_{\sigma,\tau}\Big[\sup_{m\leq n}\frac{m}{n}d_\cC(\og_m)\Big],
\]
so that, using the same arguments  and Doobs (or Hoeffding) maximal inequality
\[ v'_n  \leq \max_{\tau} \E\Big[ \sup_{m \leq n} \frac{m}{n}\left\| \og_n - \E_{\tau} [\og_n] \right\| \Big] \leq \sup_{\tg} \E\left[ \frac{1}{n}\left\|\sum_{s=1}^m \tg_s - \E[\tg_s|h^{s-1}]\right\|\right] \leq O\left(\frac{\|g\|_\infty}{\sqrt{n}}\right).
\]
Finally, the doubling trick works with this definition of $v'_n$, see the proof of Corollary~\ref{CR:AppNormInf}.

\bigskip

This technique  seems void at first sight, but might be  useful in some specific examples, as in Proposition~\ref{PR:RegBlac} in Section~\ref{SE:Equivalences} (see also Remark~\ref{RM:RegBlac}). In this case, because of the geometry of $\cC$, one has $d_\cC(z) \leq 2\|z - c\|_\infty$, for every $c \in \cC$. Then  the same tools yield  that  $v_n$ is smaller than  $O\left(\sqrt{\log(d)/n}\right)$ which is  negligible  compared to $\|g\|_\infty/\sqrt{n}\simeq \sqrt{d/n}$ as the dimension $d$ increases. 

\section{Regret minimization}\label{SE:Regret}

Hannan~\cite{Han57}  introduced the concept of \textsl{external regret} in repeated two-player games (between a player and Nature, with scalar payoff) in order to define an exogenous criterium to evaluate a strategy in a non-Bayesian framework. In words, the player has no external regret (or his strategy is \textsl{externally consistent}) if, asymptotically, he could not have gained strictly more if he had known, before the beginning of the game, the empirical distribution of moves of Nature. This notion has notably been refined by Foster \& Vohra~\cite{FosVoh97} (see also Fudenberg \& Levine~\cite{FudLev99}) into \textsl{internal regret}: a player has no internal regret (or his strategy is \textsl{internally consistent}) if he has no external regret on the set of stages where he played a specific action, as soon as this set is big enough.

\subsection{Finite action spaces}

Consider a two-person repeated game where action spaces of the player and Nature are $\cA$ and $\cB$ (of cardinalities $A$ and $B$)  and $\rho : \cA \times \cB \to [0,1]$ is a real payoff mapping. Extension of $\rho$ to $\Delta(\cA)\times\Delta(\cB)$ and strategies are defined as in the previous section.

\subsubsection{External regret}

Choices of actions $a_n \in \cA$ and $b_n \in \cB$ generate a regret $r_n \in \R^A$ defined by
\[ r_n = r(a_n,b_n) := \Big( \rho(1,b_n)-\rho(a_n,b_n), \ldots, \rho(A,b_n)-\rho(a_n,b_n)\Big) \in \R^A.
\]
Intuitively, the regret $r_n$ represents the differences between what the player could have got and what he actually got. And a player has no external regret if, asymptotically, every component of the average regret is non positive. In words, this means that the player could not think \textsl{" if I had known} [the empirical distribution of Nature's actions], \textsl{ I would have always played  action $a^*$"}, hence the terminology of \textsl{regret}. Indeed, by linearity of $\rho$,
\[
\orr_n = \Big(\rho(1,\ob_n)-\overline{\rho}_n, \ldots, \rho(A,\ob_n)-\overline{\rho}_n\Big) \in \R^A.\]

Given a vector $U \in \R^d$, the notation $U^+$ will stand for the positive part of $U$, i.e., $U^+=\left(\max \{ U^i,0 \}\right)_{1\leq i \leq d}$. Similarly, $U^-$ is the negative part of $U$.

\begin{definition}
A strategy $\sigma$ of the player has no external regret if, for all strategy $\tau$ of Nature, $\P_{\sigma,\tau}$-almost surely,
\begin{equation}\label{DF:DefRegExt}
\limsup_{n\to\infty} \max_{a^* \in \cA} \rho(a^*,\ob_n)-\overline{\rho}_n \leq 0, \ \text{or equivalently, } \ \limsup_{n \to \infty} \|\orr_n^+\|_{\infty} \leq 0.
\end{equation}
\end{definition}
The existence of externally consistent strategies goes back to Hannan~\cite{Han57}. However, the following theorem, with rates of convergence independent of Nature's strategy, is due to Cesa-Bianchi \& Lugosi~\cite{CesLug06}.
\begin{theorem}\label{TH:RegExt}
There exists an externally consistent strategy $\sigma$, such that, no matter the strategy $\tau$ of Nature and for every $n \in \N$,
\[ \E_{\sigma,\tau}\Big[\max_{a^* \in \cA} \rho(a^*,\ob_n)-\overline{\rho}_n\Big]=\E_{\sigma,\tau}\Big[\|\orr_n^+\|_\infty\Big] \leq 2\sqrt{\frac{\log(A)}{n}}.
\]
\end{theorem}
We will not yet provide  proofs of this result; instead, we will show a weaker result, following Zinkevich~\cite{Zin03}. The basic idea is to notice that the overall objective is to maximize the convex function $\rho(\cdot,\ob_n)$ and therefore to apply any convex-maximization techniques, for example a gradient descent.

\textbf{Proof:} First, we claim that for every $n \in \N$, there exists a strategy $\sigma_n$ (that depends on $n$), such that
\[ \E_{\sigma,\tau}\Big[\max_{a^* \in \cA} \rho(a^*,\ob_n)-\overline{\rho}_n\Big] \leq \sqrt{\frac{A}{n}}.\]

  Let $\eta$ be a parameter to be fixed later and define, for every $m\leq n$ the strategy $\sigma$ following an usual gradient descend:
\[ x'_{m+1}=x_m+\eta \rho(\cdot,b_m) \ \text{and}\ x_{m+1}= \Pi_{\Delta(\cA)} \left(x'_{m+1}\right), \quad \text{with}\ x_1=x'_1=\left(\frac{1}{A},\ldots,\frac{1}{A}\right),  \]
the projection step ensures that $x_{m+1}$ stays in $\Delta(\cA)$. Simple calculations show that, for every $a \in \cA$,
\begin{align*}
\E_{\sigma,\tau}\left[\sum_{m=1}^n \rho(a,b_m)-\rho_m\right] & = \E_{\sigma,\tau}\left[\sum_{m=1}^n (a-x_m)'\rho(\cdot,b_m)\right]=\sum_{m=1}^n (a-x_m)'\frac{(x'_{m+1}-x_m)}{\eta}\\
& = \frac{1}{2\eta} \sum_{m=1}^n \|a-x_m\|^2+\|x_m-x'_{m+1}\|^2-\|a-x'_{m+1}\|^2\\
&\leq \frac{1}{2\eta} \sum_{m=1}^n \|a-x'_m\|^2+\|x_m-x'_{m+1}\|^2-\|a-x'_{m+1}\|^2\\
& = \frac{1}{2\eta}   \|a-x_1\|^2+ \frac{1}{2\eta}\sum_{m=1}^n\eta^2\|\rho(\cdot,b_m)\|^2 \leq \frac{1}{2\eta}+\frac{\eta nA}{2}.
\end{align*}
Balancing the two terms by choosing $\eta=1/\sqrt{nA}$ proves the claim. We stress out the fact that this strategy ensures that, at stage $t$, the regret is bounded as
\[ \E_{\sigma,\tau}\left[\sum_{m=1}^t \rho(a,b_m)-\rho_m\right] \leq \frac{\sqrt{nA}}{2}+\frac{t\sqrt{A}}{2\sqrt{n}},\]
which might be considerably bigger than $\sqrt{tA}$ for small $t$, but this uniform guarantee allows the use of a doubling trick, as in Corollary~\ref{CR:AppNormInf}, to conclude.\qed

To get the $\log(A)$ term instead of $A$ in the upper bound, one just has to follow the algorithm known as \textsl{exponential weight algorithm}, defined by:
\[ x_{n+1}[a] = \frac{\exp\big(\eta_n \rho(a,\ob_n)\big)}{\sum_{a' \in \cA} \exp\big(\eta_n \rho(a',\ob_n)\big) } \ \text{where} \ \eta_n=\sqrt{8n\log(A)}\ ,
\]
see, e.g., Littlestone \& Warmuth~\cite{LitWar94}, Vovk~\cite{Vov90} or Auer, Cesa-Bianchi \& Gentile~\cite{AueCesGen02}.

The following corollary shows that the previous result can be extended to the compact case. Actually, the proof is exactly the same, since it did not use the fact that $\cB$ is finite, thus is omitted.

\begin{corollary}\label{CO:RegExt} Assume that Nature chooses at every stage an outcome vector $U_n$ in a compact set $\cU \subset [0 ; 1]^A$ such that the players payoff  at this stage is $U_n^{a_n}$. Then there exists a strategy $\sigma$, such that, no matter the strategy $\tau$ of Nature and for every $n \in \N$, \[ \E_{\sigma,\tau}\Big[\max_{a^* \in \cA} U_n^{a^*}-U_n^{a_n}\Big] \leq 2\sqrt{\frac{\log(A)}{n}}.
\]
\end{corollary}

Theorem~\ref{TH:RegExt} and Corollary ~\ref{CO:RegExt} can actually be proved using more complex optimization procedures, as mirror descent instead of gradient descent (see e.g., Rakhlin~\cite{Rak09} or Bubeck~\cite{Bub12} for a survey on the use of these techniques in machine learning) and without using doubling trick. We will, on the contrary, prove them using approachability theory.

\begin{remark}
In Section~\ref{SE:WeakAppr}, we claimed that we could not  use a doubling trick. It was possible here because  any strategy $\sigma_n$, although only optimal at the final stage $n$, ensures relatively good performance at all stages. For instance, at the specific stage $t=n/2$, the regret is bounded in $\frac{3\sqrt{2}}{4}\sqrt{A/t} \simeq 1.06\sqrt{A/t}$. This was not the case in the previous section, where the distance to the set could be of the order of a constant.

More specifically,  strategies $\sigma_n$ are somehow equivalent to weak approachability (only the final stage matters). If we could always concatenate strategies using a doubling trick to output a strategy that behaves well at all stages, then we could construct approachability strategy from weak approachability strategies. This would mean that  any set is either approachable or excludable, which is not true in general (in fact, as proved in Section~\ref{SE:AppImpReg}, regret corresponds more to the approachability of convex sets, on which weak and regular approachability coincide).
\end{remark}

An usual criticism to the notion of regret in games (and this could lead to long and probably unfruitful debates) is that a player compares his payoff with the payoff he would have got if he had always played the pure action $a^*$. However, if he had played something else, then Nature would (or at least could) have chosen a totally different sequence $b_n$ so the comparison is meaningless. An easy and unsatisfactory answer is to say that a player's action does not change the behavior of Nature (as in the learning with experts advices literature, see Cesa-Bianchi \& Lugosi~\cite{CesLug06}). A less unsatisfactory answer consists in stating that since there is absolutely no prior on Nature,  it is impossible to infer whatsoever on her strategy if the world had been different. So we should compare the payoff with respect to  best information available, which is the current sequence.

Let us develop a third point of view, based on game theoretic perspectives. The basic idea is that regret is \textsl{not} a criterion to compare different strategies: it does not say that  a strategy without  regret is a better strategy than always playing $a^*$. In our repeated game, the player maximizes his cumulated payoff without any structural assumption on Nature. Therefore, he can just sequentially formulate  \textsl{predictions} upon her behavior (we purposely remain vague on this subject) and play a \textsl{best response} to it. Regret is a simple measure on how much a sequence of predictions is \textsl{correct} or not. A large regret would mean that the player was most of the time wrong.

\subsubsection{Internal and $\Phi$-regret}
The notion of external regret has been refined by Foster \& Vohra~\cite{FosVoh97} into the so-called \textsl{internal regret}. In words, a player has no internal regret (or his strategy is internally consistent) if he has no external regret on the set of stages on which he chose a specific given action.

Formally,  choices of action $a_n \in \cA$ and $b_n \in \cB$ generate, besides an external regret $r_n$, an internal regret $R_n$ which is an $A \times A$-matrix whose raw are null except the $a_n$-th one which is $r_n'$; stated otherwise
\[
R_n^{a,a'}=R(a_n,b_n)^{a,a'} := \left\{ \begin{array}{cl} \rho(a',b_n)-\rho(a_n,b_n) & \text{if}\ a = a_n \\ 0 & \text{otherwise} \end{array}\right. .
\]
Let us introduce here some notations. Given two sequences $g_n \in \R^d$ and $a_n\in\cA$, recall that $\og_n$ denote the average up to stage $n$. We define, for every $a \in \cA$, the following subset of stages and conditional averages
\[ \N_n[a]:= \Big\{ m \in \{1, \ldots, n\}\ \text{s.t.}\ a_m=a\Big\} \ \text{and} \ \og_n[a]= \frac{\sum_{m\in \N_n[a]} g_m}{|\N_n[a]|}.
\]

\begin{definition}
A strategy $\sigma$ is internally consistent if, no matter the strategy $\tau$ of Nature, $\P_{\sigma,\tau}$-almost surely,
\[ \limsup_{n \to \infty} \|\oR_n^+\|_\infty \leq 0 \ \text{or equivalently} \ \limsup_{n \to \infty}  \max_{a \in \cA} \frac{|\N_n[a]|}{n}\left( \max_{a^* \in \cA} \rho\big(a^*,\ob_n[a]\big)-\overline{\rho}_n[a] \right) \leq 0.
\]
\end{definition}
It is compulsory to multiply the regret accumulated on $\N_n[a]$ by the frequency of action $a$, namely $|\N_n[a]|/n$. Otherwise internally consistent strategies would not exist. However, another possible formulation (see Lehre \& Solan~\cite{LehSol07}) is to require that
\[\limsup_{n\to\infty}\max_{a^* \in \cA} \rho\big(a^*,\ob_n[a]\big)-\overline{\rho}_n[a]\leq0, \quad \text{for every action}\ a \in \cA\ \text{s.t.}\ \lim_{n \to \infty} |\N_n[a]| = \infty,\]
but, unfortunately, this definition does not allow to measure internal regret at a given finite stage $n$.

Existence part of the following theorem is first due to Foster \& Vohra~\cite{FosVoh97}; rates of convergence (constant are not optimal, see e.g. Stoltz \& Lugosi~\cite{StoLug05}) can be inferred from  rates of external regret, as showed in Section~\ref{SE:ExtToInt} where proof is postponed.
\begin{theorem}
There exist internally consistent strategies such that, for every $n \in \N$
\[ \E_{\sigma,\tau}\left[\max_{a \in \cA} \frac{|\N_n[a]|}{n}\left( \max_{a^* \in \cA} \rho\big(a^*,\ob_n[a]\big)-\overline{\rho}_n[a] \right)\right]=\E_{\sigma,\tau}\Big[\left\|\oR_n^+\right\|_\infty\Big] \leq 3\sqrt{\frac{\log(A)}{n}}.
\]
\end{theorem}

Regret has been refined further by Blum \& Mansour~\cite{BluMan05} into swap-regret (or $\Phi$-regret). Define, for every mapping $\phi : \cA \to \cA$,  family $\Phi \subset \{ \phi : \cA \to \cA \}$ and  $n \in \N$,
\[ \overline{R^\phi}_n=\frac{1}{n}\sum_{m=1}^n \rho(\phi(a_m),b_m) - \rho(a_m,b_m) \ \text{and}\ \overline{R^\Phi}_n= \left( \overline{R^\phi}_n\right)_{\phi \in \Phi} \in \R^{|\Phi|}.\]

\begin{definition}
A strategy $\sigma$ has no $\Phi$-regret\ if, no matter the strategy $\tau$ of Nature, $\P_{\sigma,\tau}$-almost surely,
\[ \limsup_{n \to \infty} \left\|\overline{R^\Phi}_n^+\right\|_\infty  \leq 0 \ \text{or equivalently} \ \limsup_{n \to \infty}  \max_{\phi \in \Phi}\frac{\sum_{m=1}^n \rho(\phi(a_m),b_m)}{n} - \overline{\rho}_n \leq 0.
\]
\end{definition}

Existence of such strategies is due to Blum \& Mansour~\cite{BluMan05}; proofs are again delayed.
\begin{theorem}
There exists strategies without $\Phi$-regret such that, for every $n \in \N$
\[ \E_{\sigma,\tau}\left[ \max_{\phi \in \Phi}\frac{\sum_{m=1}^n \rho(\phi(a_m),b_m)}{n} - \overline{\rho}_n \right]=\E_{\sigma,\tau}\Big[\left\|\overline{R^\Phi}_n^+\right\|_\infty \Big] \leq 3\sqrt{\frac{\log(|\Phi |)}{n}}.
\]
\end{theorem}

The notion of $\Phi$-regret is a refinement of respectively external and  internal regret, because of the specific choices of families $\Phi_e:=\Big\{\phi_{a^*} ; \forall a^* \in \cA, \phi_{a^*}(a)=a^*, \forall a \in \cA\Big\}$ or $\Phi_i:=\Big\{\phi_{a',a^*} ;\ \forall a',a^* \in \cA, \phi_{a',a^*}(a')=a^* \ \text{and}\ \phi_{a',a^*}(a)=a \ \text{if}\ a \neq a' \Big\}$. Proposition~\ref{PR:PhiExtInt} links the different aforementioned quantities, and shows that minimizing internal regret is, in some sense, enough to minimize each one of them (up to the cost of a factor $A$). We will need the following notation.

Given a family $\Phi\subset \{ \phi : \cA \to \cA \}$, we define the matrix $H_\Phi$ of size $|\Phi|\times A^2$ by
\[ H_\Phi^{\phi,(a,a')}=1\ \ \text{if}\ \phi(a)=a' \ \ \text{and 0 otherwise}.
\]
\begin{proposition}\label{PR:PhiExtInt}
Given any family $\Phi\subset \{ \phi : \cA \to \cA \}$, one has $\overline{R^\Phi}_n=H_\Phi \oR_n$ where $\oR_n$ is seen as a vector of size $A^2$.  As a consequence, $\left\|\overline{R^\Phi}_n^+\right\|_\infty \leq A \left\|\oR_n^+\right\|_\infty$.

For the specific case of external regret, one also has $\orr_n=\oR_n \mathbf{1}$ (where $\oR_n$ is seen as a matrix and $\mathbf{1}$ is a vector with only ones). The converse is not true as there exist externally consistent strategies with linear internal regret.
\end{proposition}
\textbf{Proof:}
The proof of the first part follows directly from the definitions of internal and $\Phi$-regret. For the existence of externally consistent strategies with linear internal regret, we refer to Stoltz \& Lugosi~\cite{StoLug05}.
 \qed

Another refinements of these concepts can be made, following this time Fudenberg \& Levine~\cite{FudLev99} and Lehrer~\cite{Leh03}, in two different directions. The first one is to assume that regret is computed not at every stages, but only on a restricted subset of stages (that might depend on the history) and the second direction is to consider time varying switch-mapping $\phi$. Formally, let $\Chi$ be an activation function, i.e., $\Chi: H \times \cA \to \{0,1\}$ and $\Chi(h^n,a_{n+1}) = 1$ indicated that the stage $n+1$ is active. We recall that $H$ stands for the set of all finite histories.. A switch function $\phi: H \times \cA \to \cA$ indicates that, after the finite history $h^{n}$, $\rho(a_{n+1},b_{n+1})$ will be compared to $\rho(\phi(h^n,a_{n+1}),b_{n+1})$.

\begin{definition}
Given an activation mapping $\Chi$ and a switch mapping $\phi$, a strategy $\sigma$ has no $(\Chi,\phi)$-regret if, no matter the strategy $\tau$ of Nature, $\P_{\sigma,\tau}$-almost surely
\[
\limsup_{n \to \infty} \frac{\sum_{m=1}^n\Chi(h^{m-1},a_m)\left[\rho(\phi(h^{m-1},a_m),b_m)-\rho_m\right]}{\sum_{m=1}^n\Chi(h^{m-1},a_m)} \leq 0,
\]
as soon as $\sum_{m=1}^n\Chi(h^{m-1},a_m)$ converges to $+\infty$.
\end{definition}
Lehrer~\cite{Leh03} has proved that, given a probability $\lambda$ on the whole set of pairs of activations-switch mappings (embedded with the product topology), there exists a strategy without $(\Chi,\phi)$-regret, for $\lambda$-almost all pairs. However, rates of convergence are not explicit, in part because we divide the score by the number of actives stages $\sum_{m=1}^n\Chi(h^{m-1},a_m)$ and not by $n$.

\subsubsection{Reductions : form external to $\Phi$-regret}\label{SE:ExtToInt}

In this section, we show how to construct a strategy with no $\Phi$-regret based on an algorithm that only outputs externally consistent strategies, developing an idea of Stoltz \& Lugosi~\cite{StoLug05} and recovering the more general result of Blum \& Mansour~\cite{BluMan05}. Indeed, consider the following auxiliary game where  action sets of the player and Nature are respectively $\Phi$ and a compact subset $\cU \subset [0;1]^A$. Given an exogenous sequences $p_n \in \Delta(\cA)$ we define the payoff at stage $n$ of the player generated by the choices of $\phi \in \Phi$ and $U_n \in \cU$ by
\[
V_n^\phi = \sum_{a \in \cA} p_n[a] U_n^{\phi(a)}:=\sum_{a \in \cA} p_n\circ \phi^{-1} [a] U_n^a \ .
\]
Let $\theta$ be an externally consistent strategy and $\theta_n^\phi$ denote the weight put by $\theta$ on $\phi$ at stage $n$. Then the expected external regret at this stage is written as
\begin{align*}r_n^{\phi'}  =  V_n^{\phi'}-\E\left[V_n^{\phi_n}\right] & =\sum_{a \in \cA} p_n [a] U_n^{\phi'(a)} - \sum_{\phi \in \Phi} \theta_n^\phi \sum_{a \in \cA} p_n[a] U_n^{\phi(a)}\\
& = \sum_{a \in \cA} p_n [a] U_n^{\phi'(a)} - \sum_{a \in \cA} \left(\sum_{\phi \in \Phi} \theta_n^\phi  p_n \circ \phi^{-1} [a] \right) U_n^a
\end{align*}

On the other hand, the strategy that dictates to play $p_n$ at stage $n$ in the original game suffers an expected $\Phi$-regret defined by
\[ \E\left[R_n^{\phi'}\right]= \sum_{a \in \cA} p_n[a]U_n^{\phi'(a)}- \sum_{a \in \cA} p_n[a] U_n^a\ .
\]
So, as soon as  $p_n[a]=\sum_{\phi \in \Phi} \theta_n^\phi  p_n \circ \phi^{-1} [a]$ for every $a \in \cA$,   $\Phi$-regret in the original game and external regret in the auxiliary game coincide exactly (in expectation). And the latter converges to zero, at the same speed of the former, i.e., at rates indicated by Theorem~\ref{TH:RegExt} and Corollary~\ref{CO:RegExt}.

The existence of such a $p_n$ is a simple consequence of Brouwer fixed point theorem. Indeed, first, notice that  $\theta_n$ depends only on the past observations, thus is independent of $p_n$. As a consequence, $p_n$ can be taken as any fixed point of the continuous mapping $p \mapsto \sum_{\phi \in \Phi} \theta_n^\phi p\circ \phi^{-1}$ from the simplex $\Delta(\cA)$ to itself.

\medskip

We only have proved the convergence of $\Phi$-regret in expectation; as usual, almost sure convergence is a consequence of concentration inequalities (or see Theorem 2.7, page 47 and Example 1, page 19, in Hall \& Heyde~\cite{HalHey80}).
\subsection{Compact action spaces, generalizations and examples}
\subsubsection{Compact action spaces}

Although $\Phi$-regret can be seen as a consequence of external or internal regret in the finite case (when $\cA$ is finite), its introduction is more useful in the following compact case.

Assume that $\cA$,   action space of the player, is no longer finite but a compact subset of some Euclidian space.  On the other side, $\cU$,  action space of Nature, is a subset of mappings from $\cA$ to $\R$. Choices of $a_n \in \cA$ and $U_n \in \cU$ generate, at stage $n$, a payoff  of $\rho_n:=U_n(a_n)$.

External regret is defined almost exactly as before, i.e., $r_n: \cA \to \R$ is a continuous mapping defined by $r_n(a)=U_n(a)-\rho_n$. In the compact case, we must however be careful in the order of quantifiers when passing to  limits: a  strategy $\sigma$ is externally consistent if, for all strategy $\tau$ of Nature, $\P_{\sigma,\tau}$-almost surely,
\[\sup_{a^* \in \cA} \limsup_{n\to\infty} \oU_n(a^*)-\overline{\rho}_n \leq 0, \ \text{or equivalently, } \  \left\| \limsup_{n \to \infty} \orr_n^+\right\|_{\infty} \leq 0.
\]

\begin{remark} We claimed that order of quantifiers has some importance. Assume that $\cA=[0,1]$ and that for every $n \in \N$ and $a \in [0;1]$,  $U_n(a) = \mathds{1}_{a \in (0,1/n)}$. Choosing always the same fixed action $a^*$ gives zero as an asymptotic average payoff, therefore the strategy that plays  $a_n=0$ should not have any regret (neither external, internal, or $\Phi$ for that matter).

On the other hand, for every $N \in \N$, the choice of $a^*=1/2N$ gives  $\oU_N(a^*)=1$, thus $\limsup_{n\to\infty} \sup_{a^* \in \cA}  \oU_n(a^*)-\overline{\rho}_n =1$. This explains the choices in the order of quantifiers in the definition.\end{remark}

Difficulties arise to define internal regret, because  scores are multiplied by frequencies of actions in the finite case. We shall instead only focus on $\Phi$-regret, whose definition is also identical: $R_n^\Phi: \Phi \to \R$ is a mapping defined by $R_n^\Phi (\phi) = U_n(\phi(a_n))-U_n(a_n)$. And a  strategy $\sigma$ has no $\Phi$-regret if, for all strategy $\tau$ of Nature, $\P_{\sigma,\tau}$-almost surely,
\[ \sup_{\phi \in \Phi} \limsup_{n\to\infty} \frac{1}{n}\sum_{m=1}^nU_n(\phi(a_n))-U_n(a_n) \leq 0, \ \text{or equivalently, } \  \left\| \limsup_{n \to \infty}\overline{R^\Phi}_n^+\right\|_{\infty} \leq 0.
\]

If $\cA$ is not compact but $(\cA,\cF,\mu)$ is a probability space, then external and $\Phi$-regret can also be defined $\mu$-almost surely. The supremum over $\Phi$ is simply replaced by for $\mu$-almost every mappings $\phi \in \Phi$.

\subsubsection{Generalizations}\label{SE:GenReg}

The whole concept of regret minimization can be extended beyond the comparison of averages of scalar payoffs. Let $g : \cA \times \cB \to \R^d$ be a vector valued payoff mapping and define a sequence of evaluation mapping $B_n : \left(\R^d\right)^n \to \R$ and a class $\Xi$ of departure sequence $\xi[g]: \cA \times \cB \to \R^d$. Then a strategy $\sigma$ has no generalized regret if
\[ \sup_{\xi \in \Xi} \limsup_{n \to \infty} B_n\Big(\xi[g](a_1,b_1),\ldots,\xi[g](a_n,b_n)\Big)-B_n\Big(g(a_1,b_1),\ldots,g(a_n,b_n)\Big) \leq 0\ .
\]
almost surely, no matter the strategy of Nature.

Of course, without additional assumptions on the sequences $B_n$ and $\Phi$,  generalized regret cannot be minimized.  Rakhlin, Sridharan \& Tewari~\cite{RakSriTew11} have used the min-max techniques to infer the existence of such strategies (associated with rates of convergences) on specific cases:

\begin{itemize}
\item[i)]External, internal and $\Phi$-regret are obtained if $g=\rho$, $B_n(z_1,\ldots,Z_n)=\frac{1}{n}\sum_{m=1}^nz_m$ and, for every $\phi  \in \Phi$, there exists $\xi \in \Xi$ such that $\xi[g](a,b)=\rho(\phi(a),b)$.
\item[ii)]Approachability of a convex $\cC$ if $B_n(z_1,\ldots,z_n)=-d_{\cC}\left(\frac{1}{n}\sum_{m=1}^n z_m\right)$ and the departure mappings are $\xi[g](a,b) \in \cC$ for any $a \in \cA$ and $b\in \cB$.
\item[iii)]When $B$ is a function of the average, i.e., $B_n(z_1,\ldots,z_n)=G\left(\frac{1}{n}\sum_{m=1}^nz_m\right)$, an interesting (yet maybe counterintuitive) property arises even in the finite case. There might exist strategies that are not externally consistent yet internally consistent, in the sense that,
\[  \limsup_{n \to \infty}  \max_{x^* \in \Delta(\cA)}\   G\Big(\rho(x^*,\ob_n)\Big) - G\Big(\overline{\rho}_n\Big)>0 \] but for every $a \in \cA$
      \[  \limsup_{n \to \infty} \frac{|\N_n[a]|}{n} \bigg( \sup_{x^* \in \Delta(\cA)}G\Big(\rho(x^*,\ob_n\Big)-G\Big(\overline{\rho}_n[a]\Big)\bigg) \leq 0\ .
    \]
\end{itemize}
\subsubsection{Experts}
An interpretation -- which is actually also a generalization -- of these results  concerns games of predictions with expert advices, studied (almost exhaustively) by Cesa-Bianchi \& Lugosi~\cite{CesLug06}. At each stage $n \in \N$, an agent must take a decision $d_n$ in some topological convex and compact set $\cD$. He is advised by a pool $\cE$ of  experts, i.e., expert $e$ suggests to choose the decision $d_n^e$ at this stage. Once his choice his made, Nature reveals the state of the world $s_n \in \cS$ (where $\cS$ is some arbitrary space) which generate a loss $L_n:=L(d_n,s_n)$.

After $n$ stages, the agent has suffered an average loss of $\oL_n=\frac{1}{n}\sum_{m=1}^n L(d_m,s_m)$ while the best expert had incurred an average loss  of $\oL_n^\star=\frac{1}{n}\min_{e \in \cE} \sum_{m=1}^nL(d_m^a,s_m)$. An evaluation criteria of a strategy of an agent compare these two quantities, as was done by Auer, Cesa-Bianchi \& Gentile~\cite{AueCesGen02}.

\begin{corollary}
If $L$ is convex and has value in $[0;1]$, then there exists an algorithm such that
\[ \oL_n-\oL_n^\star \leq 2\sqrt{\frac{\log(|\cE|)}{n}}.
\]
\end{corollary}
\textbf{Proof:} Consider an externally consistent strategy $\sigma$ given by Theorem~\ref{TH:RegExt}, where the action set is the set of experts and the payoff at stage $n$ of choosing expert $e$ is $\rho(e,s_n)=-L\left(d_{n}^e,s_n\right)$. Denote by $x_{n+1} \in \Delta(\cE)$ the mixed action dictates by $\sigma$ at stage $n+1$. It induces the  decision $d_{n+1}=\sum_{e \in \cE} x_{n+1}[e] d_{n+1}^e$ which satisfies, by convexity of $L$:
\[ L(d_{n+1},s_{n+1}) \leq \sum_{e \in \cE} x_{n+1}[e] L(d_{n+1}^e,s_{n+1}) = \E_{\sigma,s_{n+1}} \left[\rho(e,s_{n+1})\right].\]
Therefore $\oL_n-\oL_n^\star$ is smaller than the expected regret of $\sigma$, hence the result. \qed

\subsection{Links with game theory}\label{SE:RegretLinks}
\subsubsection{Regret and sets of equilibria}
Existence of consistent strategies can be used to prove classical game theory results: non-emptiness of Hannan (or correlated) sets and min-max theorems, as noticed by Blum \& Mansour~\cite{BluMan05} and Cesa-Bianchi \& Lugosi~\cite{CesLug06}.

Consider a game between  a set of  players   $\cI$ of size $I$,   where $\cA_i$ denotes the finite action space of player $i$ and $\rho_i : \prod_{i \in \cI} \cA_i \to \R$ his payoff function (extended multi-linearly as usual). Hannan set of player $i$ is the subset of joint distributions of actions defined by
\begin{align*}
\cH_i & = \left\{ q \in \Delta\left(\prod_{i \in \cI} \cA_i\right)\ ;\ \rho_i\big(a, q_{-i}\big) \leq \rho_i(q), \forall a \in \cA_i \right\}\\
& =  \left\{ q \in \Delta\left(\prod_{i \in \cI}\cA_i\right)\ ; \  \max_{a^* \in \cA_i} \rho_i\big(a^*, q_{-i}\big) - \rho_i(q) \leq 0 \right\}\ ,
\end{align*}
where $\rho_i(q)=\E_q [\rho_i]$ and $q_{-n}$ is the marginal of $q$ on $\prod_{j \neq i} \cA_j$, i.e., the empirical joint distribution of actions played by the opponents of player $i$. Informally, a joint distribution $q$ belongs to $\cH_i$ if  player $i$ has no interest to always play a fixed action $a^* \in \cA_i$ if his opponents coordinate to play accordingly to $q_{-i}$.

By linearity of  $\rho_i$, if a strategy of player $i$ is externally consistent (independently of the behavior of its opponents), then necessarily the empirical joint distribution of actions converges to $\cH_i$. We qualify this property as  \textsl{unilateral}, as it does not make any assumption on opponents' strategies.

If every player follows unilaterally an externally consistent strategy (but not necessary output by the same algorithm), then  empirical distributions of actions will converge to the Hannan set of the game, $\overline{\cH}=\cap_{i \in \cI} \cH_i$, which is therefore guaranteed to be non empty.

The main difference between elements of Hannan set and Nash equilibria is that in the latter the distribution must be a product distribution. So set of Nash equilibria is always contained, but might be in some arbitrary game, much smaller than $\overline{\cH}$.

\bigskip

On the other hand, in zero sum game, elements of Hannan set satisfy the following property. If $q \in \Delta(\cA \times \cB)$ belongs to $\overline{\cH}$, then if we denote by $q_1 \in \Delta(\cA)$ and $q_2\in\Delta(\cB)$ its marginals, necessarily
\[ \min_{y \in \Delta(\cB)} \max_{x \in \Delta(\cA)} \rho(x,y) \leq \max_{a \in \cA} \rho(a,q_2) \leq \rho(q) \leq \min_{b \in\cB} \rho(q_1,b)\leq \max_{x \in \Delta(\cA)} \min_{y \in \Delta(\cB)} \rho(x,y)\ .
\]
Since $ \max_{x \in \Delta(\cA)} \min_{y \in \Delta(\cB)} \rho(x,y)\leq \min_{y \in \Delta(\cB)} \max_{x \in \Delta(\cA)} \rho(x,y) $ always holds, both quantities must coincide and, by definition, are equal to the value $v$ of the game. More importantly, the first and last inequality above imply that
\[ \max_{a \in \cA} \rho(a,q_2)=\max_{x \in \Delta(\cA)} \rho(x,q_2)=v\  \text{and}\ \min_{b\in\cB} \rho(q_1,b)=\min_{y \in \Delta(\cB)} \rho(q_1,y) =v,\]
thus $(q_1,q_2)$ is a pair of \textsl{optimal mixed actions}.

As a consequence, in a zero sum game, if players follows unilaterally consistent strategies, they will obtain asymptotically at least the value. And if both players have consistent strategies, their empirical mixed action converges to their set of optimal mixed actions.

\medskip

This property has been somehow generalized by Hart and Mas-Colell~\cite{HarMas03a} in potential games, see also Viossat \& Zapechelnyuk \cite{VioZap13}. They  have constructed a specific externally consistent strategy such that, if both players follows it, the product of empirical distributions of actions converges to the set of Nash equilibria (and more precisely to a subset of it whose payoff are identical). However, this is only a \textsl{global} property (as opposed to unilateral properties) as both players must follow this specific strategy. Moreover, the result does not extend to any game, even those with an unique  Nash equilibria.

\bigskip

We proved, following Cesa-Bianchi \& Lugosi \cite{CesLug06} and Sorin \cite{Sor02},  a min-max theorem due to von Neumann using externally consistent strategies. It is actually possible to get the following generalized version of Fan~\cite{Fan53}. We first recall that a  mapping $\rho$ on $\cA \times \cB$ is said to be concave-like if for every $a, a' \in \cA$ and $\alpha \in [0,1]$, there exists $a^* \in \cA$ such that $\rho(a^*,\cdot) \geq \alpha \rho(a,\cdot)+(1-\alpha)\rho(a',\cdot)$.  Convexity-like is defined similarly.

\begin{theorem} Let $\cA$ be a compact set, $\cB$ any set and $\rho$ a concave-like convex-like mapping on $\cA \times \cB$ bounded from below and such that $g(\cdot,b)$ is upper-semicontinuous for every $b \in \cB$. Then the zero-sum game on $\cA$ and $\cB$ has a value.
\end{theorem}
\textbf{Proof:} Let $\cB'$ be any finite subset of $\cB$ and consider an externally consistent strategy of the first player; its existence is ensures by the following Corollary \ref{CR:RegretUsc}. It also implies that, for every $\varepsilon >0$, there exists a sequence $\delta_n\geq0$ going to zero such that, at stage $n$, \[
\inf_{b\in\cB} \max_{a \in \cA} \rho(a,b) \leq \max_{a \in \cA} \rho(a,b^*_n) \leq  \max_{a \in \cA}\frac{1}{n}\sum_{m=1}^n \rho(a,b_m)  \leq \overline{\rho}_n +\varepsilon+\delta_n\ , \]
where $b^*_n$ is given by the definition of convexity-like  applied to $\sum_{m=1}^n b_m/n$. On Nature's side, we can assume that her strategy is such that, at stage $n$, $b_n$ is an action realizing $\inf_{b \in \cB'} \rho(a_n,b)$ up to $1/2^n$. As a consequence,
\[\overline{\rho}_n \leq \frac{1}{n}\sum_{m=1}^n   \min_{b \in \cB}\rho(a_m,b) +\frac{1}{2^m} \leq   \min_{b \in \cB} \frac{1}{n}\sum_{m=1}^n \rho(a_m,b) + \frac{1}{n}\leq \min_{b \in \cB'} \rho(a^*_n,b) + \frac{1}{n}
\]
where $a^*_n$ is given by the definition of concavity-like. As a consequence, taking $n$ and $\varepsilon$ to their limits yields that, for any finite subset $\cB'$,
\[\inf_{b\in\cB} \max_{a \in \cA} \rho(a,b) \leq \sup_{a\in\cA} \inf_{b\in\cB'}\rho(a,b).
\]
Since $\cA$ is compact and $\rho(\cdot,b)$ is upper-semicontinuous, for every $\varepsilon$ and $\cB'$ the set
\[ \cA_\varepsilon[\cB']=\left\{ a \in \cA \ \text{s.t} \ \rho(a,b) \geq \inf_{b\in\cB} \max_{a \in \cA} \rho(a,b)-\varepsilon, \forall b \in \cB'\right\}
\]
is a compact non-empty set, and this remains true for any  finite intersection over different subsets. As a consequence, the whole intersection (over every $\varepsilon$ and $\cB'$) remains compact and non-empty, and any point $a$  in it must satisfy that $\rho(a,b) \geq \inf_{b\in\cB} \max_{a \in \cA} \rho(a,b)$, for every $b \in \cB$. Stated otherwise,
\[\inf_{b\in\cB} \max_{a \in \cA} \rho(a,b) \leq \max_{a\in\cA} \inf_{b\in\cB}\rho(a,b)
\]
and the game has a value.
\qed

Stronger results can be proved using internally consistent strategies. Aumann~\cite{Aum74} defined correlated equilibria in a game as a distribution on the set of profiles of action $q \in \Delta\left(\prod_{i \in \cI} \cA_i\right)$ such that, for every player $i \in \cI$ and every action $a \in \cA_i$:
\[ \rho_i\big(a, q_{-i}[a]\big) \geq \max_{a^* \in \cA_i}  \rho_i\big(a^*, q_{-i}[a]\big), \ \text{or} \ q_i[a]\Big(\max_{a^* \in \cA_i}  \rho_i\big(a^*, q_{-i}[a]\big)-\rho_i\big(a, q_{-i}[a]\big) \Big) \leq 0
\]
where  $q_{-i}[a] \in \Delta\left(\prod_{j\neq i} \cA_j\right)$ is the probability induced by $q$ knowing that $a_i=a$ and $q_i[a]$ is the probability put on $a \in \cA_i$ by $q$ (or the relative frequency of action $a \in \cA_i$). In words, assume that a referee draws a lottery accordingly to $q$ and only tells player $i$  an action he should play. Then, a correlated equilibrium is a joint distribution such that  every player, when he is told to play action $a$ (and assuming that the others follows their recommendation), cannot gain strictly more by playing $a^*$ instead of $a$.

It is quite clear (from their very definition) that if every player follows unilaterally an internally consistent strategy then the empirical distribution of actions converges to the set of correlated equilibria (but maybe not to one specific correlated equilibrium), see Foster \& Vohra~\cite{FosVoh97}.

\subsubsection{Regret, (smooth) fictitious play and follow the perturbed leader}

Fictitious play is a classic unilateral discrete time dynamic in game theory. At stage $n$, each player computes  empirical (either joint or product) distributions of actions of his opponents and plays a best response to it. Although quite natural, this strategy is not externally consistent. On the contrary, Fudenberg \& Levine~\cite{FudLev99} have introduced a slight modification, called \textsl{smooth fictitious play} that has asymptotically a regret smaller than $\varepsilon$ (where $\epsilon>0$ is fixed), see also Hofbauer,  Sorin \& Viossat~\cite{HofSorVio09}.

Let $\rho_\varepsilon$ denotes an $\varepsilon$-perturbation of  $\rho$ (induced by $\psi: \Delta(\cA) \to \R$) defined by
\[ \rho_\varepsilon(x,y) = \rho(x,y)+ \varepsilon \psi(x), \ \forall y \in \Delta(\cB).
\]
Since we are interested in unilateral procedure, we might as well make a change of variable by defining $U = \Big(\rho(a,y) \Big)_{a \in \cA} \in [0;1]^A$ so that $\rho(x,U)= \langle x, U\rangle$. As a consequence, the mapping $\rho_\varepsilon$ can be rewritten as
\[ \rho_\varepsilon(x,U)=\langle x, U \rangle + \varepsilon \psi(x).
\]
We also define the $\varepsilon$-best response mapping by $\BR_\varepsilon(U)=\argmax_{x \in \Delta(cA)} \langle x,U\rangle + \psi(x)$.

\bigskip

We assume that the mapping  $\psi : \Delta(\cA) \to \R$ is chosen so that
\begin{itemize}
\item[i)] $\psi$ is a continuously differentiable mapping and $\|\psi\|_{\infty}\leq 1$;
\item[ii)] The $\varepsilon$-best response mapping $ \BR_\varepsilon$ is univoque and continuous;
\item[iii)] $\BR_\varepsilon(U)$ does not belong to the boundary of $\Delta(\cA)$.
\end{itemize}
Actually, point iii) ensures that $\rho_\varepsilon$ attains its maximum at a point where its first derivative vanishes. It can  therefore be weaken into one of the following
\begin{itemize}
\item[iii')] for every $U \in [0;1]^A$, $D_1 \rho_\varepsilon(\cdot,U)$ is equal to zero at $x =\BR_\varepsilon(U)$
\item[iii'')] $D_1 \rho_\varepsilon(\cdot,U)$ is orthogonal to the gradient of $\BR_\varepsilon$ at $U$.
\end{itemize}
\medskip

Study of $\sigma(h^n)=\BR_\varepsilon(\oU_n)$, the strategy associated with this perturbation, might be simpler in continuous time. First, we introduce the mapping $W: [0;1]^A \to \R$ defined by
\[
W_\varepsilon(U)= \sup_{x \in \Delta(\cA)} \rho_\varepsilon(x,U)= \langle \BR_\varepsilon(U),U\rangle+\varepsilon\psi(\BR_\varepsilon(U))\ .\]
In particular, because of point i),   regret  is asymptotically smaller than $2\varepsilon$ as soon as
 $\limsup_{n \to \infty} W_\varepsilon(\oU_n) - \overline{\rho}_n \leq \varepsilon$. As in Section~\ref{SE:AppContTime}, the  continuous-time dynamic associated  with the discrete-time dynamic of $(\oU_n,\overline{\rho}_n)$ is
 \[ \left(\dot \bU,  \dot{\boldsymbol{\rho}}\right) \in \Big\{ (V, \langle \BR_\varepsilon(\bU), V \rangle \ ; \ V \in [0;1]^A \Big\} - \Big(\bU, \boldsymbol{\rho}\Big).
 \]
 Define $\lambda(t)=W_\varepsilon(\bU(t))-\boldsymbol{\rho}(t)$ then one has $\dot \lambda + \lambda \leq \varepsilon$ thus $\lambda(t) \leq \varepsilon + Me^{-t}$ for some constant $M$.
 As a consequence, $\lambda$ is a Lyapounov function with respect to the set
 \[ \Big\{ (U,\rho) \in \R^A \times \R\ ; \ W_\varepsilon(U)-\rho\leq \varepsilon \Big\}
 \]
 which is thus a global attractor of the dynamic (see Bena\"im, Hofbauer \& Sorin~\cite{BenHofSor06}). So $(\oU_n,\overline{\rho}_n)$ converges almost surely to it and the strategy is $\varepsilon$-externally consistent. Bena\"im \& Faure~\cite{BenFau12} proved recently that external consistency can be achieve (without requiring a doubling trick argument) with a smooth fictitious play with a vanishing step size; indeed, they showed that if  $\varepsilon$ is not fixed but depends on $n \in \N$ as $\varepsilon_n=n^\gamma$, with $\gamma <1$, then asymptotically the regret converges to zero.

 \bigskip

	Smooth fictitious play (also known as \textsl{follow the regularized leader}) is a generalization of two classes of algorithms,  \textsl{exponential weight algorithms} or  its even more general version called \textsl{follow the perturbed leader} (see Cesa-Bianchi \& Lugosi~\cite{CesLug06}, Sections 4.2 and 4.3). To recover the first class of algorithms, entropy must be used as regularization, i.e., $\psi(x)=-\sum_{a \in \cA} x[a]\log(x[a])$ to get
\[ \BR_{1/\eta}(U)[a]= \frac{\exp(\eta\, U^a)}{\sum_{a' \in \cA} \exp(\eta\, U^a)}
	\]
	which is, by definition, the exponential weight algorithm.

Links with \textsl{follow the perturbed leader} (or Stochastic Fictitious Play accordingly to Fudenberg \& Kreps~\cite{FudKre93}) might be a bit more tedious. This algorithm does not choose a deterministic regularization $\varepsilon \psi$ but perturbs each component of $\oU_n$ by a random quantity $\varepsilon_n^a$, such that the joint density $f: \R^A \to \R$ of the vector $\big(\varepsilon_n^a\big)_{a \in \cA}$ is independent of $\oU_n$ and $n$. Action played at stage $n+1$ is any maximizer of $\oU_n^a+\varepsilon_n^a$. In particular, a given action $a$ is chosen at this stage with probability $X^a(\oU_n)$ where $X^a(\cdot)$ is defined by
\[ X^{a}(U) = \P\big\{\argmax_{a' \in \cA} U^{a'}+\varepsilon^{a'} = a  \big\}\ .
\]
Follow the Perturbed Leader generates a discrete stochastic process $(\oU_n,\overline{\rho}_n)$ which is an A.S.D. of the following differential inclusion
\[
\Big(\dot\bU,\dot{\boldsymbol{\rho}}\Big) \in \Big\{ \left(V, \langle X(\bU),V \rangle \right) \ ; \ V \in [0;1]^A \Big\} - \Big( \bU,\boldsymbol{\rho}\Big) \ .
\]
This is a special case of Smooth Fictitious Play since, as soon as $f$ is positive and $X$ is continuously differentiable, Hofbauer \& Sandholm~\cite{HofSan02} have shown that there exists a deterministic regularization $\varepsilon \psi$ such that $X(U)=\BR_{\varepsilon}(U)$. For example, in the case where $\varepsilon^a$ are i.i.d.\ with cumulative distribution $F(x)=\exp\big(-\exp(-\eta x -\gamma)\big)$ (where $\gamma$ the Euler constant), follow the perturbed leader coincides exactly with  exponential weight algorithm (see e.g., Lemma 1 in McFadden~\cite{McF74}).

\bigskip

As mentioned before, proofs based on A.S.D. do not exhibit rates of convergences (and this might be seen a major drawback of these techniques). However, we only considered here strategies that do not depend on the past sequence of player's actions (but only on the sequence of Nature's choices).  So the discrete process is very closed to the one induced by  procedures in law (this is not the case for approachability, see Section~\ref{SE:AppComLaw}) which is in turn close to the continuous-time process. And it is actually possible to quantify explicitly these relative differences, see e.g.,   Sorin~\cite{Sor09} or Kwon~\cite{Kwo12}, in order to recover exact rates of convergence.

\section{Calibration}\label{SE:Calibration}
We recall that calibration is a criterion introduced by Dawid~\cite{Daw82} in the following repeated games between a player and Nature. At each stage $n \in \N$, Nature chooses a state of the world $\omega_n$ in some finite set $\Omega$ and the player makes a prediction upon its law by choosing a probability distribution $p_n \in \Delta(\Omega)$. Strategies of the player and Nature are  mappings from the set of finite histories $\cup_{n \in \N} (\Omega \times \Delta(\Omega)^n$ into, respectively, $\Delta(\Delta(\Omega))$ and $\Delta(\Omega)$.

The usual example consisting of a meteorologist that predicts each day the probability of rain corresponds to  $\Omega=\{0,1\}$, with $\omega=1$ if it rains. This binary case is in fact much easier than the general case, as discussed in Section~\ref{SE:EfficientCalib}.

\subsection{Finite ($\varepsilon$ and grid) calibration}

We will need the following notations. For every $p \in \Delta(\Omega)$ -- seen as a subset of $\R^{\Omega-1}$ -- and $\varepsilon>0$, let $\N_n[p,\varepsilon]$ be the set of stages where the prediction was $\epsilon$-close to $p$, i.e.,
\[ \N_n[p,\varepsilon] = \Big\{ m \in \{1,\ldots,n\}\ \text{s.t.}\ \|p_m-p\|\leq \varepsilon \Big\} ,
\]
where $\|\cdot\|$ is an Euclidian norm of $\R^{\Omega-1}$. We denote by $\overline{\omega}_n[p,\varepsilon] \in \Delta(\Omega)$ the empirical distribution of states on $\N_n[p,\varepsilon]$ and by $\op_n[p,\varepsilon]$ the average prediction on it.

\begin{definition}\label{DF:EpsCalib}
A strategy $\sigma$ of the player is $\varepsilon$-calibrated if for every strategy $\tau$ of Nature, and for every $p \in \Delta(\Omega)$,
\[ \limsup_{n \to \infty} \frac{|\N_n[p,\varepsilon]|}{n}\bigg(\Big\|\op_n[p,\varepsilon]-\overline{\omega}_n[p,\varepsilon]\Big\| - \varepsilon \bigg) \leq 0, \quad \P_{\sigma,\tau}\text{-as}.
\]
A strategy is calibrated if it is $\varepsilon$-calibrated, for every $\varepsilon>0$.
\end{definition}
 Intuitively, a strategy is $\varepsilon$-calibrated if on the set of stages (assuming that it is big enough) where the prediction was $\varepsilon$-close to some $p \in \Delta(\Omega)$, the empirical distribution of states is close to this specific $p$. Although not stated explicitly in Definition \ref{DF:EpsCalib}, it is possible to require that rates of convergence are independent of Nature's strategy  see Section \ref{SE:REgImplCal} below. With a careful concatenation of $\varepsilon$-calibrated strategies, following the doubling trick, one can easily obtain a calibrated strategies, as did Foster \& Vohra~\cite{FosVoh98} or Fudenberg \& Levine~\cite{FudLev99a}. It remains to construct such strategies, which can be done using the slightly weaker concept of calibrated strategies with respect to an $\varepsilon$-grid of $\Delta(\Omega)$ defined below.

\bigskip

We recall that a finite subset $\big\{ x[\ell]\ ; \ \ell \in \cL\big\}$ of $\cK \subset \R^d$ is an $\varepsilon$-grid of $\cK$ if for every $x \in \cK$, there exists $\ell \in \cL$ such that $\|x-x[\ell]\|\leq \varepsilon$. Moreover, such a grid is regular if there exists $\{e_1,\ldots,e_d\}$, $d$ linearly independent vectors, such that
\[ \Big\{ x[\ell]\ ; \ \ell \in \cL\Big\} = \left \{ \sum_{k=1}^d n_ke_k\ ; \ n_k \in \Z \right\} \cap \cK\ .
\]
Assume that the player can only make predictions on a grid $\big\{ p[\ell]\ ; \ \ell \in \cL\big\}$ of $\Delta(\Omega)$, so that a strategy is a mapping from the finite histories into $\Delta(\cL)$. Empirical distribution of states on $\N_n(\ell):=\big\{ m \in \{1,\ldots, n \ \text{s.t.}\ p_m=p[\ell]\big\}$ is denoted by $\oq_n[\ell]$.
\begin{definition}
A strategy $\sigma$ of the player is calibrated with respect to  $\big\{ p[\ell]\ ; \ \ell \in \cL\big\}$ if for every strategy $\tau$ of Nature, for every $\ell \in \cL$,
\[ \limsup_{n \to \infty} \frac{|\N_n[\ell]|}{n}\bigg(\big\| \overline{\omega}_n[\ell] - p[\ell] \big\|-\min_{k \in \cL}\big\|\overline{\omega}_n[\ell]-p[k]\big\|\bigg) \leq 0,\quad \P_{\sigma,\tau}\text{-as}.
\]
\end{definition}
In words, a strategy is calibrated with respect to a grid if on the set of stages where $p[\ell]$ is predicted, the empirical distributions of states is closer to $p[\ell]$ than to any other $p[k]$.

\begin{remark}\label{RM:Voronoi}
Given a finite grid, the Vorono\"i cell associated with $p[\ell]$ is the set of points closer to $p[\ell]$ than to any other $p[k]$, i.e.,
\[ V[\ell] := \Big\{ p \in \Delta(\Omega)\ \text{s.t.}\ \big\| p - p[\ell] \big\| \leq \min_{k \in \cL}\big\| p -p[k]\big\| \Big\}.\]
Each Vorono\"i is a polytope since they are defined by a finite number of linear inequalities, their union covers $\Delta(\Omega)$ and any intersection has empty interior.
The fact that the calibration score
\[ \frac{|\N_n[\ell]|}{n}\bigg(\big\| \overline{\omega}_n[\ell] - p[\ell] \big\|-\min_{k \in \cL}\big\|\overline{\omega}_n[\ell]-p[k]\big\|\bigg)\]
is non positive means that $\overline{\omega}_n[\ell]$ belongs to (or converges to) the Vorono\"i cell $V[\ell]$.
\end{remark}
Dawid~\cite{Daw85} and Oakes~\cite{Oak85} proved that there does not exist deterministic $\varepsilon$-calibrated strategies, based on a counter example given in the following section. On the other hand, there exists random $\varepsilon$-calibrated strategies, as proved by Foster and Vohra~\cite{FosVoh97}  by exhibiting an algorithm to construct makes the so-called Brier score decrease to zero.

\begin{theorem}
For every grid, there exists a calibrated strategy with respect to it. As a consequence, for every $\varepsilon>0$, there exist $\varepsilon$-calibrated strategies, and thus calibrated strategies.
\end{theorem}

To end this section, we note that finite calibration can also be defined with respect to some weights $\{\nu[\ell] \in \R; \ell \in \cL\}$. 
A strategy $\sigma$ of the player is weighted-calibrated with respect to  $\big\{ p[\ell], \nu[\ell]\ ; \ \ell \in \cL\big\}$ if for every strategy $\tau$ of Nature, for every $\ell \in \cL$,
\[ \limsup_{n \to \infty} \frac{|\N_n[\ell]|}{n}\bigg(\Big(\big\| \overline{\omega}_n[\ell] - p[\ell] \big\|^2-\nu[\ell]\Big)-\Big(\min_{k \in \cL}\big\|\overline{\omega}_n[\ell]-p[k]\big\|^2-\nu[k]\Big)\bigg) \leq 0,\quad \P_{\sigma,\tau}\text{-as}.
\]

\begin{corollary}
For every grid and weights, there exists a calibrated strategy with respect to them.
\end{corollary}

Given $\big\{ p[\ell], \nu[\ell]\ ; \ \ell \in \cL\big\}$ , the Laguerre cell (or Power cell) associated with $p[\ell]$ and $\nu[\ell]$ is 
\[ P[\ell] := \Big\{ p \in \Delta(\Omega)\ \text{s.t.}\ \big\| p - p[\ell]\big\|^2 -\nu[\ell] \leq \min_{k \in \cL}\big\| p -p[k]\big\|^2-\nu[k] \Big\};\]
as in Remark \ref{RM:Voronoi}, a weighted-calibrated strategy ensures that $\overline{\omega}_n[\ell]$ converges, as soon as the frequency of $\ell$ is not zero, to $P[\ell]$. Because of  the  squared norms, this set is also  a polytope.

\subsubsection{Discussion on the impossibility of deterministic calibration}\label{SE:DeterCalibImp}
When $\Omega=\{0,1\}$, Oakes~\cite{Oak85} and Dawid~\cite{Daw85}  output an example of Nature's strategy  ensuring that no $\varepsilon$-deterministic calibrated strategies exist. Their idea is actually quite simple yet highly unstable. Define the strategy as follows: given the past history $h^n$,
\[ \text{if}\ p_{n+1} \geq \frac{1}{2} \ \text{then}\  \omega_{n+1}=0\ \text{and if}\ p_{n+1}< \frac{1}{2}\ \text{then}\ \omega_{n+1}=1;\]
In words, if the forecaster claims that it will rain with high probability then Nature does not make it rain and if it claims that it will not rain, Nature makes it rain.

This prevents any deterministic strategies from being $\varepsilon$-calibrated, but this is not immediate (and the proof, although quite simple will shed lights on the following discussion).  We distinguish two cases, either the predictions of $1/2$ have an asymptotic positive frequency or a null frequency, i.e., if
\[\text{either} \ \limsup_{n \to \infty} \frac{\Big|\big\{m \leq n \ \text{s.t.}\ p_m=1/2\big\}\Big|}{n}>0 \quad \text{or} \quad \lim_{n \to \infty} \frac{\Big|\big\{m \leq n \ \text{s.t.}\ p_m=1/2\big\}\Big|}{n}=0.\]
In the first case, $\op[1/2+\varepsilon,\varepsilon]\geq 1/2$ while $\oq[1/2+\varepsilon,\varepsilon]=0$ thus such a strategy is not $\varepsilon$-calibrated.

In the second case, we can assume that no prediction falls exactly at $1/2$ (since their frequency goes to zero). If the predictions bigger than $1/2$ have an asymptotic positive frequency, then necessarily, there must exist $p^*$ such that the set of stages where predictions belong to $[p^\star-\varepsilon,p^*+\varepsilon] \subset [1/2;1]$ also has a positive frequency. And since $\op_n[p^*,\varepsilon]\geq 1/2$ and  $\overline{\omega}_n[p^*,\varepsilon]=0$, the strategy is not $\varepsilon$-calibrated.

 If the predictions bigger than $1/2$ have an asymptotic null frequency, then necessarily  the predictions smaller than $1/2$ have an asymptotic positive frequency, and the same arguments hold (because we assumed that no predictions were equal to exactly 1/2). So no deterministic strategy can be $\varepsilon$-calibrated.

\bigskip

On the other hand, consider the  deterministic strategy of the player that predicts at odd stages  $p_n=1/2$ and at even stages $p_n=1/2-1/n$. The only accumulation point of the sequence of predictions if $1/2$, so for every $p \neq 1/2$ there are a finite number of prediction $\varepsilon$-close to $p$, for every $\varepsilon$ smaller than some $\varepsilon_p>0$. And on the other hand, for $p=1/2$, no matter $\varepsilon$, if $n$ is big enough, $\N_n[0.5,\varepsilon]$ contains approximatively half predictions below 1/2 and half above, so the empirical distributions is asymptotically equal to 1/2. As a consequence, no matter $\varepsilon>0$ and $p \neq 0.5$,
\[ \limsup_{n \to \infty}\frac{|\N_n[0.5,\varepsilon]|}{n}\Big\|\op_n[0.5,\varepsilon]-\overline{\omega}_n[0.5,\varepsilon]\Big\|=0\ \text{and}\ \lim_{\varepsilon' \to 0} \limsup_{n \to \infty} \frac{|\N_n[p,\varepsilon']|}{n}=0. \]

Obviously, this does not contradict Oakes~\cite{Oak85} and Dawid~\cite{Daw85} counter-example. The reason is that, on the stages when the prediction is  \textsl{$\varepsilon$-close to $p^*=1/2+\varepsilon$}, the average prediction is 1/2 while the empirical state is 0. But one might argue that predictions are actually \textsl{never close} to $p^*$ (but $\varepsilon$-away), so Oakes and Dawid argument fails if calibration was defined only with respect only to those points $p$ that are accumulation points of the sequence of predictions (i.e., there are predictions arbitrarily close to them).

This argument can be generalized to any stationary strategy of Nature (i.e., if $\omega_n=f(p_n)$ for some fixed but possibly random mapping $f$). Unfortunately, we are unable to claim that there exists deterministic ($\varepsilon$-)calibrated strategies with respect to accumulation points, but this shows how the very concept of calibration is unstable with respect to small variations in definition or objectives. This subject is somehow once again developed in Section~\ref{SE:SmoothCalib}.

\subsubsection{Efficient calibration in the binary case}\label{SE:EfficientCalib}

Foster~\cite{Fos99} has designed an algorithm that computes efficiently an $\varepsilon$-strategy in the binary case (although it seems that  it was Abernethy, Bartlett \& Hazan~\cite{AbeBarHaz11} that noticed its efficiency). The idea is to consider a calibrated strategy with respect to the regular  grid  $\big\{p[\ell]:=\varepsilon+2\ell \varepsilon\ ;\ \ell \in \cL \big\}$ where $\cL:=\big\{0,1,\ldots, (\lfloor \varepsilon^{-1}\rfloor-1)/2\big\}$

Following Foster's notation, we define, for every $\ell \in \cL$,
\[e^\ell_n=\frac{|\N_n[\ell]|}{n}\bigg(\overline{\omega}_n[\ell]-(p[\ell]+\varepsilon)\bigg)\ \text{and}\ d^\ell_n=\frac{|\N_n[\ell]|}{n}\bigg(\overline{\omega}_n[\ell]-(p[\ell]-\varepsilon) \bigg)\]
so that a strategy is calibrated if, asymptotically, every $e^\ell_n$ and $d^\ell_n$ are smaller than zero. Foster's algorithm consists in finding at stage $n$ an element $\ell^* \in \cL$ such that
\begin{itemize}
\item[--]either both $e_n^{\ell^*} \leq 0$ and $d_n^{\ell^*} \leq 0$;  in that case, predict $p[\ell^*]$
\item[--] or $e_n^{\ell^*-1}>0$    and $d_n^{\ell^*}>0$; in that case play $p[\ell^*]$ or $p[\ell^*-1]$ with a respective probability  proportional to $d_n^{\ell^*}$ and $e_n^{\ell^*-1}$.
\end{itemize}
Existence of such a $\ell^*$ is ensured by the fact that the first $d^1_n$ and the last $e^L_n$ are always non positive. Computations show that the error converges to zero.

So the tricky  remaining part  consists in finding efficiently this $\ell^*$. To this purpose, Abernethy, Bartlett \& Hazan~\cite{AbeBarHaz11}  introduced, for every $\ell \in \cL$,  the quantity
\[\theta_n^\ell= e^\ell_n\ \text{if}\ e^\ell_n > 0  , \quad \theta_n^\ell= -d^\ell_n\ \text{if}\ d^\ell_n > 0 \quad \text{and} \quad \theta_n^\ell=0 \quad \text{otherwise}  ,\]
which is well defined since $e^\ell_n$ and $d^\ell_n$ cannot be simultaneously positive. Specifically, it always holds that $\theta_n^1 \geq 0$ and $\theta_n^L \leq 0$ so if any of them is equal to zero, Foster's strategy dictates to predict it. Otherwise, one must find $\ell^*$ such that $\theta_n^{\ell^*-1}>0$ and $\theta_n^{\ell^*}<0$ and the main argument is that it can be done through a binary search, thus in $O\left(\log(1/\varepsilon)\right)$ steps.

\bigskip

Foster's strategy can be somehow generalized with more than two outputs (see e.g., Mannor \& Stoltz~\cite{ManSto10}) although, unfortunately, at the cost of efficiency since the binary search trick does not extend.

\subsection{Generalization}
Recall that, roughly speaking, a strategy is calibrated if on the set of stages where the prediction was \textsl{close} to $p$, the average prediction and the empirical distribution of outcome asymptotically coincide. General concepts of calibration are induced by a different definition of \textsl{closeness}.

Let $\cF$ be a family of Borel measurable subsets of $\Delta(\Omega)$ and denote, for every $F \in \cF$,
\[
\N_n[F]= \Big\{ m \leq n\ \text{s.t.}\ p_m \in F\Big\}, \ \overline{\omega}_n[F]=\frac{\sum_{m\in\N_n[F]}\omega_m}{|\N_n[F]|}  \ \text{and} \ \op_n[F]=\frac{\sum_{m\in\N_n[F]}p_{m}
}{|\N_n[F]|} \]
respectively the set of stages where the prediction was in $F$ (before the $n$-th), the empirical distribution of outcomes and the average prediction on it.

\begin{definition}
A strategy $\sigma$ of the player is $\cF$-calibrated if for every strategy $\tau$ of Nature,\[ \limsup_{n \to \infty} \sup_{F \in \cF} \frac{|\N_n[F]|}{n}\Big\| \overline{\omega}_n[F] - \op_n[F] \Big\| \leq 0,\quad \P_{\sigma,\tau}\text{-as}\ .
\]
\end{definition}

Several types of families have been considered by various authors. For instance, from the most to the least complicated
\begin{itemize}
\item[i)] Mannor \& Stoltz~\cite{ManSto10} treated the most difficult problem where $\cF$ is the family of all Borel measurable subsets of $\Delta(\Omega)$;
\item[ii)] Rakhlin, Sridharan \& Tewari~\cite{RakSriTew11} considered the family of every possible $\ell_1$ balls;
\item[ii)] Perchet~\cite{Per10} defined $\cF$ to be some neighborhood basis of $\Delta(\Omega)$.
\end{itemize}

In the first case, the minmax techniques of Rakhlin, Sridharan and Tewari~\cite{RakSriTew10,RakSriTew11} upper-bound the calibration error at stage $n$ (but with a strategy that depends on $n$) in $O\left(n^{-1/(|\Omega|+1)}\right)$ while for the two last cases the bound shrinks to $O\left(n^{-1/2}\right)$. On the other hand, Mannor \& Stoltz~\cite{ManSto10} and Perchet~\cite{Per10} obtained (actually before) the same results, yet in a constructive way. They are developed in Section~\ref{SE:REgImplCal}.

\bigskip

Drawbacks of these definitions of calibration (which will lead to another type of generalization) are  illustrated by the following examples.

Assume that $\Omega=\{0,1\}$ and that the sequence of outcomes is $0,1,0,1,0$... (i.e., $\omega_n=1$ iff n is even). Consider a player that predicts, at every stage, that the probability of $1$ is exactly $1/2$. Then this strategy is calibrated accordingly to any of the previous  definitions of calibration. On the other hand, on the set of even stages,  empirical distribution is 1 while  average prediction is $1/2$ which contradict  precepts of calibration.

Even more intricate: assume that $\Omega=\{0,1,2\}$ that $\omega_n=0$ with probability $1/3$ and that $1$ and $2$ alternates on the remaining set of stages. The sequence of outcomes on any fixed subset of $\N$ contains asymptotically as many 0 than 1 and 2 so predicting $1/3,1/3,1/3$ at every stage is not contradicting. On the other hand, if we consider only the set of stages where the outcome was $1$ or $2$ then the prediction is always $1/2,1/2$ while 1 and 2 alternate.

\bigskip

We introduce the following concepts of checking rules. Let $\cU$ and $\cT$ be respectively an \textsl{active universe mapping} and a \textsl{testing mapping}, i.e.,
\[
\cU : \bigcup_{n \in \N}(\Delta(\Omega)\times \Omega)^n \rightrightarrows \Delta(\Omega)\times \Omega \quad \text{and} \quad \cT : \bigcup_{n \in \N}(\Delta(\Omega)\times \Omega)^n \rightrightarrows  \Delta(\Omega)\times \Omega
\]
such that $\cT(h^n) \subset \cU(h^n)$. The interpretation is that  stage $n+1$ is \textsl{active} if $(p_{n+1},\omega_{n+1})$ belongs to the active universe $U(h^n)$; given a set of active stages, calibration compares the empirical frequency of the \textsl{tested} event with the average prediction of this event.

Such a pair $(\cU,\cT)$ forms a \textsl{checking rule} and we define as before the set of active stages
\[\N_n[\cU,\cT]= \Big\{m \leq n \ \text{s.t.}\ (p_m,\omega_m) \in \cU(h^{m-1}) \Big\} \ ,\]
the empirical probability of tested events
\[ \overline{\omega}_n[\cU,\cT]=\frac{\sum_{m \in \N_n[\cU,\cT]}\mathds{1}\{(p_{m},\omega_m) \in \cT(h^{m-1})\} }{|\N_n[\cU,\cT]|}\ ,\]
and the average predicted conditional probability of tested events
\[\op_n[\cU,\cT]=\frac{\sum_{m \in \N_n[\cU,\cT]} p_m\Big\{(p_{m},\omega_m) \in \cT(h^{m-1})\, \Big| \, \cU(h^{m-1}) \Big\}}{|\N_n[\cU,\cT]|}\ .\]

\begin{definition}
A strategy $\sigma$ is calibrated with respect to some given checking rule $(\cU,\cT)$ if, for every strategy $\tau$ of Nature,
\[ \limsup_{n \to \infty}  \frac{|\N_n[\cU,\cT]|}{n}\Big| \overline{\omega}_n[\cU,\cT] - \op_n[\cU,\cT] \Big| \leq 0,\quad \P_{\sigma,\tau}\text{-as}\ ,
\]
with the assumption that $p\{A|B\}=+\infty$ if $p\{B\}=0$.
 \end{definition}

The following theorem (a weaker version first appeared  in Lehrer~\cite{Leh01}) continues the discussion of Section~\ref{SE:DeterCalibImp} and weakens furthermore the range of the counterexample of Oakes and Dawid. It shows that if checking rules do not depend on current predictions (but possibly on  past predictions), then deterministic calibration does exist; this is quite obvious if one faces only one checking rule, but the result actually holds with an infinite number of them.

To be  formal, we embed the set of checking-rules independent of current prediction (i.e.\ pairs of mappings from $\bigcup_{n \in \N}(\Delta(\Omega)\times \Omega)^n$ into $\Omega$) with the cylinder topology.

\begin{theorem}
Let $\lambda$ be a probability distribution on the set of checking-rules independent of current predictions. Then there exists a deterministic strategy $\sigma$ that is calibrated with $\lambda$-almost every checking rules.
\end{theorem}
Actually, the result that we shall prove is  stronger as we will show that,  as soon as $|\N_n[\cU,\cT]|$ goes to infinity, $\limsup_{n \to \infty}  \Big| \oq_n[\cU,\cT] - \op_n[\cU,\cT] \Big| \leq 0$, $\P_{\sigma,\tau}$-as.

\bigskip
A similar result (that extends Foster \& Vohra~\cite{FosVoh98}) due to Sandroni, Smorodinsky \& Vohra~\cite{SanSmoVoh03} deals with checking rules depending on current predictions, under the following extra assumptions. We assume that the calibration test compares the  empirical distribution of outcomes with the average prediction on the set of \textsl{active} stages where predictions were in some given set $F \subset \Delta(\Omega)$; activeness of stages might depend on past histories. Formally, $\cU(h^n)$ is either empty (so the stage $n+1$ is not active) or $\cU(h^n)= F \times\Omega$. Mapping $\Tau$ is, on the other side, constant, i.e., $\Tau(h^n)=F\times\{\omega\}$ for some $\omega \in \Omega$ (at least on active stages).

\begin{proposition}
Consider a countable number of such checking rules. Then there exists a strategy of the player that is calibrated with every one of them.
\end{proposition}
\subsection{Smooth calibration}\label{SE:SmoothCalib}
Smooth  calibration is another criterion (close to usual calibration) that can be satisfied with a deterministic strategy, as proved by Foster and Kakade. Even more surprisingly, it can be used to output a calibrated strategy showing  again the  instability of Oakes and Dawid's result.

The idea is to \textsl{smooth}  definitions of calibrations. Indeed, notice that given $F \subset \Delta(\Omega)$, the calibration score can be written as
\[ \frac{|\N_n[F]|}{n}\Big\|\overline{\omega}_n[F]-\op_n[F]\Big\| = \frac{1}{n} \left\|\sum_{m=1}^n \mathds{1} \{ p_m \in F\} (\omega_m-p_m)\right\|
\]
and the mapping $p \mapsto \mathds{1}\{p \in F\}$ is not  continuous. Instead, given some continuous mapping $g: \Delta(\Omega) \to [0,1]$, consider the following smoothen version of the score
\[\frac{1}{n} \left\| \sum_{m=1}^n g(p_m) (\omega_m-p_m)\right\|; \]
with respect to some checking rule $(\cU,\cT)$ independent of the current predictions (so that $\cU(h^n)$ and $\cT(h^n)$ can be seen as subset of $\Omega$), this score becomes
\[\frac{1}{n} \left\| \sum_{m=1}^n g(p_m) \mathds{1}\{\omega_{m}\in\cU(h^{m-1})\Big(\mathds{1}\{\omega_m \in \cT(h^{m-1})-p_m\{\cT(h^{m-1})|\cU(h^{m-1})\}\Big)\right\| \ .
\]

A weaker version of the following Proposition has been proved independently by  Kakade \& Foster~\cite{KakFos04} and Vovk, Nouretdinov, Takemura \& Shafer~\cite{VovNouTak05};  the former named this property \textsl{weak calibration}, but we used the term weak in another meaning (i.e., when  horizon of the game is fixed and known).
\begin{proposition}
There exists a deterministic strategy $\sigma$ of the player such that, no matter  Nature's strategy,  for every continuous mapping $g:\Delta(\Omega) \to \R_+$,
\[ \limsup_{n \to \infty} \frac{1}{n} \left\| \sum_{m=1}^n g(p_m) (\omega_m-p_m)\right\| \leq 0\ .\]

If $\mu$ is a probability distribution on the set of checking rules independent of current predictions, then there exists a deterministic $\sigma$ such that
\[\lim_{n \to \infty}\frac{1}{n}  \sum_{m=1}^n g(p_m) \mathds{1}\{\omega_{m}\in\cU(h^{m-1})\Big(\mathds{1}\{\omega_m \in \cT(h^{m-1})-p_m\{\cT(h^{m-1})|\cU(h^{m-1})\}\Big) \leq 0,\]
for $\mu$-ae checking rule and every continuous mapping $g$, no matter Nature's strategy.
\end{proposition}
As noticed by Foster and Kakade, the convergence in first part of the result can be made uniform with respect to Nature's strategy.

\bigskip

Actually, the most surprising and interesting property of smooth calibration is not so much that there exist deterministic smooth calibrated algorithms, but that they can be used to construct an \textsl{almost deterministic} $\varepsilon$-calibrated strategy as follows, see   Kakade \& Foster~\cite{KakFos04} for more details

Let $\varepsilon$ be fixed and consider a finite $\varepsilon$-triangulation of $\Delta(\Omega)$ whose vertices are $\cV:=\{ v_1, \ldots, v_V\}$. Any $p \in \Delta(\Omega)$ belongs to one simplex of the triangulation  and we denote by $V(p)$ its vertices  (if there are more than one simplices, then choose one arbitrarily). The point $p$ can be written as a convex combination of vertices in $V(p)$, i.e. $p=\sum_{v\in V(p)} \mu_v(p) v$ and it is even possible to decompose $p = \sum_{v \in \cV} \mu_v(p) v$ by assuming that $\mu_v(p)=0$ for any $p$ that does not belong to the same simplex. All those mappings $\mu_v$ are continuous and  Lipschitz.

We construct an $\varepsilon$-calibrated strategy $\sigma$ using a fixed deterministic smooth calibrated strategy $\sigma_d$ in the following way. Whenever $\sigma_d$ dictates to predict $p \in \Delta(\Omega)$,  $\sigma$ predicts $v \in \cV$ with probability $\mu_{v}(p)$. Immediate calculations show that, for every $v \in \cV$,
\[ \E_{\sigma,\tau}\left[\frac{1}{n}\sum_{m=1}^n \mathds{1}\{p_m=v\} (\omega_m-v) \right]  =\frac{1}{n} \sum_{m=1}^n \mu_v(p_m) (\omega_m-p_m)+ \frac{1}{n} \sum_{m=1}^n \mu_v(p_m) (p_m-v) .\]
Since $\mu(p_m) \|p_m-v\| \leq \varepsilon$, expected calibration score (and the actual score, thanks to concentration inequalities) are $\varepsilon$-close to  the smooth calibration score, hence the result.

\bigskip

Key features of this construction are that, although it is impossible to construct an  $\varepsilon$-calibrated strategy deterministically (as proved by Oakes and Dawid), it is possible by using randomizations on arbitrarily small balls. This is why we used  the term of \textsl{almost deterministic} strategies.

\bigskip

Concerning the complexity of (weak) calibration, a recent result of Hazan \& Kakade~\cite{HazKak12}, based on an idea of  Kakade \& Foster~\cite{KakFos04},  shows that it is a hard criterion to satisfy. Indeed, an almost deterministic strategy $\sigma$ (based on some triangulation of $\Delta(\Omega)$) can be used to find $\varepsilon$-Nash equilibria of games. We sketch the proof in the following.

Consider a game between a set of players $\cI$ with actions sets $\cA_i$ and payoff mappings $\rho_i$. Define $\Omega=\prod_{i \in \cI} \cA_i$ and let $X_i : \Delta(\Omega) \to \Delta(\cA_i)$ be a smooth $\varepsilon$-best response of player $i$ (i.e., given any $p \in \Delta(\Omega)$, if $p^{-i}$ denotes the $i$-th marginal of $p$, then $\rho_i(X_i(p),p^{-i}) \geq \max_{x \in \Delta(\cA_i)} \rho_i(x,p^{-i})-\varepsilon$).

We denote by $p_n \in \Delta(\Omega)$ the prediction output at stage $n$ by the strategy $\sigma$ and we assume that  player $i$ plays accordingly to $X_i(p_n)$. The profile of actions actually played is $\omega_n \in \Omega$ and one has $\E[\omega_n]=\Big(X_1(p_n),\ldots,X_I(p_n)\Big)=: X(p_n)$. Since $\sigma$ is $\varepsilon$-calibrated,  for every vertex $v$ and with probability one,
\[ \limsup_{n \to \infty} \frac{\sum_{m=1}^n \mathds{1}\{p_m =v\}}{n}\left\|\frac{\sum_{m=1}^n \mathds{1}\{p_m =v\} (v-\omega_m)}{\sum_{m=1}^n \mathds{1}\{p_m =v\}}\right\| - \frac{\sum_{m=1}^n \mu_v(p_m)}{n} \varepsilon \leq 0.\]
Concentration inequalities, and the fact that $X(p_n)-\omega_n$ and $\mathds{1}\{p_n =v\}-\mu_v(p_n)$ are  martingale differences imply that, with probability one,
\[ \lim_{n \to \infty}\frac{\sum_{m=1}^n \mathds{1}\{p_m =v\} (X(v)-\omega_m)}{n}=0\ \text{and}\ \lim_{n \to \infty} \frac{\sum_{m=1}^n \mathds{1}\{p_m =v\} -\mu_v(p_m)}{n} =0\ . \]
As a consequence, summing  terms, for every vertex $v$ that is predicted with a positive density (i.e., such that $\limsup_{n\to \infty}  \sum_{m=1}^n \mathds{1}\{p_m =v\}/n >0$), one must have $\|v-X(v)\| \leq \varepsilon$. And so, by the very definition of $X(\cdot)$,  $v$ must be a  $2\varepsilon$-Nash equilibrium.

Therefore, not only does the empirical profile of action converge to the convex hull of $\varepsilon$-Nash equilibria, but also if  a stage $n$ chosen at random then, with arbitrarily great probability, $X(v_n)$ is an $2\varepsilon$-Nash equilibrium.

\section{Equivalences between approachability, regret and calibration}\label{SE:Equivalences}

This part is devoted mainly to describe how approachability can be used to construct consistent and calibrated strategies. We also show why calibration is an important and useful tool, as it can be used to construct general (and even approachability)  consistent strategies. Since we can also reduce approachability to regret, this complete the circle and this is the reason why we called these notions  \textsl{equivalent}.

\subsection{Using approachability to get regret}\label{SE:AppImpReg}
\subsubsection{From approachability  to regret; the finite case}
Although Blackwell~\cite{Bla56a} was the first to notice that consistent strategies can be constructed using approachability theory, we first treat Hart \& Mas-Colell~\cite{HarMas01a} idea in finite dimension.

We recall that  choices of actions $a_n \in \cA$ and $b_n \in \cB$ generate at stage $n \in \N$ an external regret $r_n$ defined by
\[ r_n = r(a_n,b_n) := \Big( \rho(1,b_n)-\rho(a_n,b_n), \ldots, \rho(A,b_n)-\rho(a_n,b_n)\Big) \in \R^A
\]
and that a strategy is externally consistent if $\|\orr_n^+\|_\infty$ goes to 0 almost surely. Actually, using approachability theory, Hart \& Mas-Colell~\cite{HarMas01a} proved the following
\begin{proposition}\label{PR:RegHMC}
The strategy $\sigma$ defined by playing, at stage $n+1$,  proportionally to $\orr_n^+$ (and arbitrarily if every component is non-positive) is externally consistent. Moreover, for every strategy $\tau$ of Nature and $n \in \N$,
\[ \E_{\sigma,\tau} \Big[\|\orr_n^+\|_\infty \Big] \leq \E_{\sigma,\tau} \Big[\|\orr_n^+\|_2 \Big] \leq \sqrt{\frac{A}{n}} \ , \]
and, for every $\eta>0$, $\P_{\sigma,\tau}\left\{ \sup_{N \geq n} \|\orr_n^+ \| \geq \eta \right\} \leq 3 \exp\left(-\frac{\eta^2n}{64A}\right)$ as soon as $\frac{\eta^2n}{32A}\geq1$
\end{proposition}
\textbf{Proof:} We simply have to prove that $\sigma$ is exactly Blackwell's approachability strategy of the negative orthant $\R_-^A$ (which is a cone) in the game where the vector payoff is $r(a,b)$. This is a consequence of the following geometric property
\begin{quotation}\label{CL:ExtReg}
No matter the choice of $x \in \Delta(\cA)$, $\Big\langle\, x\, , \E_x [r(a,b)]\, \Big\rangle =0$, for all $b \in \cB$.
\end{quotation}
Indeed, the $k$-th component of $\E_x [r(a,b)]$ is, by linearity, $r(k,b)-r(x,b)$, thus the inner product is equal to $\sum_{a \in \cA} x[a] \big(r(a,b)-r(x,b)\big)=r(x,b)-r(x,b)=0$.

\medskip

Since $x_{n+1}=\sigma(h^n)$ is proportional to $\orr_n^+$, the geometric property implies  that
\[ \Big\langle\, \orr_n^+, \E_{\sigma,\tau} [r_{n+1}]\, \Big\rangle =0 \quad \text{thus} \quad \Big\langle\, \orr_n- \orr_n^-\, , \E_{\sigma,\tau} [r_{n+1}]-\orr_n^-\, \Big\rangle =0
\]
because one always has $\langle z^+, z^- \rangle =0$. Since $\orr_n^-$ is the projection of $\orr_n$ on the negative orthant, this proves that $\sigma$ satisfies Blackwell property, hence is an approachability strategy. Bonds follow from Corollary~\ref{CR:BlackCone}. \qed

Once the reduction from external regret minimization to approachability of $\R_-^A$ has been made, the existence of externally consistent strategies is immediate because $\R_-^A$ is obviously a convex approachable set. Indeed, for every $y \in \Delta(\cB)$, there exists $x \in \Delta(\cA)$ such that $r(x,y) \in \R_-^A$: it suffices to take for $x$ any best response to $y$. The most interesting feature of Proposition~\ref{PR:RegHMC} is that the strategy is very simple and natural: the more regret a specific action induces, the more it should be played (and with a weight exactly proportional to this regret generated).

Generalizations to the compact case (when Nature chooses at stage $n \in \N$ an outcome vector $U_n \in [0,1]^A$) are immediate and omitted.

\begin{remark}\label{RM:BlackOpti} One might argue that with exponential weight algorithm, the dependency in $A$ in rates of convergence shrinks to  $\sqrt{\log(A)}$ instead of $\sqrt{A}$, so the strategy we output might not be optimal. Actually this argument is flawed, rates of convergence are indeed optimal since we minimized the $\ell_2$-norm of the regret. It is only possible to upperbound with $\sqrt{\log(A)/n}$ the $\ell_\infty$-norm of the regret.
\end{remark}
Actually, Hart \& Mas-Colell strategy is an approachability strategy of $\R_-^A$ driven by the potential $\Phi(z)=\|z^+\|^2$ (that represents the $\ell_2$-norm of the regret) while exponential weights are  driven by the soft-max potential $\Phi(z)=\frac{1}{\eta}\log \left(\sum_{a \in \cA} e^{\eta z_a}\right)$ which is a twice differentiable surrogate of $\|z^+\|_\infty$. However, minimization of the infinite norm of regret can also be reduced to approachability, see Proposition~\ref{PR:RegBlac} below (following actually an idea of Blackwell~\cite{Bla56a}).

\begin{proposition}\label{PR:RegBlac}
Assume that Nature chooses  outcome vectors $U \in [0,1]^A$ and define the game with vector payoffs  and target set defined as follows
\[ g(a,U)= \Big( U^a,U \Big) \in [0,1] \times [0,1]^A \quad \text{and}\quad \cC = \Big\{(z,V) \in [0,1]\times[0,1]^A \ \text{s.t.}\ z \geq \max_{a \in \cA} V^a \Big\}.
\]
Then  any approachability strategy of $\cC$ (which is a convex approachable set) minimizes the $\ell_\infty$ norm of the external regret since
\[d_{\cC}\big(\og_n\big) \leq \big\|\orr_n^+\big\|_{\infty} \leq \sqrt{2}\,d_{\cC}\big(\og_n\big).\]
\end{proposition}
\textbf{Proof:} Convexity of $\cC$ (which is actually a polytope, i.e., the intersection of a finite number of half-spaces and a compact set ) is a direct consequence of its definition since
\[\cC=\bigcap_{a \in \cA} \Big\{ (z,V) \ \text{s.t}\ z \geq V^a \Big\} \bigcap\, [0,1]\times[0,1]^A\ .
\]
 Approachability of $\cC$ is immediate: for every $U \in \cU$, choosing $a$ to be one of the highest component of $U$ ensures that $g(a,U)=(U^a,U)$ belongs to $\cC$. It remains to prove the inequalities.

Notice that if we denote by $(\oz_n,\oU_n)$  the average vector payoff  at stage $n$, then  $\oU_n$ is the average outcome vector and  $\oz_n$ is the average actual payoff. As a consequence, the $\ell_\infty$ norm of the regret, $\|\orr_n\|_\infty=\max_{a \in \cA} \oU_n^a-\oz_n$, is exactly equals to the distance between $\big(\oz_n,\oU_n\big)$ and $\big(\max_{a \in \cA} \oU_n^a,\oU_n\big)$. By definition, the latter belongs to $\cC$, therefore one has $d_{\cC}\big(\og_n\big) \leq \big\|\orr_n^+\big\|_{\infty}$.

Let $a^* \in \argmax_{a \in \cA} \oU_n^a$ and $\big(\oz^c_n,\oU^c_n\big)=\Pi_\cC\big(\oz_n,\oU_n\big)$, then
\[ \big\|\orr_n^+\big\|_{\infty}=  \oU_n^{a^*}-\oz_n=  \oU_n^{a^*}-\oz_n^c+\oz_n^c-\oz_n \leq \Big\|\oU_n-\oU_n^c\Big\|+\Big|\oz_n^c-\oz_n\Big| \leq \sqrt{2}d_\cC(\og_n)\]where we used the fact that $U \mapsto \max_{a \in \cA} U^a$ is 1-Lipschitz.\qed

Extensions to the case where $\cA \subset \R^A$ is a compact convex set are immediate as the finiteness of $\cA$ is not used in the proof.

Actually, Blackwell proved this result in the finite case, where Nature chooses action in $\cB$; in that case,  stage payoffs are $g'(a,b)=(\rho(a,b), \delta_b) \in \R \times \Delta(\cB)$ where, as usual, $\Delta(\cB)$ is seen as a subset of $\R^B$. The target set is \[\cC'=\Big\{ (z,y) \in \R \times \Delta(\cB) \ \text{s.t.}\ z \geq \max_{a \in \cA} \rho(a,y)\Big\}\] and since $\|g'(a,b)\| \leq \sqrt{2}$, approachability results imply that $\og_n'$ converges to $\cC'$ at the rate of $\sqrt{2/n}$ thus expected regret is bounded in the order of $\sqrt{B/n}$ (because in this framework, $y \mapsto \rho(a,y)$  is $\sqrt{B}$-Lipschitz and not 1-Lipschitz).

This shows that regret can be bounded, not only with respect to the number of player's actions (i.e. in $\sqrt{\log(A)/n}$), but also with respect to Nature's one (in $\sqrt{B/n}$). This might lead to some improvement if the former is exponentially larger than the latter.

\begin{remark}\label{RM:RegBlac}
In the compact case,  usual proofs show that Blackwell's approachability strategy  ensures that $d_\cC(\og_n) \leq \sqrt{\|g\|_\infty/n}=\sqrt{(A+1)/n}$. However, there exist a consistent strategy such that $\|\orr_n^+\|_\infty \leq 3\sqrt{\log(A)/n}$. So this is an example where  the optimal dimension dependency  of rates of approachability is not  $\sqrt{\|g\|_\infty}$, but much smaller.

There are two possible explanations: either minimizing step by step the $\ell_2$ distance (i.e.\ following Blackwell's strategy) is not optimal, or some important facts are hidden within proofs. In Remarks~\ref{RM:BlackOpti} we claimed that the answer was the first possibility: indeed, the final objective was to minimize the $\ell_\infty$-distance, so minimizing the $\ell_2$ norm must induce an additional dimension-dependent constant. This is not the case here, because the final objective is within constant of the $\ell_2$-distance.

An open and fairly question is wether the dimension dependent term should depend on the specific target set $\cC$ or not. In these examples,  respective sizes of the target sets $\cC$ within the set of feasible payoff vectors are rather intriguing. For instance, in the framework of Proposition~\ref{PR:RegHMC}, the volume of $\co\{g(a,b)\}$ is $2^A$ times the volume of $\cC$ while it is only $A+1$ times the volume of $\cC$  in the framework of Proposition~\ref{PR:RegBlac}. This has to be compared with the respective size of  dimension dependent constants which were $\sqrt{A}$ and $\sqrt{\log(A)}$. \end{remark}

We now turn to the minimization of internal regret $R_n=R(a_n,b_n)$. We recall that it is a $A\times A$-matrix  whose $(a,a')$ component is $\rho(a',b)-\rho(a,b)$ if $a=a_n$ and 0 otherwise.  The generalization of Hart \& Mas-Colell strategy will appeal to the concept of \textsl{invariant measures} of matrices.

A probability distribution $\lambda \in \Delta(\{1,\ldots,d\})$ is an invariant measure of  a some $d\times d$-matrix $M$ with non-negative coefficient, if
\[ \sum_{k=1}^d \lambda^k M^{k,i} =\lambda^i \sum_{k=1}^d M^{i,k}, \quad \forall i \in \{1, \ldots, d\}\ ,
\]
and their existence is a consequence of Perron-Frobenius theorem (this also generalizes usual invariant measure of Markov chains (see e.g.\ Seneta~\cite{Sen81}).

Sorin~\cite{Sor08}, but also Hart \& Mas-Colell~\cite{HarMas00} and Foster \& Vohra~\cite{FosVoh99} used the existence of invariant measure to output a simple internally consistent strategy.

\begin{proposition}\label{PR:RegIntHMC}
The strategy $\sigma$ that dictates to play at stage $n+1$ an invariant measure of $\oR_n^+$ (and arbitrarily if every component is non-positive) is internally consistent. Moreover, for every strategy $\tau$ of Nature and $n \in \N$,
\[ \E_{\sigma,\tau} \Big[\left\|\oR_n^+\right\|_\infty \Big] \leq \E_{\sigma,\tau} \Big[\left\|\oR_n^+\right\|_2 \Big] \leq \sqrt{\frac{A}{n}} \ , \]
and, for every $\eta>0$, $\P_{\sigma,\tau}\left\{ \sup_{N \geq n} \left\|\oR_n^+ \right\| \geq \eta \right\} \leq 3 \exp\left(-\frac{\eta^2n}{64A}\right)$ as soon as $\frac{\eta^2n}{32A}\geq1$.
\end{proposition}
\textbf{Proof:} As for external regret, we just need to prove that $\sigma$ is exactly Blackwell's approachability strategy of the negative orthant. And again, this is a consequence of a geometric property: \begin{quotation}\begin{center}\label{CL:IntReg}
Any  invariant measure $\lambda$ of any matrix $M$ with non-negative coefficient satisfies, no matter the choice of $b \in \cB$, $\Big\langle\, M\, , \E_\lambda [R(a,b)]\, \Big\rangle =0$.
\end{center}\end{quotation}
Let $U^a:=\rho(a,b)$, then the $(i,k)$-component of $\E_\lambda[R(a,b)]$ is $\lambda^i \Big(U^k-U^i\Big)$. So the inner product is equal to $\sum_{i,k}M^{i,k} \lambda^i \Big(U^k-U^i\Big)$ and the coefficient before $U^i$ in this sum is
\[ \sum_{k} \lambda^kM^{k,i}-\lambda^i\sum_k M^{i,k} = 0
\]
since $\lambda$ is an invariant measure of $M$.

\medskip

Since $x_{n+1}=\sigma(h^n)$ is an invariant measure of $\oR_n^+$, geometric properties implies that
\[ \Big\langle\, \oR_n^+, , \E_{\sigma,\tau} [R_{n+1}]\, \Big\rangle =0 \quad \text{thus} \quad \Big\langle\, \oR_n- \oR_n^-\, , \E_{\sigma,\tau} [R_{n+1}]-\oR_n^-\, \Big\rangle =0\ .
\]
This proves that $\sigma$ satisfies Blackwell property, hence is an approachability strategy and bonds follows from Corollary~\ref{CR:BlackCone}. \qed

Once again, using approachability theory to prove  existence of internally consistent strategies is immediate:  the negative orthant satisfies Blackwell's property. An interesting feature of this algorithm is the simple characterization of this optimal (for the minimization of the $\ell_2$ norm) strategy.

 \bigskip

Interestingly, the reduction from external to internal consistent strategies (see Section \ref{SE:ExtToInt} or Stoltz \& Lugosi \cite{StoLug05}) run with the algorithm of Proposition \ref{PR:RegHMC} constructs exactly the strategy of Proposition \ref{PR:RegIntHMC}.

So both Propositions~\ref{PR:RegHMC} and~\ref{PR:RegIntHMC} can be unified  into the the following theorem that deals  more generally with $\Phi$-regret. It  exhibits a strategy with the same complexity as the previous internally consistent strategy, dictating to play at each stage an invariant measure of some matrix.  Given a family $\Phi$, we recall that  $\Phi$-regret at stage $n$ is denoted by $R_n^\Phi \in \R^{|\Phi|}$ and defined by
\[ R^\Phi_n=R^\Phi(a_n,b_n):= \Big( \rho(\phi(a_n),b_n) - \rho(a_n,b_n) \Big)_{\phi \in \Phi} .\]
Finally, given $M \in \R^{|\Phi|}$, let $\Theta^\Phi(M)$ be the $A\times A$-matrix whose $(a,a')$ component is $\Theta^\Phi(M)^{a,a'} = \sum_{\phi : \phi(a)=a'} M^\phi$

\begin{theorem}\label{TH:PhiRegretInvariant}
Let $\Phi$ be a family of swap mappings. The strategy playing at stage $n+1$ accordingly to any invariant measure of $\Theta^\Phi(\overline{R^\Phi_n}^+)$ has no $\Phi$-regret. Moreover, for every strategy $\tau$ of Nature and $n \in \N$,
\[ \E_{\sigma,\tau} \Big[\left\|\overline{R^\Phi_n}^+\right\|_\infty \Big] \leq \E_{\sigma,\tau} \Big[\left\|\overline{R^\Phi_n}^+\right\|_2 \Big] \leq \sqrt{\frac{A_\Phi}{n}} \ , \text{with}\ A_\Phi=\max_{a \in \cA} \Big| \Big\{\phi \in \Phi \ \text{s.t.}\ \phi(a)\neq a \Big\}\Big|\]
and, for every $\eta>0$, $\P_{\sigma,\tau}\left\{ \sup_{N \geq n} \left\|\overline{R^\Phi_n}^+ \right\| \geq \eta \right\} \leq 3 \exp\left(-\frac{\eta^2n}{64A_\Phi}\right)$ as soon as $\frac{\eta^2n}{32A_\Phi}\geq1$.
\end{theorem}
\textbf{Proof:} The proof follows closely the ones of Propositions~\ref{PR:RegHMC} and~\ref{PR:RegIntHMC}. Indeed, one just has to prove that this strategy is an approachability strategy of $\R_-^{|\Phi|}$, using the following geometric property:
\begin{quotation}\begin{center}\label{CL:PhiReg}
Any  invariant measure $\lambda$ of any matrix $\Theta(M)$ with non-negative coefficient satisfies, no matter the choice of $b \in \cB$, $\Big\langle\, M\, , \E_\lambda [R^\Phi(a,b)]\, \Big\rangle =0$.
\end{center}\end{quotation}
Indeed, if one denote $U=\rho(\cdot,b)$, then
\begin{align*}\Big\langle M , \E_\lambda [R^\Phi(a,b)]\Big\rangle &= \sum_{\phi \in \Phi} M^\phi \sum_{a \in \cA} \lambda^a \Big(U^{\phi(a)}-U^a\Big)\\
&=\sum_{a \in\cA} \lambda^a\sum_{a' \in \cA} \sum_{\phi : \phi(a)=a'} M^\phi\Big(U^{a'}-U^a\Big)\\
&=\sum_{a \in\cA} \lambda^a\sum_{a' \in \cA} \Theta^\Phi(M)^{a,a'}\Big(U^{a'}-U^a\Big)\\
&= \sum_{a \in \cA} \left(\sum_{a'\in\cA} \lambda^{a'}\Theta^\Phi(M)^{a',a}-\lambda^a \sum_{a'\in\cA}\Theta^\Phi(M)^{a,a'} \right)U^a=0
\end{align*}
since $\lambda$ is an invariant measure of $\Theta^\Phi(M)$. As a consequence, this strategy is exactly Blackwell's approachability strategy of the negative orthant. The result comes from the fact that $R^{\Phi}(a,b)$ has at most $A_\Phi$ non-zero components, each one in $[-1,1]$,  thus $\|R^{\Phi}(a,b)\|^2 \leq A_\Phi$.
\qed

\begin{remark} As usual, if techniques from approachability in infinite dimension are used instead of regular approachability, the term in $\sqrt{A}$ for external and internal regret or  $\sqrt{A_\Phi}$ for $\Phi$-regret can be replaced by respectively $\sqrt{\log(A)}$ or $\sqrt{A\log(A)}$, up to some constant.
\end{remark}
\subsubsection{From approachability  to regret; the infinite case}
We turn in this section to the case where action set $\cA$ is no longer finite but some convex compact metric set and at stage $n \in \N$, Nature chooses a mapping $U_n : \cA \to [0,1]$ in a set $\cU$ of equicontinuous mapping. We show how previous results can be  extended to this \textsl{compact case} (indeed, Arzela-Ascoli theorem ensures that $\cU$ is relatively compact).

\begin{theorem}\label{TH:RegMinInfinCase}
In this compact case, there exists a strategy without $\Phi_c$ regret, where $\Phi_c$ is the set of continuous mapping from $\cA$ to itself.
\end{theorem}
\textbf{Proof:}
Consider an auxiliary game where action sets of player and Nature are $\cA$ and $\cU$. Choices of $a \in \cA$ and $U \in \cU$ generates a payoff $\tU[a] \in \cL_2(\Phi_c,\lambda)$, where $\lambda$ is some fixed probability distribution over $(\Phi_c,\|\cdot\|_\infty)$ embedded with the Borelian $\sigma$-field, defined by
\[ \tU[a](\phi) := U(\phi(a)) - U(a), \quad \forall \phi \in \Phi_c.
\]
The convex set $\cC=\cL_2^- (\Phi_c,\lambda):=\Big\{\tU \in \cL_2(\Phi_c,\lambda)\ \text{s.t.}\ \tU \leq 0\Big\}$ is not excludable by Nature; indeed, for any $U \in \cU$, there exists $a \in \cA$ (any global maximizer of $U$) such that $\tU[a]$ belongs to $\cC$. Thus it is approachable by the player, and any approachability strategy has no $\phi$-regret, for $\lambda$-almost all mapping $\phi \in \Phi_c$.

\medskip

However, $\Phi_c$ is separable (see Rudin~\cite{Rud74} or Stoltz \& Lugosi~\cite{StoLug07}), so there exists $\{\phi_k;\, k \in \N\}$ a countable dense subset of $\Phi_c$; the corresponding probability $\lambda$ we consider is $\lambda=\sum_{k \in \N} 2^{-k}\delta_{\phi_k}$.
Since $\cU$ is a family of equicontinuous mappings, every mapping $U \in \cU$ share the same modulus of continuity $\omega(\cdot)$; this means that, for every $\varepsilon >0$ there exists $\delta:=\omega(\varepsilon)$ such that if $d(a,a') \leq \delta$ then $|U(a)-U(a')| \leq \varepsilon$, for any mapping $U \in \cU$. Given $\phi \in \Phi_c$, there exists $\phi_k$ such that $\|\phi-\phi_k\|\leq \delta$ thus
\[ \frac{1}{n} \sum_{m=1}^n U_m[\phi(a_m)] -U_m[a_m] \leq  \frac{1}{n} \sum_{m=1}^n U_m[\phi_{k}(a_m)] -U_m[a_m] +\varepsilon\, .
\]
Since $\sigma$ has no $\phi_k$-regret, its $\phi$-regret is asymptotically smaller than $\varepsilon$, for every $\varepsilon>0$, thus it has no $\Phi_c$-regret.
\qed

\begin{corollary}\label{CR:RegretUsc}
Conclusions of Theorem \ref{TH:RegMinInfinCase} hold if  $\cU$ is the convex hull of a finite set of upper-semicontinuous mappings bounded from below and $\Phi_c$ is the set of constant mappings. 
\end{corollary}
\textbf{Proof:} Every $U \in \cU$ is upper-semicontinuous over a compact set, it admits a maximum. Therefore $\cU$ is uniformly bounded and the set $\cC$ is approachable, with respect to some probability distribution $\lambda$ that remains to be defined. 

Denote by $U_1,\ldots,U_m$ the extreme points of $\cU$. As they are upper-semicontinuous and bounded, their exists a countable subset  $\{a_k; k \in \cN\} \subset \cA$ such that, for every $\varepsilon>0$ and every $a \in \cA$, there exists $a_k$ satisfying $U_i(a_k) \geq U_i(a)-\varepsilon$, for every $i \in \{1,\ldots,m\}$. Define $\lambda$ as any probability measure whose support is exactly this countable subset.

The rest of the proof follows the one of Theorem \ref{TH:RegMinInfinCase}.\qed

In the finite case, approachability theory not only provides a quick and easy proof of consistent strategies, but also exhibit explicitly some of them. In fact,  playing somehow proportionally to the positive part of the regret is still externally consistent in the compact case. Let $\lambda$ be any positive probability measure on $\{a_k;\, k \in \N\}$, a countable dense subset of $\cA$ and denote by $\orr_n^+[a_k]$ the external regret at stage $n$ induces by action $a_k$.

Consider the strategy that chooses $a_k$ at stage $n+1$ with probability $\frac{\lambda_k \orr_n^+[a_k]}{\sum_{\ell}\lambda_\ell \orr_n^+[a_\ell]}$. Then, as in the finite case, one can easily show that the geometric property holds, i.e.,
\[\langle \E[r_{n+1}], \orr_n^+ \rangle = \sum_{k}\frac{\lambda_k \orr_n^+[a_k]}{\sum_{\ell}\lambda_\ell \orr_n^+[a_\ell]} \sum_{\ell} \Big(U_{n+1}(a_\ell)- U_{n+1}(a_k)\Big) \orr_n^+[a_\ell]\lambda_\ell =0\ .\]
Approachability in infinite dimension (along with the density argument) ensures that this strategy has no external regret.

\bigskip

Concerning $\Phi$-regret, one cannot simply play accordingly to any invariant measure of  some infinite dimensional matrix, as their existence is not ensured. However, it is still possible to discretize finitely $\cA$ to get a $\Phi$-regret smaller than $\varepsilon$, with $\varepsilon$-arbitrary small (or even equal to 0, if $\varepsilon$ is taken as a decreasing sequence, see Proposition~\ref{PR:BiaAppEps}).

Let $\omega_\cU(\cdot)$ be the common modulus of continuity of  $U \in \cU$ and $\bA$ a finite $\omega_\cU(\varepsilon)$-grid of $\cA$. For any $\phi \in \Phi$, we define $\underline{\phi} : \cA \to \bA$  by $\underline{\phi}=\argmin_{\ba' \in \bA}  d(\phi(a),\ba')$ with ties broken arbitrarily. As a consequence, for every $a \in \cA$,  $U \in \cU$ and  non negative $q \in \cL_2(\Phi,\lambda)$,
\[\left| \int_{\Phi}q(\phi) U(\underline{\phi}(a))d\lambda - \int_{\Phi}q(\phi) U(\phi(a))d\lambda \right|\leq \varepsilon\ .
\]
We define, for any  $(\ba,\ba')$,  $\Theta[q]^{\ba,\ba'}:=\int_{\Phi^{\ba,\ba'}} qd\lambda$, where $\Phi^{\ba,\ba'}:= \Big\{ \phi \in \Phi \ \text{s.t.}\ \underline{\phi}(\ba)=\ba'\Big\}$.  Let $x$ be any invariant measure of the matrix $\Theta[q]$ then one has
\[\Big\langle\, q\, , \tU(x) \Big\rangle \leq  \sum_{\ba \in \bA} x^\ba \sum_{\ba \in \bA } \Theta[q]^{\ba,\ba'}\Big(U(\ba')-U(\ba\Big)+  \left| \int_{\Phi}q(\phi) \Big(U(\underline{\phi}(a))-U(\phi(a))\Big)d\lambda \right| \leq \varepsilon.\]
This proves that $\cL_2^-(\Phi_c,\lambda)$ is a $B$-set, hence approachable. We can only claim that the strategy we exhibited has some \textsl{flavors} of invariant measures.

\subsection{Using regret to get calibration}\label{SE:REgImplCal}
We show in this section that finite calibration can easily be understood in terms of internal regret. The first idea goes to Foster \& Vohra~\cite{FosVoh97} and it has been somehow clarified by Sorin~\cite{Sor08}. Recall that, in finite calibration, Nature chooses at stage $n$ an outcome $\omega_n \in \Omega$. The player formulates a prediction on $\omega_n$ by choosing a probability distribution $p[\ell_n] \in \Delta(\Omega)$ that must belong to a finite grid $\{p[\ell]\, ; \ell \in \cL\}$.

\begin{theorem}\label{TH:RegFinCal}
There exists a strategy $\sigma$   calibrated with respect to  the grid $\{p[\ell]\, ; \ell \in \cL\}$, such that, no matter the strategy $\tau$ of Nature,
\[ \E_{\sigma,\tau}\left[\sup_{\ell \in \cL} \frac{|\N_n[\ell]|}{n}\bigg(\big\| \overline{\omega}_n[\ell] - p[\ell] \big\|^2-\min_{k \in \cL}\big\|\overline{\omega}_n[\ell]-p[k]\big\|^2\bigg)\right]\leq 6 \sqrt{\frac{\log(L)}{n}}, \ \text{so}
\]
\[ \E_{\sigma,\tau}\left[\sup_{\ell \in \cL} \frac{|\N_n[\ell]|}{n}\bigg(\big\| \overline{\omega}_n[\ell] - p[\ell] \big\|-\min_{k \in \cL}\big\|\overline{\omega}_n[\ell]-p[k]\big\|\bigg)\right]\leq \frac{6}{\delta(\cL)} \sqrt{\frac{\log(L)}{n}},
\]
where $\delta(\cL)=\inf_{\ell \neq k \in \cL} \big\|p[\ell]-p[k]\big\|$ is the diameter of the grid.
\end{theorem}
\textbf{Proof:} The proof uses the  fact (simply obtained by expanding sums) that, for any sequence $q_m$ and every $\ell, k \in \cL$,
\[ \sum_{m \in \N_n[\ell]}\frac{\|\omega_m-p[\ell]\|^2}{|\N_n[\ell]|}-\frac{\|\omega_m-p[k]\|^2}{|\N_n[\ell]|}=\big\| \overline{\omega}_n[\ell] - p[\ell] \big\|^2-\big\|\overline{\omega}_n[\ell]-p[k]\big\|^2\ .\]

Now consider the game with action space $\cL$ and $\Omega$ where  choices of $\ell$ and $\omega$ generate the payoff $\rho(\ell,\omega)=-\|\omega-p[\ell]\|^2$. An internally consistent strategy satisfies, by definition,
\[ \limsup_{n\to \infty} \sup_{\ell,k}  \frac{|\N_n[\ell]|}{n}\bigg(\sum_{m\in\N_n[\ell]}\frac{\big\| \omega_m - p[\ell] \big\|^2}{|\N_n[\ell]|}-\frac{\big\|\omega_m-p[k]\big\|^2}{|\N_n[\ell]|}\bigg)\leq0.
\]
So this, along with the basic fact, shows that any internally consistent strategy is calibrated with respect to the grid $\{p[\ell]; \ell \in \cL\}$.  Rates of convergences follows from those of internal consistency.
\qed

We stress out that we proved a stronger result than require; the calibration score converges almost surely to zero,  at a rate independent of Nature's strategy.

\begin{remark}
This proof of calibration highlights the following fact. It does not really matter that $\omega_m$ belongs to a finite set $\Omega$ and that $p_n$ are probability distributions over $\Omega$. Indeed, one can just assume that sequences $\omega_n$ and $p_n$ belong to some compact set of an Euclidian space $\R^d$. Similarly, given two finite families of predictions $\{p[\ell] \in \R^d; \ell \in \cL\}$ and weights $\{\nu[\ell] \in \R; \ell \in \cL\}$, we recall that  weighted calibration sis defined as
\[ \sup_{\ell \in \cL}\frac{|\N_n[\ell]|}{n}\bigg(\left(\big\| \overline{\omega}_n[\ell] - p[\ell] \big\|^2-\nu[\ell]\right)-\left(\min_{k \in \cL}\big\|\overline{\omega}_n[\ell]-p[k]\big\|^2-\nu[k]\right)\bigg),
\]
and the exact same proof (yet with $\rho(\ell,\omega)=-\|\omega-p[\ell]\|^2+\nu[\ell]$) gives the existence of weighted calibrated strategies. Rates of convergence are identical except that the constant $6$ is replaced with $6+3\max_{\ell \in \cL}|\nu[\ell]|$. \end{remark}

Notice that we defined finite calibration with respect to the uniform norm of the positive part of $\Big(\big\| \overline{\omega}_n[\ell] - p[\ell] \big\|-\min_{k \in \cL}\big\|\overline{\omega}_n[\ell]-p[k]\big\|\Big)_{\ell,k}$. And this quantity is upper-bounded optimally  by the exponential weight algorithm. We could as well have defined calibration in terms of the $\ell^2$ norm of this vector and  as in regret minimization, playing an invariant measure could then improve bounds.

Next proposition states that, quite surprisingly, there exist $\varepsilon$-calibrated strategies with rates of convergence independent of $\varepsilon$ (and even of $\Omega$, for a slightly weaker notion).

\begin{proposition}\label{PR:EpsCalInde}
For every $\varepsilon>0$, there exists a grid $\{p[\ell] ; \ell \in \cL\}$ and a  strategy $\sigma$ such that, no matter the strategy $\tau$ of Nature and for every $n \in \N$,
\[ \E_{\sigma,\tau} \left[\sup_{\ell  \in \cL}\frac{|\N_n[\ell]|}{n}\bigg(\Big\|\op_n[\ell]-\overline{\omega}_n[\ell]\Big\| - \varepsilon \bigg)\right] \leq  \sqrt{\frac{1}{n}}\ .\]
Moreover, this strategy is $\varepsilon$-calibrated, with a rate of convergence independent of $\varepsilon$,  since one also has, for every $n \in \N$,
\[   \E_{\sigma,\tau} \left[\sup_{p \in \Delta(\Omega)} \frac{|\N_n[p,\varepsilon]|}{n}\bigg(\Big\|\op_n[p,\varepsilon]-\overline{\omega}_n[p,\varepsilon]\Big\| - \varepsilon \bigg)\right] \leq \sqrt{\frac{\gamma(\Omega)}{n}}, \ \text{with}\ \gamma(\Omega)\leq(2\Omega^\Omega).
\]
\end{proposition}
\textbf{Proof:} Let $\varepsilon$ be fixed; the strategy considered is simply a calibrated strategy with respect to some well chosen grid of $\Delta(\Omega)$. Recall that $\Delta(\Omega)$ is written as the following subset of $\R^{d}$ with $d=\Omega-1$:
\[ \Delta(\Omega) := \Big\{ q=(q_1,\ldots,q_{d}) \in \R^{d}\ \text{s.t.}\ q_1,\ldots,q_{d} \geq 0 \ \text{and} \ \sum_{k=1}^dq_k\leq 1\Big\}.
\]
Denote by $\be_k$ the unit vector of $\R^d$ whose components are all zero except the $k$-th which is one. The regular grid  considered is indexed by $\cL_\varepsilon$ and defined by
\[ \left\{ \sum_{k=1}^d \frac{2n_k}{\sqrt{d}}\varepsilon.\be_k \in \Delta(\Omega) ; n_k \in \N \right\}=:\left\{p[\ell]= \sum_{k=1}^d \frac{2n_k[\ell]}{\sqrt{d}}\varepsilon.\be_k\, ; \, \ell \in \cL_\varepsilon\right\}
\]
Given a point  $p[\ell]$ of the grid, its neighbors are points $p[\ell']$ such that $n_k[\ell]=n_k[\ell']$ for every $k \in \{1,\ldots,d\}$ except for exactly one $k_0$  which is such that $\big|n_{k_0}[\ell]-n_{k_0}[\ell']\big|=1$.  So if we denote by $\cN[\ell] \subset \cL_\varepsilon$ the neighbors of $p[\ell]$, it contains at most $2d$ elements.

The basic idea behind the specific geometry of this grid is that
\begin{quotation}\begin{center} any point $q \in \Delta(\Omega)$ is closer to $p[\ell]$ than to any other  $p[\ell']$ if and only if it is closer to $p[\ell]$ than to any of its neighbors.\end{center}\end{quotation}
Consider the game introduced in the proof of Theorem~\ref{TH:RegFinCal}, except that  choices of $\ell_n \in \cL_\varepsilon$ and $\omega_n$ generate an internal regret $R_n'$ whose $(\ell,\ell')$-th component is
\[ (R_n')^{\ell,\ell'} = \left\{\begin{array}{ll} \Big\| \omega _n-p[\ell]\Big\|^2-\Big\| \omega _n-p[\ell']\Big\|^2 & \ \text{if} \ \ell=\ell_n \ \text{and} \ \ell' \in \cN[\ell]\\ 0 &\ \text{otherwise} \end{array}\right. .
\]
As a consequence, using the simple fact concerning averages of norms,
\[ \|R_n'\| \leq 4\frac{\varepsilon}{\sqrt{d}} \ \text{and}\ \oR'_n = \frac{|\N_n[\ell]|}{n}\bigg( \big\| \overline{\omega}_n[\ell]-p[\ell]\big\|^2-\big\| \overline{\omega}_n[\ell]-p[\ell']\big\|^2\bigg)_{l \in \cL_\varepsilon,k\in\cN[\ell]}\ .
\]
Same arguments as in the proof of Proposition~\ref{PR:RegIntHMC} yield that playing, at stage $n+1$, any invariant measure of $(\oR'_n)^+$ ensures that $\E_{\sigma,\tau}\left[\Big\|(\oR'_n)^+\Big\|^2\right]\leq 16\varepsilon^2/(dn)$.

It remains to relate $\big\|\overline{\omega}_n[\ell]-p[\ell]\big\|$ to $\|\oR_n'\|$. First, we write $\overline{\omega}_n[\ell]=p[\ell]+\sum_{k=1}^d x_k \be_k$ where we can assume (up to a change of signe) that every $x_k$ is positive and even $x_k\geq \varepsilon/\sqrt{d}$ (otherwise $\overline{\omega}_n[\ell]$ is even closer to $p[\ell]$).

We  denote by $p[\ell_k]=p[\ell]+2\varepsilon\be_k/\sqrt{d}$ the neighbor of $p[\ell]$ in the direction of $\be_k$, so that $\be_k=(p[\ell_k]-p[\ell])/\|p[\ell_k]-p[\ell]\|$. Triangle inequality implies that
\begin{align*} \big\|p[\ell]-\overline{\omega}_n[\ell]\big\| &\leq \left\|p[\ell]-\left(p[\ell]+\sum_{k=1}^d\frac{\varepsilon}{\sqrt{d}}\be_k\right)\right\|+\left\|\left(p[\ell]+\sum_{k=1}^d\frac{\varepsilon}{\sqrt{d}}\be_k\right)-\overline{\omega}_n[\ell]\right\|\\
& = \varepsilon + \left\|p[\ell]+\sum_{k=1}^d\frac{p[\ell_k]-p[\ell]}{2}-\overline{\omega}_n[\ell]\right\|\\
&= \varepsilon +  \sqrt{\sum_{k=1}^d \left\langle  p[\ell]+\sum_{k=1}^d\frac{p[\ell_k]-p[\ell]}{2}-\overline{\omega}_n[\ell], \be_k\right\rangle^2}\\
&= \varepsilon +  \sqrt{\sum_{k=1}^d \left\langle  \frac{p[\ell_k]+p[\ell]}{2}-\overline{\omega}_n[\ell], \frac{p[\ell_k]-p[\ell]}{\|p[\ell_k]-p[\ell]\|}\right\rangle^2}\\
&=\varepsilon +  \sqrt{\sum_{k=1}^d \frac{\Big(\big\|\overline{\omega}_n[\ell]-p[\ell]\big\|^2-\big\|\overline{\omega}_n[\ell]-p[\ell_k]\big\|^2\Big)^2}{4\big\|p[\ell_k]-p[\ell]\big\|^2}}\\
&=\varepsilon+\frac{\sqrt{d}}{4\varepsilon}\sqrt{\sum_{k=1}^d \Big(\big\|\overline{\omega}_n[\ell]-p[\ell]\big\|^2-\big\|\overline{\omega}_n[\ell]-p[\ell_k]\big\|^2\Big)^2}.
\end{align*}
To sum up, we have prove that, for every $\ell \in \cL_\varepsilon$,
\[ \big\|p[\ell]-\overline{\omega}_n[\ell]\big\| -\varepsilon\leq  \frac{\sqrt{d}}{4\varepsilon}\sqrt{\sum_{k\in \cN[\ell]} \left[\Big(\big\|\overline{\omega}_n[\ell]-p[\ell]\big\|^2-\big\|\overline{\omega}_n[\ell]-p[\ell_k]\big\|^2\Big)^+\right]^2}\ .
\]
Multiplying both sides of this inequality by $\frac{|\N_n[\ell]|}{n}$, taking the square and summing over $\ell \in \cL_\varepsilon$, one obtains
\[ \left\| \tR_n\right\|_2^2:= \sum_{\ell \in \cL_\varepsilon} \left[\frac{|\N_n[\ell]|}{n}\bigg(\big\|p[\ell]-\overline{\omega}_n[\ell]\big\| -\varepsilon\bigg)\right]^2\leq \frac{d}{16\varepsilon^2} \left\|\left(\oR_n'\right)^+\right\|^2_2\ ,
\]
therefore the strategy $\sigma$ ensures that
\[ \E_{\sigma,\tau}\left[ \sup_{\ell \in \cL_\varepsilon} \frac{|\N_n[\ell]|}{n}\bigg(\big\|p[\ell]-\overline{\omega}_n[\ell]\big\| -\varepsilon\bigg) \right]\leq  \E_{\sigma,\tau}\left[ \left\| \tR_n\right\|_2 \right]
\leq \sqrt{\frac{1}{n}}\ .\]
This gives the first part of the proof. The last part is due to the fact that there are less than $(2\sqrt{d})^d$ points in the $\varepsilon$-ball centered at some $p \in \Delta(\Omega)$.
\qed

Proposition~\ref{PR:EpsCalInde} also allows to recover the following result of Mannor \& Stoltz~\cite{ManSto10}

\begin{theorem}\label{TH:ManSto}
When $\cF$ is the family of all Borel subsets of $\Delta(\Omega)$, there exists a $\cF$-calibrated  strategy $\sigma$ such that, for every strategy $\tau$
of Nature,
\[ \E_{\sigma,\tau}\left[\frac{|N_n[F]|}{n}\bigg\|\op_n[F]-\overline{\omega}_n[F]\bigg\|\right] \leq 7n^{-\frac{1}{\Omega+1}}, \quad \P_{\sigma,\tau}\text{-a.s.},
\]
and, for every $\delta >0$, with probability at least $1-\delta$, one also has
\[ \frac{|N_n[F]|}{n}\bigg\|\op_n[F]-\overline{\omega}_n[F]\bigg\| \leq  \frac{9}{n^{\frac{1}{\Omega+1}}}+ 2\sqrt{\frac{\log\left(\frac{2}{\delta}\right)}{n}}, \quad \P_{\sigma,\tau}\text{-a.s.}.
\]
\end{theorem}
\textbf{Proof:} The result is a consequence of  a doubling trick applied to strategies constructed in Proposition~\ref{PR:EpsCalInde}. Assume that the strategy adapted to some $\varepsilon$ is played during $N$ stages. On those stages, one has
\[\frac{|\N_n[F]|}{n}\bigg\|\op_n[F]-\overline{\omega}_n[F]\bigg\| -\varepsilon\leq \sum_{\ell \in \cL_{\varepsilon_k}} \frac{|\N_n[\ell]|}{n}\left(\bigg\|p[\ell]-\overline{\omega}_n[\ell]\bigg\|-\varepsilon\right) \leq \sqrt{L_\varepsilon}\left\|\widetilde{R}_n\right\|_2.\]
Taking expectation  and using the fact that $L_\varepsilon \leq \varepsilon^{-d}$ yield that
\[ \E_{\sigma,\tau}\left[\frac{|\N_n[F]|}{n}\bigg\|\op_n[F]-\overline{\omega}_n[F]\bigg\| \right] \leq \varepsilon + \sqrt{\frac{1}{\varepsilon^{d}}} \sqrt{\frac{1}{n}}
\]
Hence, the doubling trick adapted to the sequences $\varepsilon_k=\left(\frac{d}{2^k}\right)^{\frac{1}{d+2}}$, played during $2^k$ stages ensures that, denoting $n=2^{k_0}+m <  2^{k_0+1}$,
\begin{align*}\E_{\sigma,\tau}\left[\frac{|\N_n[F]|}{n}\bigg\|\op_n[F]-\overline{\omega}_n[F]\bigg\| \right] &\leq \frac{1}{n}\left(\sum_{k=0}^{k_0-1}2^k2\varepsilon_k+m\varepsilon_{k_0}+\sqrt{\frac{1}{\varepsilon^d_{k_0}}m}\right)\\
& \leq  2 \frac{1}{2^{k_0}}\sum_{k=0}^{k_0}2^{k\frac{d+1}{d+2}}\leq \frac{4}{2^{\frac{d+1}{d+2}}-1} \frac{1}{2^{\frac{k_0+1}{d+2}}}\leq 7\frac{1}{n^{\frac{1}{d+2}}}.
\end{align*}

High probability bounds are classics consequences of concentration inequalities, since $|\N_n[F]|\Big\|\overline{\omega}_n[F]-\op_n[F]\Big\|/n=\Big\|\oY_n\Big\|$ where the sequence $Y_n= \left(\omega_n-p_n\right)\mathds{1}\{p_n \in F\}$ is such that $\|Y_n\|^2\leq 2$ and, by Jensen inequality, $\|\oY_n\|\leq \|\oY_n-\E[\oY_n]\|+E[\|\oY_n\|]$. \qed

In fact, Theorem~\ref{TH:ManSto} slightly improves the result of Mannor \& Stoltz~\cite{ManSto10} since it implies  that
\[
\limsup_{n \to \infty} n^{\frac{1}{\Omega+1}}\frac{|N_n[F]|}{n}\bigg\|\overline{\omega}_n[F]-\op_n[F]\bigg\| \leq  7,\quad \P_{\sigma,\tau}\text{-as}.
\]

Rakhlin, Sridharan \& Tewari~\cite{RakSriTew11} wrote the calibration problem in terms of a generalized regret, see Section~\ref{SE:GenReg}. Formally, assume that actions spaces are respectively $\Delta(\Omega)$ and $\Omega$ and that the stage game payoff is null, i.e. $g(p,\omega)=0$. The class of departure function considered are $\Big\{ \xi_{p,\lambda}  ; \ p \in \Delta(\Omega), \lambda >0 \Big\}$ where $\xi_{p,\lambda} : \Delta(\Omega) \times \Omega \to \R^\Omega $ and the evaluation mappings $B_n : \left(\R^\Omega\right)^n \to \R$ are defined by, for every $n \in \N$,
\[  \xi_{p,\lambda}[g](p_n,\omega_n) = \mathds{1}\{ \|p_n-p\|_1 \leq \lambda\} (p_n-\delta_{\omega_n}) \ \text{and}\ B_n(Z_1,\ldots,Z_n)=\left\|\frac{1}{n}\sum_{m=1}^nZ_m\right\|.\]
As a consequence, one easily has that regret is upper bounded by calibration score, as
\begin{align*} &\sup_{p, \lambda}B_n\Big(\xi_{p,\lambda}[g](p_1,\omega_1),\ldots,\xi_{p,\lambda}[g](p_n,\omega_n)\Big) - B_n\Big(g(p_1,\omega_1),\ldots,g(p_n,\omega_n)\Big)\\
=&\sup_{p,\lambda} \frac{\N_n[p,\lambda]}{n}\bigg\|\overline{\omega}_n[p,\lambda]-\op_n[p,\lambda]\bigg\|\ .
\end{align*}
The max-min formulation of the regret minimization problem (see Section~\ref{SE:GenReg}) proves that $\sup_{p,\lambda} \bigg\|\overline{\omega}_N[p,\lambda]-\op_N[p,\lambda]\bigg\|\N_N[p,\lambda]/N$ is upper bounded at the final stage $N$ by $c\Omega^2 \sqrt{\log(N)/N}$ where $c$ is a universal constant. An alternative (and actually more general) proof is given in the next section.

\subsection{Using Approachability to get (smooth and generalized) Calibration}
In this section, we show that recent results in calibration can be rewritten solely as the existence or construction of some approachability strategy.  The first result we exhibit is a generalization of both a previous one of Perchet~\cite{Per10} (since the strategy is calibrated with respect to much larger families) and Rakhlin, Sridharan \& Tewari~\cite{RakSriTew11} (because the proof is constructive and not horizon dependent).

\begin{theorem}\label{TH:CalibViaAppro}
Let $\cF:=\Big\{B[p,\lambda]_\infty ; p\in\Delta(\Omega), \lambda >0\Big\}$ be the family of $\ell_\infty$-balls. Then there exists a calibrated strategy $\sigma$ such that, no matter the strategy $\tau$ of Nature and for every $n \in \N$,
\[ \E_{\sigma,\tau}\left[ \sup_{p \in \Delta(\Omega), \lambda >0} \frac{|\N_n[p,\lambda]|}{n} \bigg\|\omega_n[p,\lambda]-\op_n[p,\lambda] \bigg\|_\infty\right]\leq12\sqrt{2\frac{\Omega}{n}\log\left(\frac{2n}{\Omega}\right)}.
\]
\end{theorem}
\textbf{Proof:} Let $\varepsilon >0$ be fixed. As in the proof of Proposition~\ref{PR:EpsCalInde}, the set $\Delta(\Omega)$ is represented as a subset of $\R^d$ (with $d=\Omega-1$),
\[ \Delta(\Omega) = \Big\{ p = (p_1,\ldots,p_d) \in \R^d \ \text{s.t.}\ p_1,\ldots,p_d \geq 0 \ \text{and} \ \sum_{k=1}^d p_k=1\Big\}
\]
and we consider the regular $\varepsilon$-grid $\cL_\varepsilon$ defined by
\[ \left\{ \sum_{k=1}^d 2n_k\varepsilon.\be_k \in \Delta(\Omega) ; n_k \in \N \right\}=:\left\{p[\ell]= \sum_{k=1}^d 2n_k[\ell]\varepsilon.\be_k\, ; \, \ell \in \cL_\varepsilon\right\}.
\]
Although the family of $\ell_\infty$-balls is infinite, the number of different possible intersections of such a ball with the grid $\cL_\varepsilon$ is obviously finite (it is trivially bounded by its  number of subsets, $2^{L_\varepsilon}$). However, an $\ell_\infty$-ball $B_\infty(p,\lambda)$ is rectangular and can be described by two extreme points: the lowest corner $p-\sum_{k=1}^d \lambda \be_k$ and the highest corner (in every direction) $p+\sum_{k=1}^d \lambda \be_k$

The grid $\cL_\varepsilon$ is regular, so this characterization holds for intersections with $\ell_\infty$ balls: they are  characterized by two extreme points. As a consequence, they are at most $\cL_\varepsilon \leq \varepsilon^{-2d}$ different possible intersections. Consider a fixed family of $\ell_\infty$-balls that induce exactly these different intersections, and denote it  $\Big\{ B_\infty(\bp[k],\lambda_k) ; \ k \in \cK \Big\}$.

We introduce an auxiliary game with action space $\cL_\varepsilon$ and $\Omega$, payoff mapping
\[ g(\ell,\omega)= \bigg( \mathds{1}\big\{\|p[\ell]-\bp[k]\|\leq \lambda_k\big\} (p[\ell]-\delta_\omega) \bigg)_{k \in \cK}
\]
and consider the  closed and convex target set $\cC:=B_\infty(0,\varepsilon) \subset \left(\R^d\right)^K$.

Given $q \in \Delta(\Omega)$,  the pure action $\ell$ corresponding to a point of the grid $p[\ell]$ such that $\|p[\ell]-q\|_\infty\leq \varepsilon$ ensures that $g(\ell,q)$ belongs to $\cC$ which is therefore approachable. Moreover, since $\cC$ is rectangular, the approachability strategy of Corollary~\ref{CR:AppNormInf}, adapted to the potential $\Phi(z)=\frac{1}{\eta}\log\left(\sum_{k \in \cK}\sum_{i=1}^de^{\eta(z^k_i-\varepsilon)}+e^{-\eta(z^k_i-\varepsilon)}\right)$, ensures that
\[\E_{\sigma,\tau}\Big[\|\og_n\|_\infty\Big]\leq \varepsilon+4\eta+\frac{\log(2dK)}{\eta n}\leq \varepsilon+4\eta+\frac{\log(2d)+2d\log(1/\varepsilon)}{\eta n}.\]
Therefore, given $N \in \N$ such that $N \geq 2ed$, the choice of $\varepsilon/4=\eta=\sqrt{\frac{d}{8N}\log\left(\frac{N}{2d}\right)}$ ensures in particular that
\[\E_{\sigma,\tau}\Big[\|\og_N\|_\infty\Big]\leq  6\sqrt{\frac{d}{N}\log\left(\frac{N}{2d}\right)}.\]
As usual,  when playing by blocks of increasing size $2^m$ (starting at $\underline{m}$ such that $2^{\underline{m}}\geq ed$), the last two displays ensure that, for every $n \in \N$,
\[\E_{\sigma,\tau}\Big[\|\og_n\|_\infty\Big]\leq  12\sqrt{2\frac{d}{n}\log\left(\frac{2n}{d}\right)}.\]
The result comes from the fact that, by construction, for every $n \in \N$,
\[ \sup_{p \in \Delta(\Omega), \lambda >0} \frac{|\N_n[p,\lambda]|}{n} \bigg\|\omega_n[p,\lambda]-\op_n[p,\lambda] \bigg\|_\infty=\|\og_n\|_\infty.
\]
\qed

If $d \geq 3$, since $\cL_\varepsilon\leq \varepsilon^{2d}/d!$, constants in  Theorem~\ref{TH:CalibViaAppro}  can be lowered if one is only interested in the asymptotic behavior. This result holds almost surely since, using concentration inequalities,  with $\P_{\sigma,\tau}$ probability at least $1-\delta$,
\[ \big\|\og_n\big\|_\infty \leq 12\sqrt{2\frac{d}{n}\log\left(\frac{2n}{d}\right)} + 2\sqrt{\frac{2d}{n}\log\left(\frac{2n}{d}\right)+\frac{1}{n}\log\left(\frac{1}{\delta}\right)}.\]
Statement concern $\ell_\infty$ balls; however, it is also possible to show that for other $\ell_p$-balls, the number of possible intersection with the grid is bounded by $O\left(\frac{1}{\varepsilon}\right)^{O(\Omega^2)}$ (see e.g. Rakhlin, Sridharan \& Tewari~\cite{RakSriTew11}). Thus the results holds, up to some polynomial term in $\Omega$, for any other $\ell_p$-norm.

This technique could actually have been used to proved Theorem~\ref{TH:ManSto}, a similar result with respect to the family of Borel sets. The difference is that the number of possible intersection between Borel sets and our grid would have been in the order of  $2^{1/\varepsilon^d}$. After taking the logarithm, equalizing the three remaining terms in regret  $\varepsilon$, $\eta$ and $1/(\varepsilon^{d}\eta n)$ yields that $\varepsilon=\eta=n^{-1/(d+2)}$. This would have been the bound on expected regret.

\bigskip

We now turn to calibration with checking rules and smooth calibration, and we show that they can be reduced to  approachability problems. We recall that given a pair of mappings $\cU$ and $\cT$, we defined
\[\N_n[\cU,\cT]= \Big\{m \leq n \ \text{s.t.}\ (p_m,\omega_m) \in \cU(h^{m-1}) \Big\} \ ,\]
the empirical probability of tested events
\[ \overline{\omega}_n[\cU,\cT]=\frac{\sum_{m \in \N_n[\cU,\cT]}\mathds{1}\{(p_{m},\omega_m) \in \cT(h^{m-1})\} }{|\N_n[\cU,\cT]|}\ ,\]
and the average predicted conditional probability of tested events
\[\op_n[\cU,\cT]=\frac{\sum_{m \in \N_n[\cU,\cT]} p_m\Big\{(p_{m},\omega_m) \in \cT(h^{m-1})\, \Big| \, \cU(h^{m-1}) \Big\}}{|\N_n[\cU,\cT]|}\ .\]
If a checking rule is independent of current predictions, then the same definition hold with $(p_m,\omega_m)\in \cU(h^n)$ (resp.\ in $\cT(h^n)$) replaced by  $\omega_m \in \cU(h^n)$ (resp.\ in $\cT(h^n)$).

\begin{theorem}\label{TH:DetCalChe}
Let $\lambda$ be a probability distribution on the set of checking-rules independent of current predictions. Then there exists a deterministic strategy $\sigma$ that is calibrated with $\lambda$-almost every checking rules such that,  $\P_{\sigma,\tau}$-almost surely,
\[ \limsup_{n \to \infty}\Big| \overline{\omega}_n[\cU,\cT] - \op_n[\cU,\cT] \Big| \leq 0,
\]
as soon as $|\N_n[\cU,\cT]|$ increases to infinity.
\end{theorem}
\textbf{Proof:}  Proof relies essentially on  approachability with activation in infinite dimension. We define an auxiliary game where payoff is a random variable over the set of checking rules independent of current predictions. Action set of the player is  reduced to $\Delta(\Omega)_0$, the interior of $\Delta(\Omega)$ -- so that conditional probabilities are well defined -- and payoff at stage $n$ is $\mathds{1}\{\omega_n \in \cT(h^{n-1})\}-p\{ \cT(h^{n-1}) |  \cU(h^{n-1})\}$ if the coordinates $(\cU,\cT)$ is active, i.e., if $n \in \N_n[\cU,\cT]$.

By definition, average payoff at stage $n$ is exactly $\overline{\omega}_n[\cU,\cT]-\op_n[\cU,\cT]$ and  we shall construct a strategy $\sigma$ that approaches the convex set $\{0\}$, that is, using Theorem~\ref{TH:AppActiv}, find $p_{n+1} \in \Delta(\Omega)_0$ such that, for every $\omega \in \Omega$,
\[ \int  \Big(\overline{\omega}_n[\cU,\cT]-\op_n[\cU,\cT] \Big).\frac{\Big(\mathds{1}\{\omega \in \cU(h^n)\} \Big(\mathds{1}\{\omega \in \cT(h^n)\} - p_{n+1}\left\{\cT(h^n) | \cU(h^n)\right\} \Big)}{|\N_{n+1}[\cU,\cT]|}d \lambda
\]
is less or equal to zero (or at least smaller than $\varepsilon_n=1/n^2$).

To construct this $p_{n+1}$, we consider the game with payoff   defined on $\Delta(\Omega)_0$ and $\Omega$  by
\[ g(p,\omega)= \int \frac{\overline{\omega}_n[\cU,\cT]-\op_n[\cU,\cT]}{1+|\N_n[\cU,\cT]|}\mathds{1}\{\omega \in \cU(h^n)\} \Big(\mathds{1}\{\omega \in \cT(h^n)\} - p\left\{\cT(h^n) | \cU(h^n)\right\} \Big)d \lambda
\]
and $g$ is extended linearly in its second variable on $\Delta(\Omega)$. Since one always has
\[ \mathds{1}\{\omega \in \cU(h^n)\} \frac{\overline{\omega}_n[\cU,\cT]-\op_n[\cU,\cT]}{1+|\N_n[\cU,\cT]|}=\mathds{1}\{\omega \in \cU(h^n)\} \frac{\overline{\omega}_n[\cU,\cT]-\op_n[\cU,\cT]}{|\N_{n+1}[\cU,\cT]|}
\]
the integrals in the last two displayed equations coincide, so we just need to prove that there exists $p_{n+1} \in \Delta(\Omega)_0$ such that $g(p_{n+1},\omega)\leq \varepsilon_n$, for every $\omega \in \Omega$ or, more generally, that
\[ \inf_{p \in \Delta(\Omega)_0} \sup_{\omega \in \Omega} g(p,\omega) \leq 0.
\]
And this is a consequence of Lemma~\ref{LM:Fan}, since $g(p,p)=0$ for every $p \in \Delta(\Omega)_0$, $g(p,\cdot)$ is affine and $g(\cdot,p)$ is continuous on $\Delta(\Omega)_0$.
\qed

When checking rules might depend on current predictions (see Sandroni, Smorodinsky \& Vohra~\cite{SanSmoVoh03} or Foster, Rakhlin, Sridharan \& Tewari~\cite{FosRakSri11}), the result and proof  are almost identical.
\begin{proposition}
Let $\lambda$ be a probability distribution on the set of checking-rules. Then there exists a  strategy $\sigma$ that is calibrated with $\lambda$-almost every checking rules such that,  $\P_{\sigma,\tau}$-almost surely,
\[ \limsup_{n \to \infty}\Big| \overline{\omega}_n[\cU,\cT] - \op_n[\cU,\cT] \Big| \leq 0,
\]
as soon as $|\N_n[\cU,\cT]|$ increases to infinity.
\end{proposition}
\text{Proof:} the proof is almost identical to the case of checking rule independent of predictions. The only difference lies in the definition of the payoff $g(p,\omega)$ which is
\[  \int \frac{\overline{\omega}_n[\cU,\cT]-\op_n[\cU,\cT]}{1+|\N_n[\cU,\cT]|}\mathds{1}\{(\omega,p) \in \cU(h^n)\} \Big(\mathds{1}\{(\omega,p) \in \cT(h^n)\} - p\left\{\cT(h^n) | \cU(h^n)\right\} \Big)d \lambda.
\]
Since $g(\cdot,\omega)$ might not be continuous, Lemma~\ref{LM:Fan} does not apply. However, $g$ is bounded and defined over $\Delta(\Omega)_0$ and $\Omega$, the former being measurable and the latter finite. Therefore, see Sorin~\cite{Sor02} Theorem A.9, this game has a value in mixed action. And this value has to be smaller than 0 since $g(p,p)=0$ for every $p\in \Delta(\Omega)_0$. \qed

The last similar reduction to approachability concerns smooth calibration.

\begin{theorem}
There exists a deterministic strategy $\sigma$ of the player such that, no matter  Nature's strategy,  for every continuous mapping $g:\Delta(\Omega) \to \R_+$,
\[ \limsup_{n \to \infty}  \frac{1}{n}\left\| \sum_{m=1}^n g(p_m) (\omega_m-p_m) \right\| \leq 0\ .\]

The same result holds if one adds checking rules independent of current predictions.\end{theorem}
\textbf{Proof:} The set of  continuous mappings from $\Delta(\Omega)$ to $\R_+$  is separable  and   we denote by   $\lambda$ a probability distribution with support   $\{ g_k ; k \in \N \}$,  a dense countable family.

Following the lines of the proof of Therorem~\ref{TH:DetCalChe}, we define
\[\overline{\omega}_n[g_k] = \frac{\sum_{m=1}^n g_k(p_m) \omega_m}{n}  \ \text{and} \ \op_n[g_k] = \frac{\sum_{m=1}^n g_k(p_m) p_m}{n} .
\]
Then, Corollary~\ref{TH:AppInfDim} ensures the existence of  an approachability strategy such that, for every $k \in \N$, $\big\|\overline{\omega}_n[g_k]-\op_n[g_k]\big\|$ converges to zero. Indeed, one just has to prove that $\{0\} \subset \cL_2$ is approachable,  thus that for every $n \in \N$, there exists $p_{n+1} \in \Delta(\Omega)$ such that, no matter $\omega \in \Omega$,
\[ \int \Big(\overline{\omega}_n[g_k]-\op_n[g_k]\Big) \Big(g_k[p_{n+1}](\omega-p_{n+1})\Big) d\lambda \leq 0,
\] where we assumed that $0/0=0$. The existence of such $p \in \Delta(\Omega)$ is again a  consequence of Ky Fan's inequality generalized in Lemma~\ref{LM:PsConv}.

Since $\{g_k ; k \in \N\}$ is a dense family, necessarily $\big\|\overline{\omega}_n[g]-\op_n[g]\big\|$ must converges to zero, for every continuous mapping $g$.
\qed

A close look to the first proof of existence of deterministic smooth calibrated strategies, due to  Kakade \& Foster~\cite{KakFos04}, shows that they  also have constructed  an $\varepsilon$-approachability strategy (and then used a doubling trick). We proposed here a direct (and maybe more intuitive) proof.

\subsection{Using calibration to get regret and approachability}

Calibration in some auxiliary game can be seen as a useful tool to construct strategies that satisfies another criterion as approachability, no internal regret and so on. This idea goes back to Foster \& Vohra~\cite{FosVoh97} and was used,  recently,  by Perchet~\cite{Per09,Per11a,Per11b}; in particular, it is useful in a specific case of general regret (see Section~\ref{SE:GenReg}) defined below.

\medskip

But first, we focus on  usual  internal regret in the finite case (although it can be generalized immediately when $\cB$ is any compact set). Recall that a strategy is internally consistent if the supremum limit of 
\[  \frac{|\N_n[a]|}{n}\left( \max_{a^* \in \cA} \rho\big(a^*,\ob_n[a]\big)-\rho\big(a,\ob_n[a]\big)\right)
\]
is non positive. By linearity of $\rho(a,\cdot)$, this quantity can be immediately  rewritten into 
\[  \frac{|\N_n[a]|}{n}\left( \left(\big\|\rho\big(a,\cdot\big)-\ob_n[a]\big\|^2-\|\rho(a,\cdot)\|^2\right)-  \min_{a^* \in \cA}  \left(\big\|\rho\big(a^*,\cdot\big)-\ob_n[a]\big\|^2-\|\rho(a^*,\cdot)\|^2\right)\right),\]
up to a factor 2. As a consequence, any weighted-calibrated strategy with respect to $\{\rho(a,\cdot), \|\rho(a,\cdot)\|^2; a \in \cA\}$ is internally consistent. Since scores are actually exactly the same,  rates of convergence of weighted calibration give rates for regret minimization.

\bigskip

We now turn to generalized regret. Assume that $\cA$ and $\cB$ are two compact and convex sets and let  $G : \cA \times \cB \to \R$ be any fixed \textsl{evaluation} mapping that might not be linear in any of its coordinates.  In this framework, a strategy has no $G$-external regret if
\[ \limsup_{n \to \infty} \sup_{a^* \in \cA} G(a^*,\ob_n)-G(\oa_n,\ob_n) \leq 0.
\]
To define internal regret, assume that a strategy  only uses a finite number of actions in $\cA_\cL=\{ a[\ell] ; \ell \in \cL\}$, so that $\sigma$ is actually a mapping from the set of finite histories into $\cL$, and $\ell_n=\ell$ means that action $a[\ell]$ is played at stage $n$. Define
\[ \N_n[\ell] = \{ m \leq n\ \text{s.t.}\ \ell_m=\ell\}, \  \text{and}\ \ob_n[\ell]=\frac{1}{|\N_n[\ell]|}\sum_{m\in\N_n[\ell]}b_m .
\]
A strategy has no $(\cL,\varepsilon)$-internal regret if, no matter the strategy $\tau$ of Nature, $\P_{\sigma,\tau}$-almost surely,
\[ \limsup_{n \to \infty} \frac{|\N_n[\ell]|}{n}\bigg(\sup_{a^* \in \cA} G\big(a^*,\ob_n[\ell]\big)-G\big(a[\ell],\ob_n[\ell]\big)-\varepsilon\bigg)  \leq 0, \quad \forall \ell \in \cL.
\]
\begin{proposition} If $G$ is continuous, then for every $\varepsilon>0$ there exists a $(\cL,\varepsilon)$-internally consistent strategy. However, their might not exist any ($\varepsilon$-)externally consistent strategies.
\end{proposition}
\textbf{Proof:} Since $G$ is continuous, for every $\varepsilon>0$, there exists some $\delta$ such that $\|(a,b)-(a',b')\|\leq \delta$ implies that $|G(a,b)-G(a',b')|\leq\varepsilon/2$. Consider $\sigma'$  any calibrated strategy with respect to $\Big\{b[\ell]; \ell \in \cL\Big\}$, a $\delta/2$ grid of $\cB$. Assume that when $\sigma'$ predicts $b[\ell]$, then $\sigma$ dictates to play $a[\ell] \in \argmax_{a \in \cA} G(a,b[\ell])$.

Since $\sigma'$ is calibrated, for every $\eta >0$,  one has that, $\P_{\sigma',\tau}$-as after some stage $N$,
\[\sup_{\ell  \in \cL}\frac{|\N_n[\ell]|}{n}\bigg(\Big\|b[\ell]-\ob_n[\ell]\Big\|^2 - \frac{\delta^2}{4} \bigg)\leq \eta.\]
In particular, as soon as  $\eta\leq \delta^2/4$, either $\eta\frac{n}{|\N_N[\ell]|}$ is smaller than $\delta^2/4$ and then $\|b[\ell]-\ob_n[\ell]\Big\|^2\leq \delta^2$, or $\frac{|\N_N[\ell]|}{n}$ is smaller than $4\eta/\delta^2$.

The first case implies  that $ G(a,b[\ell])-G(a,\ob_n[\ell]) \leq \varepsilon/2$ for every $a \in \cA$, thus in both cases one has that, after stage $N$,
\[\sup_{\ell  \in \cL}\frac{|\N_n[\ell]|}{n}\bigg(\sup_{a^*\in \cA}G(a^*,b[\ell])-G(a[\ell],\ob_n[\ell])-\varepsilon \bigg) \leq \frac{8 \|G\|_\infty}{\delta^2} \eta,
\]
which characterizes a $(\cL,\varepsilon)$-calibrated strategy.

\bigskip

It remains to prove that there might not exist externally consistent strategies. Define $G(a,b)=(1-4b)a$, for every $a \in [0,1]$ and $b\in[0,1]$ and  assume that during the first $N$ stages (with $N$ is large enough) $b_n=0$. Necessarily $\oa_N$ is arbitrarily close to $1$. During the next $N$ stage, define $b_n=1$  then $\oa_{2N}$ is at most $1/2$ thus the external regret is of at least $1/2$.
\qed

\bigskip

We now prove how to construct an $\varepsilon$-approachability strategy via calibration. Given a closed and compact set $\cC \subset \R^d$ and a vector payoff mapping $g : \Delta(\cA) \times \Delta(\cB) \to \R^d$, define $G(x,y) = - d_\cC ( g(x,y) ) $ for every $x \in \Delta(\cA)$ and $y \in \Delta(\cB)$. If $\cC$ is approachable, then Blackwell's condition ensures that $\sup_{x^* \in \Delta(\cA)} G(x^*,y)=0$ for every $y \in \Delta(\cB)$. By convexity of $d_\cC$ and the triangle inequality,
\begin{align*}
d_\cC(\og_n) & \leq \sum_{\ell \in \cL} \frac{|\N_n[\ell]|}{n}d_\cC(\og_n[\ell])\\
& \leq \varepsilon +\sum_{\ell \in \cL} \frac{|\N_n[\ell]|}{n} \bigg(d_\cC\big(g(a[\ell],\ob_n[\ell])\big)-\varepsilon\bigg)+\sum_{\ell \in \cL} \frac{|\N_n[\ell]|}{n} \bigg\|g(x[\ell],\ob_n[\ell])-\og_n[\ell]\bigg\|.
\end{align*}
Both sums converges almost surely to zero, respectively because $\sigma$ has no internal regret (with respect to $G$) and because of concentration inequalities since $g(x[\ell],b_n)=\E[g(a_n,b_n)]$. One can resort to the doubling trick (since we can easily derive uniform speed of convergence) to get an approachability strategy.

\subsection{Using regret to get  approachability}

We proved in the last section how calibration and generalized regret can be used to construct approachability strategy, as noticed by Perchet~\cite{Per09} or Rakhlin, Sridharan \& Tewari~\cite{RakSriTew11}. A completely different link  can also be formulated between regret and approachability, as discovered recently by Abernathy, Bartlett \& Hazan~\cite{AbeBarHaz11}. We recall that Blackwell's strategy consists in playing, at stage $n+1$, optimally in the zero-sum projected game $\langle g(x,y) - \pi_\cC(\og_n) , \og_n-\pi_\cC(\og_n) \rangle$. Abernathy, Bartlett \& Hazan~\cite{AbeBarHaz11} proposed to use a regret minimization scheme to determine, stage by stage, in which projected game to play (i.e., not necessarily along the direction $\og_n-\pi_\cC(\og_n)$).

\medskip

The formulation is rather simple when $\cC=\{0\} \subset \R^d$, so we will focus only on this case. It can however be  generalized to any convex cone and therefore to any convex set in $\R^d$ (seen as a section of a convex cone in $\R^{d+1}$). The basic idea is to notice that, for $\cC=\{0\}$ and every $n \in \N$,
\[d_\cC(\og_n)=\|\og_n\|=\sup_{\theta \in B(0,1)} \langle \theta , \og_n \rangle, \ \text{where}\ B(0,1)=\Big\{\theta \in \R^d, \|\theta\|_2\leq 1\Big\}.\]
Assume that at stage $m$, the player played optimality in the projected game along the direction $\theta_{m-1}$. Since $\cC$ is approachable, this zero-sum game has a negative value, hence $\langle \theta_{m-1}, \E[g_{m}] \rangle \leq 0$.  As a consequence,
\[ \E\Big[d_\cC(\og_n)\Big] = \E\Big[\|\og_n\|\Big]\leq \E \left[ \sup_{\theta \in B(0,1)} \langle \theta, \og_n\rangle - \frac{1}{n}\sum_{m=1}^n \langle \theta_{m-1}, g_{m}\rangle\right].
\]
The term inside the expectation can be written as the external regret if player and Nature's action set are respectively $B(0,1)$  and $\Big\{g(a,b) ; (a,b) \in \cA \times \cB \Big\}$. As a consequence, an approachability strategy can indeed be described as a two step procedure.  At any stage $n$, choose, in a first step, a direction $\theta_n \in B(0,1)$ following any regret minimization algorithm.  Then, in a second step, play optimally in the projected zero-sum game on $\theta_n$.

\medskip

Blackwell's strategy dictates to choose (in the first step) the direction $\theta_n$ that maximizes $\langle \theta, \og_n \rangle$ ; in other words, this is precisely the \textsl{follow the leader} algorithm that does not guarantee a  shrinking regret (in full generality). The key point to understand this feature is that, by definition of the second step, $\langle \theta_{m}, \E[g_{m+1}] \rangle$ is always non-positive (no-matter the choice of $\theta_{m}$) ; so, in this auxiliary game, Nature is in fact very restricted on her choice of actions and what is even more intricate, these restrictions depend on the player's move.

 \section{Appendix}

\subsection{Game Theory lemma}
The following Lemma generalizes  Ky Fan's inequality~\cite{Fan72} recalled below:

Let $\cK$ be a convex compact set of some Euclidian space and $g : \cK \times \cK \to \R$ such that, for every $y \in \cK$, $g(\cdot,y)$ is concave over $\cK$ and for every $x \in \cK$, $g(x,\cdot)$ is continuous over $K$. If $g(x,x)=0$ for every $x \in \cK$, then there exists $x_0 \in \cK$ such that $\sup_{x \in \cK} g(x,x_0) \leq 0$.
\begin{lemma}\label{LM:Fan} Let $g$ be a mapping on some  compact and convex  set $\cX \subset \R^d$ such that $g(x,x)=0$ for every $x \in \cX_0$ the interior of $\cX$ (such a mapping is called anti-symmetric).

If for every $x \in \cX$, $g(\cdot, x)$ is concave  and $g(x,\cdot)$ is continuous and uniformly bounded by some $M >0$ on $X_0$, then
\[ \inf_{x \in \cX_0} \sup_{x' \in \cX} g(x',x) \leq 0.
\]
\end{lemma}
\textbf{Proof:}
Without loss of generality, we assume that 0 belongs to $\cX_0$ and we denote, for every $\varepsilon>0$ small enough, the convex compact set $\cX_\varepsilon:= \big\{ (1-\varepsilon)x ; x \in \cX \big\}$. Then $g$ and $\cX_\varepsilon$ satisfy assumptions of Ky Fan's inequality. Thus, there exists $x_\varepsilon$ such that $g(x,x_\varepsilon)\leq0$ for every $x \in \cX_\varepsilon$.

Given $x \in \cX$, we denote by $x_{-}$ the point on the boundary of $\cX$ on the opposite direction of $x$, i.e., such that $\frac{x_-}{\|x_-\|}=-\frac{x}{\|x\|}$. We also define $\|\cX_-\|=\inf_{x \in \cX} \|x\|$.

 Since $g(\cdot,x_\varepsilon)$ is concave, for every $x$ in $\cX$ that is not in $\cX_\varepsilon$, one has
\[ \frac{g\big(x,x_\varepsilon\big)-g\big((1-\varepsilon)x,x_\varepsilon\big)}{\varepsilon\|x\|} \leq \frac{g\big((1-\varepsilon)x,x_\varepsilon\big)-g\big(x_-,x_\varepsilon\big)}{(1-\varepsilon)\|x\|+\|x_-\|},
\]
therefore, since $(1-\varepsilon)x\in \cX_\varepsilon$ and $g(\cdot,x_\varepsilon)\leq 0$ on $\cX_\varepsilon$, one has
\[g\big(x,x_\varepsilon\big) \leq -\frac{\varepsilon\|x\|}{(1-\varepsilon)\|x\|+\|x_-\|}g(x_-,x_\varepsilon)\leq \varepsilon\frac{M\|\cX\|}{(2-\varepsilon) \|\cX_-\|}.
\]
Hence the result, since the right hand term goes to $0$ as $\varepsilon$ decreases to $0$.
\qed
\subsubsection{Uniform concentration inequalities}
The following lemmas are central in different proofs.  We recall that a process $Z_t \in \R^d$ is a  martingale difference sequence if $\E\big[Z_{t+1}\big|Z_1,\dots,Z_t\big]=0$. Moreover, if  $  \|Z_t\|_2 \leq K$ then  Hoeffding-Azuma's inequality in Euclidian spaces (see Corollary 3.5 in Kallenberg \& Sztencel~\cite{KalSzt91}) yields  that, for every integer $T \ge 1$,
\begin{equation}
\label{EQ:hoeffdingInfDim}
\P\left\{ \left\|\oZ_T\right\| \geq \varepsilon \right\} \leq \left(1+ \sqrt{\frac{T}{K^2}}\varepsilon\right) \exp\left(-\frac{1}{2}\frac{T}{K^2}\varepsilon^2\right)\leq 2 \exp\left(-\frac{1}{4}\frac{T}{K^2}\varepsilon^2\right),
\end{equation}
or, $\P\left\{ \left\|\oZ_T \right\| \geq \phi^{-1}(\delta)/\sqrt{T} \right\} \leq \delta$
with $\phi(x):=(1+x/K)\exp\left(-x^2/2K^2\right)$. Actually, a weak maximal version of this inequality holds:
\[\P\left\{\exists\, t \leq T\,,\  \left\| \oZ_t \right\| \geq \frac{T}{t} \varepsilon \right\} \leq \phi\left(\sqrt{T}\varepsilon\right) \quad \mbox{or} \quad \P\left\{\exists\, t \leq T\,,\  \left\|\oZ_t \right\| \geq \frac{\sqrt{T}}{t} \phi^{-1}(\delta) \right\} \leq \delta.\]
For $d=1$, one can define $\phi(x)=2\exp\left(-x^2/2\right)$ and  $\phi(x)=2\exp\left(-x^2/4\right)$ otherwise.

Stronger maximal inequalities for averages of martingale differences exist:\begin{lemma}\label{LM:peelingDebut}
Let $Z_t$ be  a martingale difference sequence with $ \|Z_t\| \leq K$ then, for every $\delta>0$ and every integer $T \ge 1$,
$$\P\left\{\exists\ t \leq T,\ \left\|\oZ_t\right\| \geq \frac{2}{\sqrt{t}}\phi^{-1}\left(\frac{\delta}{4}\frac{t}{T}\right) \right\} \leq \delta.$$
\end{lemma}
\textbf{Proof:} Define $\varepsilon_t=2\phi^{-1}\left(\delta t / 4T\right)/\sqrt{t} $.
Using a peeling argument, one obtains
\begin{align*}
\P\left\{\exists\ t \leq T,\ \|\oZ_t\| \geq \varepsilon_t\right\}  &\leq \sum_{m=1}^{\lfloor \log_2(T)\rfloor} \P\Big\{\bigcup_{ t =2^m}^{ 2^{m+1}-1}\{ \|\oZ_t\| \geq \varepsilon_t\}\Big\} \\
&\leq\sum_{m=1}^{\lfloor \log_2(T)\rfloor} \P\Big\{\bigcup_{ t =2^m}^{ 2^{m+1}}\{  \|\oZ_t\| \geq \varepsilon_{2^{m+1}}\}\Big\}\\
&  \leq \sum_{m=1}^{\lfloor \log_2(T)\rfloor} \P\Big\{\bigcup_{ t =2^m}^{ 2^{m+1}}\{  t\|\oZ_t\| \geq 2^m\varepsilon_{2^{m+1}}\}\Big\}&\\
& \leq \sum_{m=1}^{\lfloor \log_2(T)\rfloor} \P\Big\{\bigcup_{ t =2^m}^{ 2^{m+1}}\{  t\|\oZ_t\| \geq \sqrt{2^{m+1}}\phi^{-1}\left(\frac{\delta}{4}\frac{2^{m+1}}{T}\right)\}\Big\}\\
& \leq \sum_{m=1}^{\lfloor \log(T)\rfloor} \frac{2^{m+1}}{T}\frac{\delta}{4}\le \frac{2^{\log_2(T)+2}}{T}\frac{\delta}{4}\le \delta.&
\end{align*}
Hence the result.\qed

\bigskip

Similarly, maximal inequalities can  be derived for  tail events:
\begin{lemma}\label{LM:peelingFin}
Let $Z_t \in \R^d$ be  a martingale difference sequence with $ \|Z_t\| \leq K$ then, for every $\varepsilon>0$ and every integer $T \ge 1$,
$$\P\Big\{\exists\ t \geq T,\ \left\|\oZ_t\right\| \geq \varepsilon \Big\} \leq 4\exp\left(-\frac{T\varepsilon^2}{8K^2}\right).$$

The exponential  dependency in $T$  can be reduced since one has, as soon as $\frac{T\varepsilon^2}{2K^2}\geq1$,
$$\P\Big\{\exists\ t \geq T+1,\ \left\|\oZ_t\right\| \geq \varepsilon \Big\} \leq \frac{2K^2}{\varepsilon^2}\left(1+\sqrt{\frac{T\varepsilon^2}{2K^2}}+\frac{1}{\sqrt{\frac{T\varepsilon^2}{2K^2}}} \right)\exp\left(-\frac{T\varepsilon^2}{2K^2}\right).$$
\end{lemma}
\textbf{Proof:}
Again, using a peeling argument, one obtains
\begin{align*}
\P&\left\{\exists\ t \geq T,\ \|\oZ_t\| \geq \varepsilon\right\}  \leq \sum_{m=0}^{+\infty} \P\Big\{\bigcup_{ t =2^mT}^{ 2^{m+1}T-1}\{ \|\oZ_t\| \geq \varepsilon\}\Big\} \\
&\leq\sum_{m=0}^{+\infty} \P\Big\{\bigcup_{ t =2^mT}^{ 2^{m+1}T}\{  t\|\oZ_t\| \geq 2^mT\varepsilon\} \Big\}  \leq \sum_{m=0}^{+\infty} \phi\left(\sqrt{2^{m+1}T}\frac{\varepsilon}{2}\right) \leq \int_0^\infty \phi\left(\sqrt{2^{x}T}\frac{\varepsilon}{2}\right)dx&\\
&=\frac{1}{\log(2)}\int_{\frac{T\varepsilon^2}{8K^2}}^\infty \frac{\phi(K\sqrt{2v})}{v}dv=\frac{1}{\log(2)}\int_{\frac{T\varepsilon^2}{8K^2}}^\infty \frac{(1+\sqrt{2v})}{v}e^{-v}dv&
\end{align*}
hence
$$\P\Big\{\exists\ t \geq T,\ \left\|\oZ_t\right\| \geq \varepsilon \Big\} \leq \frac{1}{\log(2)}\left(\frac{1}{\frac{T\varepsilon^2}{8K^2}}+\frac{\sqrt{2}}{\sqrt{\frac{T\varepsilon^2}{8K^2}}}\right)\exp\left(-\frac{T\varepsilon^2}{8K^2}\right)$$
and the first part of the result follows.
\medskip

The second part of the proof follows from the facts that
\begin{align*}
\P\left\{\exists\ t \geq T+1,\ \|\oZ_t\| \geq \varepsilon\right\}  &\leq \sum_{t=T+1}^{+\infty} \P\Big\{ \|\oZ_t\| \geq \varepsilon\Big\}\\ &
 \leq \sum_{t=T+1}^{+\infty}  \left(1+ \sqrt{\frac{t}{K^2}}\varepsilon\right) \exp\left(-\frac{1}{2}\frac{t}{K^2}\varepsilon^2\right)\\
&\leq \int_{T}^{\infty}\left(1+\sqrt{\frac{x\varepsilon^2}{K^2}}\right)\exp\left(-\frac{x\varepsilon^2}{2K^2}\right)dx\\ & \leq  \frac{2K^2}{\varepsilon^2}\int_{\sqrt{\frac{T\varepsilon^2}{K^2}}}^\infty (u+u^2)\exp(-u^2/2)du,\end{align*}
and
\[\int_x^\infty (u+u^2)e^{-\frac{u^2}{2}}du  = (1+x)e^{-\frac{x^2}{2}}+\int_x^\infty e^{-\frac{u^2}{2}}du \leq \left(1+ x+\frac{1}{x}\right)e^{-\frac{x^2}{2}}.\]
\qed

\subsection{Probability lemmas}
\begin{lemma}\label{LM:PsConv}
Let $f_n \in \LT$ such that  $\sum_{n \in \N} \|f_n\|^2/n < \infty$ and $\|f_{n+1}-f_n\| \leq \frac{1}{n}$, then $f_n$ converges to 0, $\mu$-as.
\end{lemma}
\textbf{Proof:} First, we prove a weaker version when $\|f_n\| <1/\sqrt{n}$ for every $n \in \N$.

Let $M_n=\lceil n^{6/5}\rceil$ and $k_n$ be the integer  minimizing $\|f_k\|$ over $[M_n+1, M_{n+1}]$. Then,
\begin{align*}\|f_{k_n}\|^2 &\leq \frac{\sum_{k=M_n+1}^{M_{n+1}} \|f_k\|^2}{M_{n+1}-M_n}  \leq \frac{M_{n+1}}{M_{n+1}-M_n}  \sum_{k=M_n+1}^{M_{n+1}}\frac{\|f_k\|^2}{k}\\ & \leq \sum_{k=M_n+1}^{M_{n+1}}2n\frac{\|f_k\|^2}{k}\leq  2 \sum_{k=M_n+1}^{M_{n+1}}\frac{\|f_k\|^2}{k^{1/6}} \leq 2 \sum_{k=M_n+1}^{M_{n+1}} \frac{1}{k^{6/5}}.
\end{align*}
Therefore, $\sum_{n \in \N} \|f_{k_n}\| < \infty$ and Fatou's lemma ensures that $f_{k_n}$ converges to 0, $\mu$-as.

Let us define $h_k=f_k-f_{k_n}$, then for evert $k> k_n$ (and similarly for $k<k_n$)
\[ \|h_k\|= \left\|\sum_{j=k_n+1}^k (f_j-f_{j-1})\right\| \leq \frac{k-k_n}{M_n}\leq \frac{M_{n+1}-M_n}{M_n}.
\]
Summing over $k$, one gets
\[ \sum_{k \in \N}\|h_k\|^2 \leq \sum_{n \in \N}(M_{n+1}-M_n) \left(\frac{M_{n+1}-M_n}{M_n}\right)^2 < \infty \quad \text{since} \quad \frac{M_{n+1}-M_n}{M_n} \leq O\left(\frac{1}{n}\right).
\]
So both $h_k$ and $f_{k_n}$ converge $\mu$-as to 0 and thus, so is $f_k=h_k+f_{k_n}$.

\medskip

In the general case, one just need to notice that there exists an increasing sequence $\beta_n >1$ such that $\sum_{n \in \N} \beta_n\|f_n\|^2/n < \infty$ and to define $M_{n+1} = \left\lceil\frac{\beta_{M_n}}{\beta_{M_n}-1}\right\rceil M_n+1$. The proof follows as before, since $M_{n+1}-M_n \sim \frac{M_n}{\beta_{M_n}-1}$ and thus $M_{n+1}/(M_{n+1}-M_n) \sim \beta_{M_n}$
\qed

\begin{lemma}\label{LM:AppInfDimAct}
Let $\cC$ be a product set in $\LT$ and assume that for every $n \in \N$
\begin{itemize}
\item[i)]  $g_n\in \LT$ is bounded by $B \in \LT$
\item[ii)] $\Chi_n \in \LT$ takes value in $\{0,1\}$
\item[iii)] $\og_{\Chi,n}=\sum_{k=1}^n \Chi_m g_m/S_n$ where $\cS_n=\sum_{m=1}^n\Chi_m$
\item[iv)] $\displaystyle \left\langle \Chi_{n+1}(\og_{\Chi,n}-\Pi_\cC(\og_{\Chi,n}), \frac{g_{n+1} - \Pi_{\cC}(\og_{\Chi,n})}{\cS_{n+1}} \right\rangle \leq \varepsilon_n$, for a sequence  of non-negative $\varepsilon_n$ such that $\sum_{n \in \N} \varepsilon_n < \infty$.

\end{itemize}
Then $\og_{\Chi,n}$ converges to $\cC$, $\mu_\infty$-as, where $\mu_\infty(A)=\mu\left(A\cap\{\lim \cS_n= \infty\}\right)$.
\end{lemma}
\textbf{Proof:}
Let $f_n=\overline{g}_{n}-\Pi_\cC(\overline{g}_{n})$ then point  \textsl{iv)} implies that:
\[\left\|f_{n+1}\right\|^2\leq \left\|f_n\right\|^2-2\left\langle\mathcal{X}_{n+1}\frac{f_n}{\overline{\mathcal{X}}_{n+1}},f_n\right\rangle+\left\|\frac{\mathcal{X}_{n+1}}{\cS_{n+1}}\left(g_{n+1}-\overline{g}_{n}\right)\right\|^2+\varepsilon_n.\]
Since both $g_{\Chi,n+1}$ and $\overline{g}_{\Chi,n}$ are bounded by  $B$ and $\mathcal{X}_n \in \{0,1\}$, one has:
\[2\left\langle\mathcal{X}_{n+1}\frac{f_n}{\cS_{n+1}},f_n\right\rangle \leq\left\|f_{n}\right\|^2- \left\|f_{n+1}\right\|^2+4\int_{\Omega}\frac{\mathcal{X}_{n+1}(\omega)}{\cS^2_{n+1}(\omega)}B^2(\omega)d\mu(\omega)+\varepsilon.\]
Notice that, for every $\omega$,  $\sum_{n \in \mathds{N}}\mathcal{X}_{n+1}(\omega)/\cS^2_{n+1}(\omega)\leq \sum_{n \geq 1} 1/n^2=\pi^2/6$. For every $n \in \N$ and $\omega \in \Omega$, let   $j(n,\omega):=\inf\left\{m \in \mathds{N}, \cS_m(\omega)=n\right\}$ be the first time such that $\cS_m(\omega)$ is bigger than $n$ (if it exists, otherwise it is  $\infty$). Define $\widetilde{f}_n(\omega):=f_{j(n,\omega)}(\omega)$, with $f_\infty(\omega)=0$, so that   \[\sum_{n \in \mathds{N}}\frac{\left\|\widetilde{f}_n\right\|^2}{n+1}=\sum_{n \in \mathds{N}}\left\langle\mathcal{X}_{n+1}\frac{f_{n}}{\cS_{n+1}},f_n\right\rangle\leq \frac{\|f_0\|^2}{2}+\frac{\pi^2}{3}\|B\|^2 + \sum_{n \in \N}\varepsilon_n.\]

Since $\cC$ is a product set, projection on $\cC$ is a coordinate-wise projection, thus
\begin{align*}\left|\widetilde{f}_{n+1}(\omega)-\widetilde{f}_n(\omega)\right|&=\left|\left[\overline{g}_{j_{n+1,\omega}}(\omega)-\overline{g}_{j_{n,\omega}}(\omega)\right]-\left[\Pi_C\left(\overline{g}_{j_{n+1,\omega}}\right)(\omega)-\Pi_C\left(\overline{g}_{j_{n,\omega}}\right)(\omega)\right]\right|\\
&\leq 2\left|\overline{g}_{j_{n+1,\omega}}(\omega)-\overline{g}_{j_{n,\omega}}(\omega)\right|= 2\left|\frac{g_{j_{n+1,\omega}}(\omega)-\overline{g}_{j_{n,\omega}}(\omega)}{n+1}\right|\leq \frac{4B(\omega)}{n+1}\end{align*}
and so $\|\widetilde{f}_{n+1}-\widetilde{f}_{n}\|^2=16\|B\|^2/(n+1)^2$.

Using  Lemma~\ref{LM:PsConv}, with $\beta_n$ an increasing sequence such that $\sum \beta_n \|\widetilde{f}\|^2/n$ converges  and $M_{n+1}=\left\lceil\frac{\beta_{M_n}}{\beta_{M_n}+1}\right\rceil$, we obtain the   $\mu$-a.s.\ convergence of $\widetilde{f}_n$. As a consequence,  after restriction to the event $\{\lim S_n = \infty\}$, $f_n$  converges $\mu$-as to zero.
\qed

Being a product set is used to bound $\left\|\widetilde{f}_{n+1}-\widetilde{f}_n\right\|$;  convexity of $\cC$ (nor actually its $\mathcal{L}_2$-boundedness, as defined by Lehrer~\cite{Leh02}) is not enough for this proof. The reason is that if we define  the mapping $\widetilde{g}_n \in \LT$ by $\widetilde{g}_n(\omega)=g_{j(n,\omega)}(\omega)$ and let $N:=j(n,\omega)$. Then, without the product property which induces a coordinate-wise projection, $\Pi_\cC\left(\widetilde{g}_n\right)(\omega)$ has no reason to be equal to  $\Pi_\cC\left(g_N\right)(\omega)$.

\begin{corollary}
Same results hold if $\cX_n \in [0,1]$ does not necessarily take values in $\{0,1\}$. \end{corollary}
\textbf{Proof:} Assume that $g_0=0$ and $\Chi_0=\cS_0=1$, then $\sum_{n=0}^\infty \Chi_n^2/\cS_n^2 $ is uniformly bounded. Indeed, define $k_n=\min\{ m\ \text{s.t.}\ S_m\geq n \}$, so that
\[\sum_{n=0}^\infty \frac{\chi_{n}^2}{\cS_n^2}-1=\sum_{n=1}^\infty \sum_{m=k_n}^{k_{n+1}-1} \frac{\chi_{n}^2}{\cS_n^2}\leq \sum_{n=1}^\infty  \frac{1}{n^2}\sum_{m=k_n}^{k_{n+1}-1}\chi_{n}^2\leq \sum_{n=1}^\infty  \frac{1}{n^2}\sum_{m=k_n}^{k_{n+1}-1}\chi_{n} \leq  \sum_{n=1}^\infty  \frac{1}{n^2}.
\]
Therefore, on $\{\lim S_n \to\infty  \}$, $\widetilde{g}_{\chi,n}=\sum_{m=0}^n\Chi_mg_m/\cS_m$ converges to $\cC$, and $\widetilde{g}_{\chi,n} - \og_{\Chi,n}$ converges to zero.

\qed

\bibliographystyle{plain}
\bibliography{biblio}
\end{document}